\documentclass[10pt,aps,pre,twocolumn]{revtex4-2}
\usepackage{hyperref}
\usepackage{graphicx, float}
\usepackage{amsfonts, amssymb, amsmath, wasysym, mathrsfs}
\usepackage{tabularx}
\usepackage{multirow}
\usepackage{xcolor}

\begin{document}

\title{Rogue waves from noise-induced modulational instability of a plane wave}

\author{D.\,S.~Agafontsev}
\affiliation{Shirshov Institute of Oceanology of RAS, Moscow, Russia}
\affiliation{Skolkovo Institute of Science and Technology, Moscow, Russia}
\affiliation{P.N. Lebedev Physical Institute of RAS, Moscow, Russia}

\begin{abstract}
In the framework of the one-dimensional nonlinear Schr{\"o}dinger equation (1D-NLSE) of the focusing type, the present paper studies numerically rogue waves (RWs) that emerge in the nonlinear stage of noise-induced modulational instability of a plane wave. 
For a large ensemble of simulations, all sufficiently large local maximums of the wavefield (i.e., RWs) are systematically collected and analyzed. 
It is shown that the frequency of occurrence of such maximums begins to increase from zero at the time, when the fourth-order moment of amplitude reaches its first (largest) local maximum, and then grows in an oscillatory manner approaching its asymptotic value at long time. 
Near the statistically stationary state, the 1D-NLSE generates a much larger number of RWs than a comparable linear system, but, in average, one RW affects a much smaller spatiotemporal area. 
The distribution of these RWs by maximum intensity represents an exponential-like function without noticeable changes in behavior, indicating a similar origin for RWs of significantly different amplitudes. 
The collected RWs are compared within a sufficiently large spatiotemporal window with nine exact solutions, of which two models reproduce RW dynamics much better than the others: a general collision of two solitons and a general collision of two Tajiri-Watanabe breathers; RWs are fitted with these collisions using pre-compiled databases of such solutions. 
The parameters of the two models turn out to be quite similar to the direct scattering transform spectrum of the whole wavefield, supporting a hypothesis that RWs may emerge due to synchronization of a few nonlinear modes (solitons or breathers) in presence of many other nonlinear modes of the wavefield.
\end{abstract}

\maketitle


\section{Introduction}
\label{Sec:Intro}

The phenomenon of rogue waves (RWs) -- unusually large waves that appear suddenly from moderate wave background -- has been intensively studied in the recent years. 
A number of mechanisms, both linear and nonlinear, have been suggested to explain their emergence, see e.g. reviews~\cite{kharif2003physical,dysthe2008oceanic,onorato2013rogue,dudley2019rogue}. 
One of these mechanisms is the appearance of RWs as a manifestation of breather-type solutions of the underlying nonlinear evolution equations~\cite{dysthe1999note,osborne2000nonlinear,osborne2010nonlinear,shrira2010makes}. 
A particular problem often considered in the context of RW research is the modulational instability (MI) of the plane wave solution within the framework of the one-dimensional nonlinear Schr{\"o}dinger equation (1D-NLSE) of the focusing type,
\begin{equation}
	i\psi_t + \frac{1}{2}\psi_{xx} + |\psi|^2 \psi = 0,
	\label{NLSE}
\end{equation}
which is widely applicable in different fields of physics ranging from nonlinear optics to hydrodynamics and Bose-Einstein condensates~\cite{kivshar2003optical,pelinovsky2008book,OsborneBook2010}. 
Several exact solutions of this equation, together with their various superpositions (collisions), have been suggested as candidates for RWs, including the Peregrine~\cite{peregrine1983water}, higher-order rational~\cite{akhmediev2009rogue}, Akhmediev~\cite{akhmediev1986modulation} and Kuznetsov~\cite{kuznetsov1977solitons} (also known as the Kuznetsov-Ma~\cite{kawata1978inverse,ma1979perturbed}) breathers. 

Note that, in the context of the 1D-NLSE~(\ref{NLSE}), RWs are commonly understood as sufficiently large local maximums of the wavefield simultaneously in space and time, together with the corresponding high-amplitude regions around them; the standard criterion is that the maximum RW intensity $\max|\psi|^{2}$ must exceed the average intensity $\langle|\psi|^{2}\rangle$ by $8$ times. 
This paper uses the same definition. 

The dynamics of breathers is not the only possible point of view on the nonlinear mechanisms of RW formation. 
The 1D-NLSE is integrable in terms of the inverse scattering transform (IST) method~\cite{zakharov1972exact,novikov1984theory}, as it allows transformation to the so-called \textit{scattering data}. 
This data includes the eigenvalue spectrum of a certain auxiliary linear system, which does not change during the time evolution in complete analogy with the Fourier spectrum in the linear waves theory. 
The eigenvalue spectrum consists of a discrete (breathers or solitons) and continuous (nonlinear dispersive waves) parts. 
Breathers are represented by discrete eigenvalues when a nonzero background (e.g., a plane wave) is present within a spatially non-localized wavefield. 
When the background is absent and the wavefield is localized, discrete eigenvalues correspond to solitons. 
By introducing a thermodynamic limit, solitons can also be used to describe spatially non-localized wavefields with the help of the \textit{soliton gas} concept, see review~\cite{suret2024soliton} and the references wherein. 

Thereby, a spatially extended wavefield, including RWs that emerge within it, can be represented as a large set of nonlinear modes -- nonlinear dispersive waves together with breathers or solitons, and in some cases, such as the noise-induced MI of a plane wave, in both ways as a collection of breathers~\cite{soto2016integrable,akhmediev2016breather,grinevich2018finite,grinevich2018exact,grinevich2019finite} or solitons~\cite{gelash2019bound,gelash2021solitonic}. 
Meanwhile, breather solutions themselves can be accurately approximated by certain combinations of solitons with carefully chosen parameters~\cite{agafontsev2024multisoliton}, suggesting that RWs as well can be understood as complex superpositions of solitons.
Importantly, solitons can dominate even in weakly nonlinear wave fields~\cite{agafontsev2023bound} if their velocity distribution is wide~\cite{gelash2018strongly}, which significantly expands the cases where the application of the soliton gas concept can be useful.

Note that there has been recently developed a method~\cite{randoux2016inverse} to examine the IST portraits of RWs by cutting them out of the wavefield, periodizing them, and then applying the direct scattering transform. 
While this procedure looks promising and its application to fundamental soliton and breather solutions led to correct results, the obtained IST portraits of RWs are difficult to interpret as they turned out to be very different from the IST spectrum of the whole wavefield. 

This paper studies numerically RWs that emerge in the nonlinear stage of noise-induced MI of the plane wave solution. 
For this purpose, all sufficiently large local maximums of the wavefield are systematically collected and analyzed for a large ensemble of MI simulations that differ from each other only in the random realization of the initial noise. 
From the earlier studies~\cite{agafontsev2015integrable,kraych2019statistical} it is known that, in the nonlinear stage of MI, the statistical functions, such as the moments of amplitude, Fourier spectrum, spatial correlation functions and distribution of intensity, perform damped oscillations with time around their asymptotic values. 
In this study, it is shown that the frequency of RW occurrence begins to increase from zero simultaneously with the fourth-order moment of amplitude $\kappa_{4} = \langle|\psi|^{4}\rangle/\langle|\psi|^{2}\rangle^{2}$ reaching its first (largest) local maximum $\kappa_{4}\approx 2.9$, and then grows in an oscillatory manner approaching its asymptotic value at long time. 
Thus, the first few local maximums of the fourth-order moment correspond, in fact, to a significantly less frequent occurrence of RWs compared to the statistically stationary state of MI (when $\kappa_{4}\approx 2$), presenting a counterexample to the common assertion that values of $\kappa_{4}$ greater than $2$ correspond to more frequent RWs. 

The study then focuses on RWs that appear near the statistically stationary state of MI. 
The 1D-NLSE generates a much larger number of such RWs than a comparable linear system: by $3.7$ times larger if the linear system has the same Fourier spectrum as near the stationary state of MI, and by $8.2$ times larger if the linear system has Gaussian spectrum of the same characteristic width. 
However, in average, one RW affects by the same times smaller spatiotemporal area, so that the resulting distributions of wavefield intensity practically coincide for these systems. 
The distribution of the 1D-NLSE RWs by maximum intensity represents a function which decays slower than exponentially and does not exhibit regions with a noticeable change in behavior. 
This suggests that the mechanism of RW formation should be universal for RWs of significantly different amplitudes, e.g., for RWs with $\max|\psi|\simeq 3$ and $\simeq 5$. 
In space-time, the positions of RW maximums are distributed almost indistinguishably ``by eye'' from random uniformly distributed positions. 
The distribution of distance to the nearest neighbor for RWs is very similar to that for random positions, except that RWs do not emerge too close to each other. 

Then, more than seven million of collected RWs near the stationary state of MI are compared within a sufficiently large spatiotemporal window with nine exact solutions of the 1D-NLSE, of which two models turn out to be by far the best in reproducing RW dynamics: (i) a general collision of two solitons (SS2v) and (ii) a general collision of two Tajiri-Watanabe breathers (BS2v). 
Note that such collisions are difficult to work with since the positions of their local maximums in space-time are not known; for this reason, RWs are fitted with them using pre-compiled databases of such solutions. 
The SS2v and BS2v models represent ad-hoc approximations involving only two nonlinear modes (solitons or breathers), which, near the RW maximums, replace the whole wavefield composed of a large number of nonlinear modes. 
However, the parameters of the two approximating modes turn out to be surprisingly similar to the IST spectrum of the whole wavefield, indicating a hypothesis that RWs may emerge due to synchronization of a few nonlinear modes in presence of many other nonlinear modes of the wavefield. 
This hypothesis is corroborated by the observation that the larger the deviations of the SS2v and BS2v models from RWs in the physical space, the larger the deviations of nonlinear modes of these models from the IST spectrum of the whole wavefield (assuming that both kinds of deviations are caused by distortions from the other nonlinear modes of the wavefield). 

The paper is organized as follows. 
The next Section~\ref{Sec:Theory} discusses the dressing method, as well as multi-soliton and multi-breather solutions. 
Section~\ref{Sec:NumMethods} describes the numerical methods together with the model solutions for RWs and the fitting procedure. 
Section~\ref{Sec:Results1} examines the frequency of RW occurrence and the distributions of RWs by intensity, spatial phase slope (chirp) and in space-time. 
Section~\ref{Sec:Results2} compares collected RWs with the model exact solutions and studies the distributions of parameters of the latter. 
Sections~\ref{Sec:Discussions} and~\ref{Sec:Conclusions} are devoted to discussions and conclusions, respectively. 
The paper has also two Appendixes, of which Appendix~\ref{Sec:App1} explains the transformations of parameters of multi-soliton and multi-breather solutions under the scaling, Galilean, gauge and mirror transformations, and Appendix~\ref{Sec:App2} discusses RWs in a comparable linear system.

\section{Dressing method, multi-soliton and multi-breather solutions}
\label{Sec:Theory}


\subsection{IST and the dressing method}
\label{Sec:Theory-1}

The 1D-NLSE belongs to a class of nonlinear partial differential equations (PDEs) integrable by means of the IST method. 
The latter is based on the representation of a PDE as a compatibility condition of an over-determined auxiliary linear system -- the Lax pair~\cite{ablowitz1981solitons,novikov1984theory}. 
For the case of 1D-NLSE, the Lax pair is known as the Zakharov-Shabat (ZS) system~\cite{zakharov1972exact} for a two-component vector wave function $\mathbf{\Phi}(x,t,\lambda) = (\phi_1,\phi_2)^{T}$,
\begin{eqnarray}
	\mathbf{\Phi}_{x} &=& \begin{pmatrix} -i \lambda & \psi \\ -\psi^* & i \lambda \end{pmatrix}\mathbf{\Phi},
	\label{ZSsystem1}\\
	\mathbf{\Phi}_t &=& \begin{pmatrix}\ -i\lambda^2 + \frac{i}{2} |\psi|^2 & \lambda \psi + \frac{i}{2} \psi_x \\ -\lambda \psi^* + \frac{i}{2} \psi^*_x & i\lambda^2 - \frac{i}{2} |\psi |^2 \end{pmatrix} \mathbf{\Phi},
	\label{ZSsystem2}
\end{eqnarray}
where the star stands for the complex conjugate and Eq.~(\ref{NLSE}) is obtained as a compatibility condition $\mathbf{\Phi}_{xt} = \mathbf{\Phi}_{tx}$. 
The first equation of the ZS system can be rewritten as an eigenvalue problem for a complex-valued spectral parameter $\lambda = \xi + i \eta$, 
\begin{equation}\label{ZSsystem1-eigenvalue}
	\widehat{\mathcal{L}}\mathbf{\Phi} = \lambda \mathbf{\Phi},
	\quad
	\widehat{\mathcal{L}} = i \begin{pmatrix}\ 1 & 0 \\ 0 & -1 \end{pmatrix}\frac{\partial}{\partial x} - i\begin{pmatrix}\ 0 & \psi \\ \psi^* & 0 \end{pmatrix}.
\end{equation}
Similarly to the Schr{\"o}dinger operator in quantum mechanics~\cite{landau1958quantum}, the scattering problem~(\ref{ZSsystem1-eigenvalue}) for the ZS operator $\widehat{\mathcal{L}}$ and wave function $\mathbf{\Phi}$ can be introduced, in which the wavefield $\psi$ of the 1D-NLSE plays the role of a potential. 
For localized potentials, this problem has bounded solutions $\mathbf{\Phi}$ for real-valued spectral parameter $\lambda = \xi\in\mathbb{R}$ (continuous spectrum), and also for a finite number of discrete points $\lambda_{n} = \xi_{n} + i\eta_{n}$, $\eta_{n}>0$, $n=1,...,N$ (discrete spectrum). 
Note that, without loss of generality, in this paper only the upper half of the $\lambda$-plane, $\eta = \mathrm{Im}\,\lambda \ge 0$, is considered. 

Most importantly, the potential $\psi(x,t)$ turns out to be in one-to-one correspondence with the so-called \textit{scattering data} -- a combination $\{\lambda_{n}, \rho_{n}(t), r(\xi,t)\}$ of the discrete spectrum points $\lambda_{n}$, associated with them coefficients $\rho_{n}(t)$, and reflection coefficient $r(\xi,t)$ representing the continuous spectrum -- and this scattering data changes trivially over time. 
In particular, the 1D-NLSE evolution preserves the discrete spectrum, $\partial_{t}\lambda_{n}=0$, while $\rho_{n}(t)$ and $r(\xi,t)$ evolve exponentially like the Fourier harmonics in the linear waves theory. 
These properties make it possible to fundamentally solve the Cauchy initial value problem for the 1D-NLSE by finding the scattering data from the solution of the scattering problem~(\ref{ZSsystem1-eigenvalue}), calculating its evolution in time, and recovering the potential $\psi$ with the inverse scattering transform. 
Note, however, that both the direct transformation to the scattering data $\{\lambda_{n}, \rho_{n}(t), r(\xi,t)\}$ and the inverse transformation to the potential $\psi(x,t)$ represent highly nontrivial nonlinear problems. 
In particular, the IST is done by solving the integral Gelfand -- Levitan -- Marchenko equations~\cite{novikov1984theory} and can be calculated analytically only in special cases, asymptotically at large time $t\to\pm\infty$, or in the semi-classical approximation~\cite{lewis1985semiclassical,jenkins2014semiclassical}. 

Like the Fourier spectrum in the linear waves theory, the scattering data can be used to characterize the potential $\psi(x,t)$: the reflection coefficient $r(\xi,t)$, representing the continuous spectrum $\lambda = \xi\in\mathbb{R}$, describes the nonlinear dispersive waves, while the discrete eigenvalues $\lambda_{n}$ together with the coefficients $\rho_{n}(t)$ correspond to solitons. 
In particular, the eigenvalues $\lambda_{n}=\xi_{n}+i\eta_{n}$ contain information about the (invariant in time) soliton amplitudes $a_{n}=2\eta_{n}$ and velocities $v_{n}=-2\xi_{n}$, while the coefficients $\rho_{n}(t)$ -- about their (evolving) positions in space and complex phases. 
In what follows, only the reflectionless potentials $r(\xi,t)=0$ are considered, i.e., solutions of the 1D-NLSE that contain only solitons. 

The dressing method (DM)~\cite{novikov1984theory,matveev1991darboux}, also known as the Darboux transformation~\cite{akhmediev1991extremely,matveev1991darboux}, is a simplified version of the conventional IST approach. 
It represents an algebraic scheme for constructing new exact solutions to integrable PDEs by ``dressing'' a bare solution, and is often used to find general multi-soliton and multi-breather solutions to the 1D-NLSE, see e.g.~\cite{zakharov1978relativistically,akhmediev1991extremely,akhmediev2009extreme,gelash2014superregular}. 
In the present paper, a variant of the DM is discussed, which is optimized for dressing with solitons. 
Note that for this task the DM requires far fewer arithmetic operations than the standard IST method, which makes it favorable when the number of solitons is large~\cite{gelash2018strongly,gelash2019bound,tarasova2023properties}.

The DM starts from an initial background potential $\psi_{0}(x,t)$ (bare solution) and the corresponding $2\times 2$ matrix solution $\mathbf{\Phi}_{0}(x,t,\lambda)$ of the ZS system (wave function) constructed from two linearly independent solutions of Eqs.~(\ref{ZSsystem1})-(\ref{ZSsystem2}). 
At the $n$-th step of the recursive procedure, the potential $\psi_{n}(x,t)$ dressed with $n$ solitons is constructed from $\psi_{n-1}(x,t)$ and the corresponding wave function $\mathbf{\Phi}_{n-1}(x,t,\lambda)$ as
\begin{eqnarray}
	\psi_{n}(x,t) = \psi_{n-1}(x,t) + 2i(\lambda_n-\lambda^*_n)\frac{q^*_{n1}q_{n2}}{|\mathbf{q_n}|^2},
	\label{psi_n}
\end{eqnarray}
where vector $\mathbf{q}_{n}=(q_{n1},q_{n2})^{T}$ is determined via $\mathbf{\Phi}_{n-1}$, the soliton eigenvalue $\lambda_{n}$, and an arbitrary constant $C_{n}\in\mathbb{C}\backslash \{0\}$ called the norming constant,
\begin{eqnarray}
	\mathbf{q}_{n}(x,t) = [\mathbf{\Phi}_{n-1}(x,t,\lambda^*_n)]^{*} \cdot \left(\begin{array}{c} C_n^{-1/2} \\C_n^{1/2} \end{array}\right).
	\label{qn}
\end{eqnarray}
The corresponding wave function $\mathbf{\Phi}_{n}(x,t,\lambda)$ is calculated using the dressing matrix $\boldsymbol{\sigma}^{(n)}(x,t,\lambda)$, 
\begin{eqnarray}
	\mathbf{\Phi}_{n}(x,t,\lambda) &=& \boldsymbol{\sigma}^{(n)}(x,t,\lambda)\cdot \mathbf{\Phi}_{n-1}(x,t,\lambda), \label{dressing-Psi}\\
	\sigma^{(n)}_{ml}(x,t,\lambda) &=& \delta_{ml} + \frac{\lambda_n-\lambda_n^*}{\lambda - \lambda_n}\frac{q_{nm}^{*}q_{nl}}{|\mathbf{q_n}|^2},
	\label{dressing-matrix}
\end{eqnarray}
where $m,l=1,2$ and $\delta_{ml}$ is the Kronecker delta. 
The outcome of the dressing by $N$ solitons can be written via the ratio of two determinants, 
\begin{eqnarray}
	&&\psi_{N}(x,t) = \psi_{0} - 2i\frac{\mathrm{det} \mathbf{P}}{\mathrm{det} \mathbf{Q}}, 
	\quad Q_{kj}=\frac{(\mathbf{\tilde{q}}_{k}\cdot \mathbf{\tilde{q}}^*_{j})}{\lambda_{k} - \lambda^*_j}, \nonumber\\
	&&\mathbf{P}=
	\left(\begin{array}{cc}
	        0 & \begin{array}{ccc}
	              \tilde{q}_{12} & \cdots & \tilde{q}_{N2}
	            \end{array}
	         \\
	        \begin{array}{c}
	          \tilde{q}^*_{11} \\
	          \vdots \\
	          \tilde{q}^*_{N1}
	        \end{array}
	         &  \begin{array}{c}
	              \mathbf{Q}^{T}
	            \end{array}
	\end{array}\right),
	\label{Ndet_SS}
\end{eqnarray}
where the new vectors $\mathbf{\tilde{q}}_{n}=(\tilde{q}_{n1},\tilde{q}_{n2})^{T}$ depend on the initial matrix $\mathbf{\Phi}_0$; these vectors are discussed below separately for the multi-soliton and multi-breather cases. 
Note that (i) rescaling $\mathbf{\tilde{q}}_{n} \to \alpha_{n}\mathbf{\tilde{q}}_{n}$ by factors $\alpha_{n}\in \mathbb{C}\backslash \{0\}$ does not change the outcome as both determinants are multiplied by $\prod_{j=n}^{N}|\alpha_{n}|^{2}$, and (ii) transformation $(\tilde{q}_{n1},\tilde{q}_{n2})^{T} \to \big(e^{-i\phi/2}\tilde{q}_{n1}, \, e^{i\phi/2}\tilde{q}_{n2}\big)^{T}$, where $\phi=\phi(x,t)\in\mathbb{R}$, leads to $\mathrm{det}\,\mathbf{P}/\mathrm{det}\,\mathbf{Q}\to e^{i\phi}\cdot\mathrm{det}\,\mathbf{P}/\mathrm{det}\,\mathbf{Q}$.


\subsection{Multi-soliton solutions}
\label{Sec:Theory-2}

An exact $N$-soliton potential $\psi_{N}^{\mathrm{S}}$ is constructed from the zero background $\psi_{0}^{\mathrm{S}}=0$ and the corresponding wave function
\begin{eqnarray}
	\mathbf{\Phi}_{0}^{\mathrm{S}}(x,t,\lambda) = \begin{pmatrix}\ e^{-i\lambda x - i \lambda^2 t} & 0 \\ 0 & e^{i\lambda x + i \lambda^2 t} \end{pmatrix},
	\label{Psi0}
\end{eqnarray}
with the outcome written as
\begin{eqnarray}
	\psi_{N}^{\mathrm{S}}(x,t) = -2i\frac{\mathrm{det} \mathbf{P}}{\mathrm{det} \mathbf{Q}},
	\label{NSsolution}
\end{eqnarray}
and the matrices $\mathbf{P}$ and $\mathbf{Q}$ defined via vectors
\begin{eqnarray}
	\mathbf{\tilde{q}}_{n}(x,t) = \left(\begin{array}{c}\tilde{q}_{n1} \\\tilde{q}_{n2}\end{array}\right) =
	\left(\begin{array}{c} C_n^{-1/2}\, e^{i\lambda_n x +  i \lambda_{n}^2 t} \\ C_n^{1/2}\, e^{-i\lambda_n x -  i \lambda_{n}^2 t} \end{array}\right).
	\label{tildeq-N-soliton}
\end{eqnarray}
Each step of the DM leads to a new potential $\psi_{n}^{\mathrm{S}}$ and a new wave function $\mathbf{\Phi}_{n}^{\mathrm{S}}$, and consists in adding one soliton $\lambda_{n}$ to $\psi_{n-1}^{\mathrm{S}}$ and one pole at $\lambda=\lambda_{n}$ to $\mathbf{\Phi}_{n-1}^{\mathrm{S}}$ (only the upper half of the $\lambda$-plane is considered). 

Note that the time dependence in Eq.~(\ref{NSsolution}) appears only due to the exponentials in the right hand side of Eq.~(\ref{tildeq-N-soliton}) and can be transferred into the definition of the norming constants, 
\begin{eqnarray}
	C_{n}(t) = C_{n}(0)\,e^{-2i\lambda_{n}^{2}t},
	\label{C_param_S-evolution}
\end{eqnarray}
for which Eq.~(\ref{tildeq-N-soliton}) is evaluated as if $t=0$. 
The norming constants, in turn, can be parameterized via the soliton positions $x_{n}\in\mathbb{R}$ and phases $\theta_{n}\in\mathbb{R}$ as
\begin{eqnarray}
	C_{n} &=& \exp\big[i\pi + 2i\lambda_{n}x_{n} + i\theta_{n}\big]. \label{C_param_S}
\end{eqnarray}
Combining Eqs.~(\ref{C_param_S-evolution})-(\ref{C_param_S}), one can get the linear evolution for the soliton positions $x_{n}$ and phases $\theta_{n}$,
\begin{eqnarray}
	x_{n}(t) = x_{0n} - 2\xi_{n}t, \quad \theta_{n}(t) = \theta_{0n} + 2(\xi_{n}^{2} + \eta_{n}^{2})t, \label{x0theta-evolution}
\end{eqnarray}
where $x_{0n}$ and $\theta_{0n}$ are the positions and phases at $t=0$. 

The one-soliton potential takes the following form,
\begin{eqnarray}
	\psi_{1}^{\mathrm{S}}(x,t) = a_{1} \frac{\mathrm{exp}\big[iv_{1}(x-x_{1}) + \frac{i(a_{1}^{2}-v_{1}^{2})t}{2} + i\theta_{1}\big]}{\cosh a_{1} \big(x-v_{1}t-x_1\big)},
	\label{1-SS}
\end{eqnarray}
where $a_{1}=2\eta_{1}$ and $v_{1}=-2\xi_{1}$ are the soliton amplitude and velocity. 
Parametrization~(\ref{C_param_S}) serves only for a convenient and physically intuitive representation of multi-soliton solutions, with $x_{n}$ and $\theta_{n}$ coinciding with the observed in the physical space position and phase only for one-soliton potential. 
In presence of other solitons or dispersive waves, the observed position and phase may differ considerably from these parameters~\cite{gelash2021solitonic}.


\subsection{Multi-breather solutions}
\label{Sec:Theory-3}

An exact $N$-breather potential $\psi_{N}^{\mathrm{B}}$ is constructed from the plane wave background $\psi_{0}^{\mathrm{B}} = a\,e^{ia^{2}t + i\Theta}$, where $a>0$ is its amplitude and $\Theta\in\mathbb{R}$ is the initial phase, and the corresponding wave function,
\begin{eqnarray}
	&& \mathbf{\Phi}_{0}^{\mathrm{B}}(x,t,\lambda) = \begin{pmatrix}\ e^{\frac{ia^{2}t}{2}-\phi}  &  p\, e^{i \Theta + \frac{ia^{2}t}{2}+\phi}  \\  p\, e^{-i \Theta - \frac{ia^{2}t}{2} - \phi}  & e^{-\frac{ia^{2}t}{2} + \phi} \end{pmatrix}, 	\label{Psi0cond}\\
	&& \phi = i\zeta(x + \lambda t), \quad p = \frac{i(\lambda-\zeta)}{a}, \quad \zeta = \sqrt{a^{2}+\lambda^2}. \nonumber
\end{eqnarray}
Note that, as is usually done in the breather theory, the branch cut $[-ia,ia]$ is used for calculation of $\zeta(\lambda)$ with the choice of the Riemann sheet such that $\mathrm{Im}\,\zeta>0$ at $\mathrm{Im}\,\lambda > 0$. 
This differs from the branch cut $i(-\infty,-a]\,\cup\, i[a,\infty)$ implied in software compilers. 
Since in this paper only the upper half of the $\lambda$-plane is considered, from now on the branch cut is denoted as $[0,ia]$; also, in the remaining part of this subsection, it is assumed that the plane wave has unit amplitude, $a=1$.

In the outcome of the DM, the plane wave $e^{it+i\Theta}$ can be extracted as a common factor, 
\begin{eqnarray}
    \psi_{N}^{\mathrm{B}}(x,t) =  e^{it+i\Theta} \bigg(1 - 2i\frac{\mathrm{det} \mathbf{P}}{\mathrm{det} \mathbf{Q}} \bigg),
    \label{NBsolution}
\end{eqnarray}
where the matrices $\mathbf{P}$ and $\mathbf{Q}$ are defined via vectors
\begin{eqnarray}
	&& \mathbf{\tilde{q}}_{n}(x,t) = \left(\begin{array}{c} C_n^{-1/2} e^{\phi_n + \frac{i\Theta}{2}} - C_n^{1/2} p_n e^{-\phi_n-\frac{i\Theta}{2}} \\ -C_n^{-1/2} p_n e^{\phi_n + \frac{i\Theta}{2}} + C_n^{1/2} e^{-\phi_n-\frac{i\Theta}{2}} \end{array}\right), \nonumber\\
	&& \phi_{n} = i\zeta_{n}(x + \lambda_{n}t), \quad\quad p_{n} = i(\lambda_{n}-\zeta_{n}), \label{qn_breathers}
\end{eqnarray}
and $\zeta_{n} = \zeta(\lambda_{n})$. 
The dressing procedure preserves non-analytic behavior of the wave function on the branch cut, adding at each step one breather to the potential $\psi_{n-1}^{\mathrm{B}}$ and one pole at $\lambda=\lambda_{n}$ to the wave function $\mathbf{\Phi}_{n-1}^{\mathrm{B}}$. 
A general one-breather solution (also known as the Tajiri-Watanabe breather~\cite{its1988exact,tajiri1998breather}) is written as 
\begin{eqnarray}
	\psi_{1}^{\mathrm{B}}(x,t) =  e^{it+i\Theta} \bigg(1 - 4\eta_1 \frac{\tilde{q}_{11}^*\tilde{q}_{12}}{|\tilde{q}_{11}|^2+|\tilde{q}_{12}|^2} \bigg).
	\label{1Bsolution}
\end{eqnarray}
Note that the plane wave and the resulting $N$-breather potentials are not localized and need to be considered by the quasi-periodic analogue of the IST called the finite-gap theory~\cite{novikov1984theory,osborne2010nonlinear}. 
However, the DM is based on a more general Lax pair representation and can still be used for constructing such solutions. 

Depending on the positioning of the soliton eigenvalue relative to the branch cut $[0,i]$, a general one-breather potential represents one of the four different cases (i) $\lambda_{1} = i$, (ii) $\lambda_{1} = i\eta_{1}$ with $\eta_{1}<1$, (iii) $\lambda_{1} = i\eta_{1}$ with $\eta_{1}>1$ and (iv) $\mathrm{Re}\,\lambda_{1}\neq 0$, corresponding to the Peregrine breather (also known as the rational breather of the first order, or RB1), Akhmediev (AB), Kuznetsov (KB) and Tajiri-Watanabe (TWB) breathers, respectively. 
In the following, different parametrizations of the norming constant $C_{1}$ are used for different breathers. 
These parametrizations are needed to represent the corresponding solutions in a convenient and physically intuitive form, and also to place the breathers at a given position in space and time.

The AB is obtained by dressing a bare solution $\psi_{0}^{\mathrm{B}} = e^{it + i\Theta}$ with soliton $\lambda_{1}=i\eta_{1}$, $\eta_{1}<1$. 
If one parametrizes the norming constant as
\begin{eqnarray}
	C_{1}^{\mathrm{AB}} = \exp\big[i(\pi+\Theta) + \Omega_{\mathrm{A}}t_{0} + i\theta_{0}\big], \label{C_param_AB}
\end{eqnarray}
then, this breather takes the following form,
\begin{eqnarray}
    &&\psi^{\mathrm{AB}}(x,t) = e^{it+i\Theta+i\pi}\times \label{A-breather}\\ 
    &&\times\biggl(1 - \frac{2\kappa_{1}^2 \cosh[\Omega_{\mathrm{A}}(t-t_0)] + i\Omega_{\mathrm{A}} \sinh[\Omega_{\mathrm{A}}(t-t_0)]}
    {\cosh[\Omega_{\mathrm{A}}(t-t_0)] - \eta_{1} \cos[2\kappa_{1} x+\theta_{0}]}\biggr), \nonumber
\end{eqnarray}
where $\kappa_{1} = \sqrt{1-\eta_{1}^2}$ and $\Omega_{\mathrm{A}} = 2\eta_{1}\kappa_{1}$, while $t_{0}\in\mathbb{R}$ is the time shift and $\theta_{0}\in[0,2\pi)$ is the space phase shift.
The AB describes a spatially unbounded periodic perturbation with period $\pi/\kappa_{1}$ on the background of the plane wave $e^{it+i\Theta+i\pi}$; the perturbation becomes most pronounced at the time $t=t_{0}$ and then vanishes within the characteristic time $\Omega_{A}^{-1}$. 
Note that the additional phase shift $e^{i\pi}$ relative to the bare solution $\psi_{0}^{\mathrm{B}} = e^{it + i\Theta}$ appears at every new step of the dressing procedure. 

The KB is constructed by dressing with soliton $\lambda_{1}=i\eta_{1}$, $\eta_{1}>1$. 
With parametrization
\begin{eqnarray}
	C_{1}^{\mathrm{KB}} = \exp\big[i(\pi+\Theta) - 2\nu_{1} x_{0} + i\theta_{0}\big], \label{C_param_KMB}
\end{eqnarray}
this solution can be written as
\begin{eqnarray}
    && \psi^{\mathrm{KB}}(x,t) = e^{it+i\Theta+i\pi}\times \label{KM-breather}\\
    && \times \biggl(1 - \frac{2\nu_{1}^2 \cos[\Omega_{\mathrm{K}}t + \theta_{0}] + i\Omega_{\mathrm{K}} \sin[\Omega_{\mathrm{K}}t + \theta_{0}]}
    {\eta_{1}\cosh[2\nu_{1}(x-x_0)] - \cos[\Omega_{\mathrm{K}}t + \theta_{0}]} \biggr),\nonumber
\end{eqnarray}
where $\nu_{1} = \sqrt{\eta_{1}^2-1}$ and $\Omega_{\mathrm{K}} = 2\eta_{1}\nu_{1}$, while $x_{0}\in\mathbb{R}$ and $\theta_{0}\in[0,2\pi)$ represent the space shift and time phase shift, respectively.
This solution describes a standing localized perturbation of the plane wave $e^{it+i\Theta+i\pi}$, which has characteristic width $(2\nu_{1})^{-1}$ in space and oscillates with period $2\pi/\Omega_{K}$ in time.

The RB1 is obtained from the KB solution in the limit $\epsilon = (\eta_{1}-1) \to 0^{+}$, corresponding to the dressing with soliton $\lambda_{1}=i$. 
Expanding Eq.~(\ref{KM-breather}) in Taylor series with respect to $\epsilon$ for $|x-x_{0}|\ll (8\epsilon)^{-1/2}$ and $|t|\ll (8\epsilon)^{-1/2}$, and using the following parametrization of the norming constant,
\begin{eqnarray}
    C_{1}^{\mathrm{RB1}} = \exp\bigg[i(\pi+\Theta) - \sqrt{8\epsilon}\,\big(x_{0} + i\,t_{0}\big)\bigg], \label{C_param_PB}
\end{eqnarray}
one can obtain the KB that behaves locally as the RB1,
\begin{eqnarray}
    && \psi^{\mathrm{RB1}}(x,t) = e^{it+i\Theta+i\pi}\times \label{P-breather}\\
	&& \times \biggl(1 - \frac{4 (1 + 2 i [t-t_{0}])}{1 + 4 [x-x_{0}]^2 + 4 [t-t_{0}]^2} \biggr). \nonumber
\end{eqnarray}
Here $x_{0}\in\mathbb{R}$ and $t_{0}\in\mathbb{R}$ are the position and time shifts respectively. 
Note that formally, at $\epsilon=0$, there is only one allowed value of the norming constant $C_{1} = e^{i\pi+i\Theta}$, which corresponds to the RB1~(\ref{P-breather}) for all possible $x_{0}$ and $t_{0}$. 
Any other value of $C_{1}$ results in the plane wave solution $\psi^{\mathrm{RB1}} = e^{it+i\Theta+i\pi}$ describing the RB1 located at infinity. 
The specific values of $x_{0}$ and $t_{0}$ are determined by how the norming constant~(\ref{C_param_PB}) tends to $e^{i\pi+i\Theta}$ in the limit $\epsilon\to 0^{+}$. 
The RB1 describes a rational perturbation of the plane wave $e^{it+i\Theta+i\pi}$, which is localized both in space and in time, and leads to the maximum amplitude $|\psi^{\mathrm{RB1}}|=3$ at the point $(x_{0},t_{0})$.

The TWB corresponds to a general case when the soliton eigenvalue $\lambda_{1}$ has nonzero real and imaginary parts; for this breather, it is more practical to work with Eq.~(\ref{1Bsolution}) without simplifications. 
The parametrization of the norming constant,
\begin{eqnarray}
	C_{1}^{\mathrm{TWB}} = \exp\big[i(\pi+\Theta) + 2i\zeta_{1} x_{1} + i\theta_{1}\big], \label{C_param_TWB}
\end{eqnarray}
allows one to find the breather's position in space $x_{1}$ together with its phase $\theta_{1}$. 
The TWB describes a coherent wave group having characteristic width $|\mathrm{Im}[\zeta_{1}]|^{-1}$, which moves on the background of a plane wave with a constant velocity $V_{1}^{\mathrm{TW}}$ and oscillates with a frequency $\Omega_{1}^{\mathrm{TW}}$~\cite{xu2019breather},
\begin{eqnarray}
	V_{1}^{\mathrm{TW}} &=& -\mathrm{Im}\,[\lambda_{1}\zeta_{1}]/\mathrm{Im}\,\zeta_{1} \label{TWB_velocity} \\
	\Omega_{1}^{\mathrm{TW}} &=& -2\,\mathrm{Re}\,[\lambda_{1}\zeta_{1}] - 2\, V_{1}^{\mathrm{TW}}\, \mathrm{Re}\,\zeta_{1}. \label{TWB_frequency}
\end{eqnarray}
Note that, for the TWB, the phases of the plane wave background are different to the left $x\to -\infty$ and to the right $x\to +\infty$ of it, see e.g.~\cite{gelash2014superregular}. 
Similarly to Eq.~(\ref{x0theta-evolution}), one can find evolution of the breather's position and phase as
\begin{eqnarray}
	x_{n}(t) = x_{0n} + V_{n}^{\mathrm{TW}}t, \quad\quad \theta_{n}(t) = \theta_{0n} + \Omega_{n}^{\mathrm{TW}}t, \label{TWB-x0theta-evolution}
\end{eqnarray}
where the subscript $n$ means that these formulas are valid also for a multi-breather solution, for which $V_{n}^{\mathrm{TW}}$ and $\Omega_{n}^{\mathrm{TW}}$ are defined as in Eqs.~(\ref{TWB_velocity})-(\ref{TWB_frequency}) with $\lambda_{1}$ and $\zeta_{1}$ replaced by $\lambda_{n}$ and $\zeta_{n}$, respectively. 
Note that, with definition of the norming constants $C^{\mathrm{TW}}_{n}$ and vectors $\mathbf{\tilde{q}}_{n}$ as in Eqs.~(\ref{C_param_TWB}) and~(\ref{qn_breathers}), the plane wave's phase $\Theta$ cancels out within the brackets of Eq.~(\ref{NBsolution}) and remains only outside of them. 
This means that rotating $\Theta$ while holding $x_{n}$ and $\theta_{n}$ fixed will only result in the corresponding rotation of phase for the entire multi-breather solution~(\ref{NBsolution}).

Among different variants of multi-breather solutions, there is a special case when the (bare) plane wave $\psi_{0}^{\mathrm{B}} = e^{it + i\Theta}$ is repeatedly dressed with the same soliton $\lambda_{1}=i$ using the same norming constant $C_{1} = e^{i\pi + i\Theta}$. 
With these parameters, the dressing scheme cannot be applied straightforwardly due to the division by zero in Eq.~(\ref{qn}) during the repeated dressings, see Eqs.~(\ref{dressing-Psi})-(\ref{dressing-matrix}); however, this difficulty can be overcome by applying the L'Hospital's rule. 
The resulting solutions are known as higher-order rational breathers; in particular, the second-order rational breather (RB2) is obtained by double dressing, the third-order breather (RB3) -- by triple dressing, and so on. 
These breathers describe localized both in space and in time rational perturbations of the plane wave background and lead to the maximum amplitude $2M+1$, where $M$ is the order of the breather.
The exact analytic relations for them are too cumbersome and can be found in~\cite{akhmediev2009rogue} where they were first discovered.


\section{Numerical methods}
\label{Sec:NumMethods}


\subsection{Numerical scheme and initial conditions}
\label{Sec:NumMethods-1}

The 1D-NLSE~(\ref{NLSE}) is solved numerically in a large box $x\in[-L/2, L/2]$, $L\gg 1$, with periodic boundary conditions using the pseudo-spectral Runge-Kutta fourth-order method. 
The box $L$ is resolved with a uniform and adaptive grid with the step $\Delta x$ determined from the analysis of the Fourier spectrum of solution $\psi$~\cite{agafontsev2015integrable}. 
As an integrable equation, the 1D-NLSE conserves an infinite series of integrals of motion~\cite{novikov1984theory}. 
The first three of them are the wave action (in notations below, it equals the average intensity), 
\begin{equation}
	N = \overline{|\psi|^{2}} = \frac{1}{L}\int_{-L/2}^{L/2}|\psi|^{2}\,dx = \sum_{k}|\psi_{k}|^{2},
	\label{wave-action}
\end{equation}
momentum, 
\begin{equation}
	P = \frac{i}{2L}\int_{-L/2}^{L/2}(\psi_{x}^{*}\psi-\psi_{x}\psi^{*})\,dx = \sum_{k}k|\psi_{k}|^{2},
	\label{momentum}
\end{equation}
and total energy, 
\begin{eqnarray}
	&& E = H_{l} + H_{nl}, \label{energy-1}\\
	&& H_{l} = \frac{\overline{|\psi_{x}|^{2}}}{2} = \frac{1}{L}\int_{-L/2}^{L/2}\frac{|\psi_{x}|^{2}}{2}dx = \sum_{k}\frac{k^{2}|\psi_{k}|^{2}}{2}, \label{energy-2}\\
	&& H_{nl} = -\frac{\overline{|\psi|^{4}}}{2} = -\frac{1}{L}\int_{-L/2}^{L/2}\frac{|\psi|^{4}}{2}dx. \label{energy-3}
\end{eqnarray}
Here and below the overline means averaging over the simulation box $L$, while $H_{l}$ is the kinetic energy (linear contribution), $H_{nl}$ is the potential energy (nonlinear contribution), and $\psi_{k}$ is the Fourier-transformed wavefield, 
$$
	\psi_{k}(t) = \mathcal{F}\big[\psi(x,t)\big] = \frac{1}{L}\int_{-L/2}^{L/2}\psi(x,t)\,e^{-ikx}\,dx. 
$$
It has been verified that the implemented numerical scheme conserves the first ten integrals of motion up to the relative errors from $10^{-10}$ (the first three invariants) to $10^{-6}$ (the tenth invariant) orders. 

The initial conditions correspond to the noise-induced MI of a plane wave of unit amplitude, 
\begin{eqnarray}
	\psi|_{t=0} = 1 + \epsilon(x),
	\label{IC-PW}
\end{eqnarray}
where $\epsilon(x)$ represents a small initial noise. 
This noise is statistically homogeneous in space and is modeled as a sum of harmonics, 
\begin{equation}
	\epsilon(x)=a_{0}\bigg(\frac{\sqrt{8\pi}}{\delta k\, L}\bigg)^{1/2} \sum_{m}e^{-k_{m}^{2}/\delta k^{2} + ik_{m}x + i\phi_{m}},
	\label{noise}
\end{equation}
with a Gaussian Fourier spectrum, $|\epsilon_{k}| \propto e^{-k^{2}/\delta k^{2}}$. 
Here $a_{0}$ is the noise amplitude, $k_{m}=2\pi m/L$ is the wavenumber, $m\in\mathbb{Z}$ is integer, $\delta k$ is the characteristic noise width in the $k$-space, and $\phi_{m}$ are random phases for each $k_{m}$ and each realization of the initial conditions. 
The mean intensity of such noise in the $x$-space equals $a_{0}^{2}$, $\langle\overline{|\epsilon|^{2}}\rangle = a_{0}^{2}$, where $\langle...\rangle$ means averaging over the ensemble of initial conditions. 
The numerical experiment uses the parameters $L=128\sqrt{2}\,\pi$, $a_{0}=10^{-5}$ and $\delta k=5\sqrt{2}$, which match those used in~\cite{agafontsev2015integrable} (after rescaling to a different normalization of the 1D-NLSE). 
Simulations are performed for $20\,000$ initial conditions, which differ only in the random realization of the initial noise~(\ref{noise}). 

Note that the usage of unit amplitude of the plane wave in Eq.~(\ref{IC-PW}) together with specific coefficients of Eq.~(\ref{NLSE}) does not lead to a loss of generality, since the problem can be rescaled in time, space and amplitude of the wavefield. 


\subsection{Identifying and recording rogue waves}
\label{Sec:NumMethods-2}

In the present paper, RWs are defined as sufficiently large local maximums of amplitude $|\psi|$ in the two-dimensional $(x,t)$-space. 
For their preliminary identification, each point of wavefield on the numerical grid is checked if the following three conditions are satisfied. 
First, the amplitude must be large enough to be in line with the commonly accepted RW criterion, $|\psi|^{2}/\langle \overline{|\psi|^{2}}\rangle > 8$. 
Second, the Hessian matrix $H = \partial^{2}|\psi|/\partial y_{i}\partial y_{j}$, where $i,j = 1, 2$ and $y_{1,2} = x, t$, must have two negative eigenvalues. 
Third, the Newton's method correction $\boldsymbol{\Delta y} = -\mathbf{H}^{-1}\mathbf{g}$ to a more accurate position of the local maximum, where $\mathbf{g} = \boldsymbol\nabla|\psi|$ is the gradient in the $(x,t)$-space, must be small enough for such a point to be considered a good approximation to the true position of the local maximum. 
These conditions determine a cloud of grid nodes around each local maximum, and each cloud is then filtered by selecting a single node $(x_{0}^{\mathrm{(n)}}, t_{0}^{\mathrm{(n)}})$ with the largest amplitude $A_{0}^{\mathrm{(n)}} = |\psi(x_{0}^{\mathrm{(n)}}, t_{0}^{\mathrm{(n)}})|$. 
The position and amplitude of the local maximum are then interpolated in between the grid nodes as $x_{0} = x_{0}^{\mathrm{(n)}} + \Delta y_{1}$, $t_{0} = t_{0}^{\mathrm{(n)}} + \Delta y_{2}$ and $A_{0} = A_{0}^{\mathrm{(n)}} + g_{i}\Delta y_{i} + \frac{1}{2}H_{ij}\Delta y_{i}\Delta y_{j}$ using the correction $\boldsymbol{\Delta y}$ calculated at the previous steps. 
For more details, see~\cite{agafontsev2015development}, where the same method has been used for strongly anisotropic three-dimensional flows.

The 1D-NLSE is invariant with respect to the scaling, gauge and Galilean transformations. 
In particular, if $\psi(x,t)$ is a solution of Eq.~(\ref{NLSE}), then
\begin{equation}
	\tilde{\psi}(x,t) = \alpha\,e^{i\big(\alpha v x - \frac{\alpha^{2}v^{2}t}{2} + \Theta_{0}\big)}\psi\bigg(\alpha\, x - \alpha^{2}vt,\, \alpha^{2}t\bigg),
	\label{transformations}
\end{equation}
is also a solution of the same equation. 
Here $\alpha$ is the scaling coefficient, $\Theta_{0}$ is the phase rotation, and $v$ is the velocity shift. 
This property must be taken into account when comparing RWs with exact solutions, since, for instance, a RW may represent a moving rescaled Peregrine breather with a rotated phase. 
Thereby, it is more convenient to perform the comparison not directly, but by making a ``snapshot'' of the RW in question in the form of an auxiliary wavefield, 
\begin{eqnarray}
	&& \Psi(\chi,\tau) = \frac{1}{A}\exp\bigg(-i\bigg[\frac{\vartheta}{A}\chi + \frac{\vartheta^{2}}{2 A^{2}}\tau + \Phi_{0}\bigg]\bigg)\times \nonumber\\
	&& \times \psi\bigg(x_{0} + \frac{1}{A}\chi + \frac{\vartheta}{A^{2}}\tau, \,\,\, t_{0} + \frac{1}{A^{2}}\tau \bigg).
	\label{wavefield-auxiliary}
\end{eqnarray}
Here $\chi$ and $\tau$ specify the amplitude-scaled spatial and temporal shifts relative to the coordinates $(x_{0},t_{0})$ of the local maximum, $A = A_{0}/3$ is the scaling coefficient, while $\Phi_{0} = \mathrm{arg}\,\psi$ and $\vartheta = \partial\,\mathrm{arg}\,\psi/\partial x$ are the phase and spatial phase slope (hereinafter called chirp) of the original RW at $(x_{0},t_{0})$. 
The auxiliary wavefield represents a solution of Eq.~(\ref{NLSE}) in the $(\chi, \tau)$-variables and is obtained from the original solution $\psi$ by using Eq.~(\ref{transformations}) with $\alpha = 1/A$, $\Theta_{0} = -\Phi_{0}$ and $v = -\vartheta$. 
At its local maximum at $(0,0)$, this solution has the same amplitude $|\Psi| = 3$ as the RB1, while its phase and chirp equal zero, $\mathrm{arg}\,\Psi = 0$ and $\partial\,\mathrm{arg}\,\Psi/\partial \chi = 0$. 
Note that, in Eq.~(\ref{transformations}), the Galilean transformation is done first to remove the non-zero chirp, followed by the rescaling and phase rotation. 

For comparison with various exact solutions, the auxiliary wavefield~(\ref{wavefield-auxiliary}) is recorded in a rectangular region 
\begin{eqnarray}
	(\chi,\tau)\in W_{0} = [-\Delta\chi_{0}, \Delta\chi_{0}]\times[-\Delta\tau_{0}, \Delta\tau_{0}],
	\label{region-W0}
\end{eqnarray}
using bicubic interpolation to determine the original wavefield $\psi$ in between the original grid nodes. 
The comparison window $W_{0}$ must not be too small, since any local maximum in its small vicinity is well approximated by a parabola. 
Also, it should not be too large, since in this case a deviation from an otherwise well-fitting exact solution will accumulate at the edges where the background wavefield dominates. 
The present paper uses the parameters $\Delta\chi_{0} = \sqrt{3/4} \approx 0.87$ and $\Delta\tau_{0} = \sqrt{27/20} \approx 1.16$: with them, if a RW is a Peregrine breather, then at $(\pm\Delta\chi_{0}, 0)$ its amplitude becomes zero, while at $(0,\pm\Delta\tau_{0})$ it decreases by half compared to the maximum value. 
The window $W_{0}$ is resolved with $49\times 25$ grid nodes along the $\chi$- and $\tau$-directions, respectively.

\begin{figure}[t]\centering
	\includegraphics[width=8.5cm]{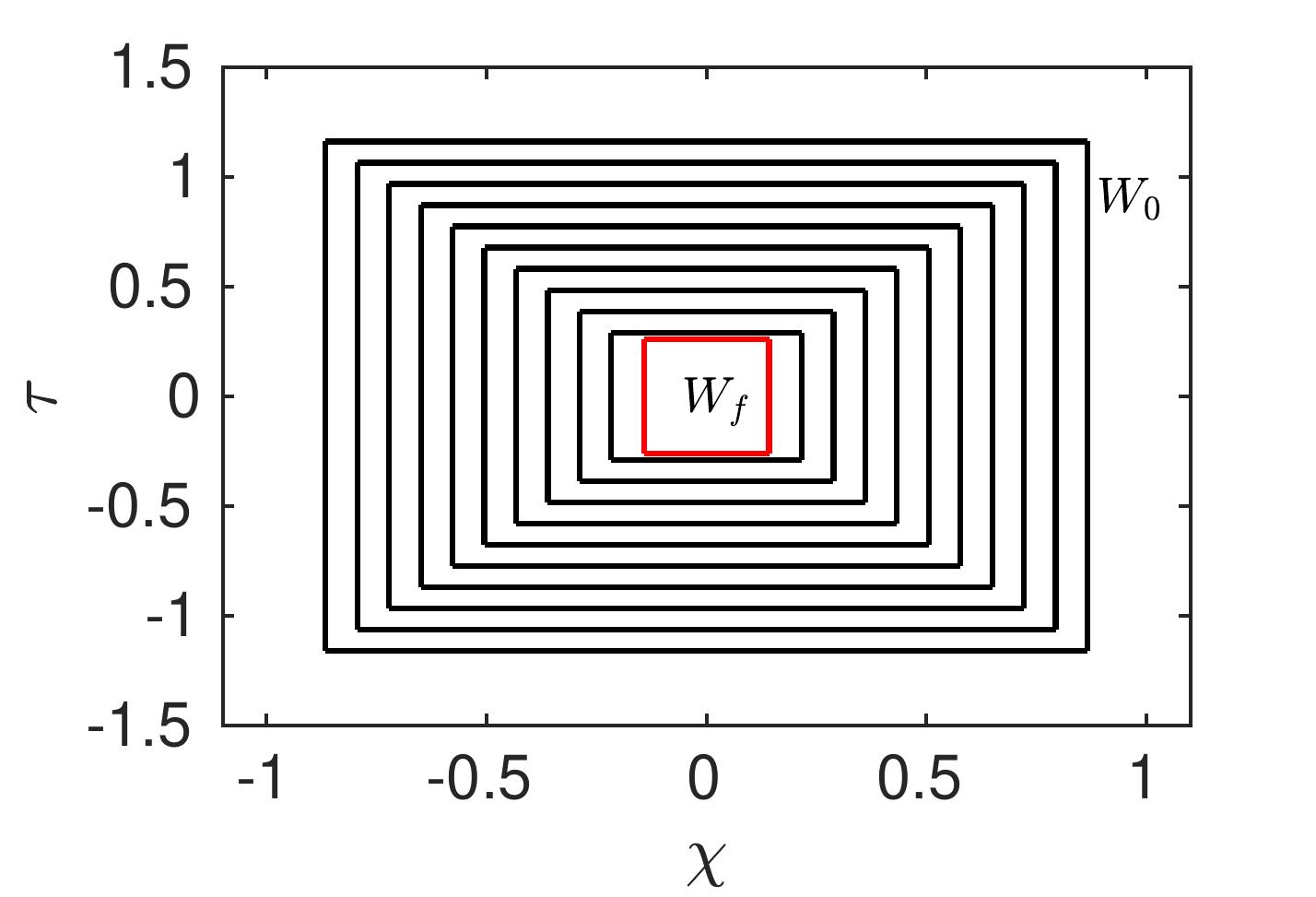}
	
	\caption{\small {\it (Color on-line)} 
	Rectangular regions in the $(\chi,\tau)$-variables, over which the deviations~(\ref{deviation-intergral}) between RWs and their fits with various exact solutions are calculated. 
	Black lines show the nested rectangles $W_{m}$, $m = 0, ..., 9$, while red lines indicate the rectangular region $W_{\mathrm{f}}$, in which the parameters of the fits with exact solutions are calculated.
	}
	\label{fig:fig01}
\end{figure}

As an integral measure reflecting the deviation between a RW and its fits with exact solutions, one can consider a quantity
\begin{eqnarray}
	\mathcal{D}_{\Lambda} = \bigg[\frac{\iint_{\Lambda} |\Psi - \Psi_{\mathrm{s}}|^{2}\,d\chi d\tau}{\iint_{\Lambda} |\Psi|^{2}\,d\chi d\tau}\bigg]^{1/2}, \label{deviation-intergral}
\end{eqnarray}
where $\Psi$ and $\Psi_{\mathrm{s}}$ are the auxiliary wavefields~(\ref{wavefield-auxiliary}) of, respectively, the RW in question and its fit with an exact solution, while $\Lambda$ is a region in the $(\chi,\tau)$-variables over which the deviation is calculated. 
To evaluate how the quality of the fitting deteriorates with increasing distance from the RW maximum, in the present paper the deviation~(\ref{deviation-intergral}) is calculated over $10$ nested rectangles $W_{m}$ whose sides are proportional to the sides of $W_{0}$,
\begin{eqnarray}
	&& (\chi,\tau)\in W_{m} = [-\Delta\chi_{m}, \Delta\chi_{m}]\times[-\Delta\tau_{m}, \Delta\tau_{m}], \nonumber\\
	&& \frac{\Delta\chi_{m}}{\Delta\chi} = \frac{\Delta\tau_{m}}{\Delta\tau} = \frac{12 - m}{12}, \quad\quad m = 0, ..., 9;
	\label{region-Wm}
\end{eqnarray}
these rectangles are shown in Fig.~\ref{fig:fig01}. 
Additionally, the deviations are calculated over $9$ bands between the rectangles, $bW_{m} = W_{m}\backslash W_{m+1}$, $m = 0, ..., 8$, where the backslash means the subtraction of sets. 


\subsection{Model solutions for rogue waves}
\label{Sec:NumMethods-3}

The recorded RWs are compared with the following nine exact solutions of the 1D-NLSE~(\ref{NLSE}): (1-3) the Peregrine (RB1), second-order (RB2) and third-order (RB3) rational breathers, (4-5) the Akhmediev (AB) and Kuznetsov (KB) breathers, (6) a simplified bound state of two solitons (SS2), (7) a simplified superposition of two Akhmediev breathers (AB2), (8) a general collision of two solitons (SS2v), and (9) a general collision of two Tajiri-Watanabe breathers (BS2v). 
The first five of these solutions have been discussed in Section~\ref{Sec:Theory-3}, while the last four solutions are described below. 

In the SS2 variant of two-soliton solution, the solitons have zero velocities, identical positions, and identical initial phases. 
Note that if, with equal velocities, soliton positions are significantly different, then the larger soliton oscillates weakly due to the influence of the smaller one, which does not resemble the typical RW dynamics. 
With equal velocities and positions, the maximum amplitude is reached when the soliton phases coincide, and this moment of time can be chosen as the initial time. 
These restrictions make it possible to treat such a solution analytically, representing it as
\begin{eqnarray}
    && \psi^{\mathrm{SS2}}(x,t) = 2\,\Omega\, \exp\bigg(i[\eta_{1}^2 + \eta_{2}^2]t\bigg)\times \nonumber\\
    && \times \frac{\eta_{1}e^{i\frac{\Omega t}{2}} \cosh 2\eta_{2}x - \eta_{2}e^{-i\frac{\Omega t}{2}} \cosh 2\eta_{1}x}{p^{2}\cosh 2 q x + q^{2} \cosh 2 p x - 4\eta_{1}\eta_{2} \cos\Omega t}, 
\label{2-SS}
\end{eqnarray}
where $\eta_{2}<\eta_{1}$ are the imaginary parts of soliton eigenvalues $\lambda_{1,2} = i\eta_{1,2}$, while $p = \eta_{1} + \eta_{2}$, $q = \eta_{1} - \eta_{2}$ and $\Omega = 2pq$. 
The maximum amplitude $|\psi^{\mathrm{SS2}}| = 2(\eta_{1}+\eta_{2})$ is reached at the point $x,t = 0$, where the phase and chirp equal zero, $\mathrm{arg}\,\psi^{\mathrm{SS2}} = 0$ and $\partial\,\mathrm{arg}\,\psi^{\mathrm{SS2}}/\partial x = 0$. 

Similarly, the AB2 variant of two-breather interaction represents a simplified situation when a plane wave $e^{it}$ is dressed with two solitons having imaginary eigenvalues $i\eta_{1,2}$, where $0<\eta_{2}\le \eta_{1} < 1$, and identical norming constants $C_{1,2} = -1$. 
In terms of parametrization~(\ref{C_param_AB}), this means that two ABs emerge at the same time $t_{0}=0$ with the same spatial phase shift $\theta_{0}=0$ (i.e., at the same place). 
The AB2 solution is too cumbersome to be written explicitly; it reaches maximum amplitude $|\psi^{\mathrm{AB2}}| = 1 + 2(\eta_{1}+\eta_{2})$ at the point $x,t = 0$, where its phase and chirp equal zero, $\mathrm{arg}\,\psi^{\mathrm{AB2}} = 0$ and $\partial\,\mathrm{arg}\,\psi^{\mathrm{AB2}}/\partial x = 0$. 
In the limit $\eta_{2}\to 0$, the AB2 model transforms into the AB (if also $\eta_{1}\to 1$, then it becomes the RB1), while in the limit $\eta_{1,2}\to 1$ the AB2 solution asymptotically converges to the RB2. 

The SS2v variant of two-soliton solution represents a general collision of two solitons, when the difference in their velocities is non-zero and no other constraints apply. 
Such solutions are too cumbersome for analytical treatment and are calculated numerically using the dressing method as discussed in Section~\ref{Sec:Theory-2}. 
The main difficulty in working with the SS2v model is that the points in the $(x,t)$-space where it reaches local maximums are unknown. 
For this reason, a database of pre-calculated SS2v local maximums is created by varying SS2v parameters over a certain numerical grid, and the closest match is selected from it during the fitting procedure.

More specifically, two solitons with eigenvalues $\xi_{1,2} + i\eta_{1,2}$ are placed at a distance $\ell = (\eta_{1}^{-1}+\eta_{2}^{-1})/2$ equal to the sum of their characteristic widths and the evolution of wavefield is observed over the time $t = \ell/2|\xi_{2}-\xi_{1}|$ during which they swap their positions relative to each other, see Eqs.~(\ref{x0theta-evolution})-(\ref{1-SS}). 
During the observation, all local maximums of the wavefield with an amplitude exceeding that of the larger soliton by more than $20$\% are collected; note that one collision can produce several suitable local maximums. 
The position of each maximum in space and time is further refined using the Newton's method. 
In addition to parameters $x_{01,02} = \pm\ell/2$, which determine the initial positions of solitons, the SS2v solution depends on six more real parameters: $\eta_{1,2}$, $\xi_{1,2}$, $\theta_{01}$ and $\theta_{02}$. 
Taking into account the scaling, gauge and Galilean transformations, one can set three of them as $\eta_{1} = 1$ (with $\eta_{2} \le \eta_{1}$), $\theta_{01} = 0$ and $\xi_{1} = -\xi_{2}$. 
When compiling the database, the remaining three parameters are iterated as follows: $\eta_{2}$ -- from $0.2$ to $1$ with $0.01$ step, $\xi_{2}$ -- from $0.01$ to $1$ with $0.01$ step, and $\theta_{02}$ -- from $0$ to $2\pi$ with $0.02\pi$ step. 
In total, $862\,809$ suitable local maximums are collected that do not transform into each other under the symmetry transformations $\psi(x,t)\to\psi(-x,t)$, $\psi(x,t)\to\psi^{*}(x,-t)$ and $\psi(x,t)\to\psi^{*}(-x,-t)$. 
The parameters of the corresponding SS2v solutions are transformed with Eqs.~(\ref{x0theta-evolution}) to make these maximums appear at the point $x,t = 0$, and the auxiliary wavefield~(\ref{wavefield-auxiliary}) around each local maximum is recorded within a ``fitting'' window $W_{\mathrm{f}}$, which will be discussed in Section~\ref{Sec:NumMethods-4} (see also Fig.~\ref{fig:fig01}). 
It is verified that iterating soliton parameters $\eta_{2}$, $\xi_{2}$ and $\theta_{02}$ over a denser grid does not lead to a significant improvement in the results.

Note that the SS2v database compiled in this way does not contain solutions in the form of bound states of two solitons, in which soliton velocities coincide exactly but soliton positions are slightly different (such solutions are not covered by the SS2 model as well). 
It is assumed that these solutions are approximated by those SS2v solutions, for which the difference in soliton velocities is small.

Similarly to the SS2v model, the BS2v variant of two-breather interaction represents a general collision of two TWBs with the only constrain that the difference in their velocities is non-zero. 
Such solutions are even more cumbersome, and the points in the $(x,t)$-space where they reach local maximums are also unknown. 
For this reason, another database of pre-calculated BS2v local maximums is created in a similar way as discussed for the SS2v solutions. 
Note that, in the BS2v case, the two colliding breathers can have either opposite or the same signs of real parts $\xi_{1,2}$ of their eigenvalues (assuming the branch cut is fixed at $[0,i]$), whereas a collision of two solitons can always be transformed so that $\xi_{1} = -\xi_{2}$. 
The BS2v database contains only those collisions, in which the breathers move in the opposite directions (having opposite signs of $\xi_{1,2}$): even this incomplete database already turned out to be very large and the calculation of collisions for breathers moving in the same direction takes a much longer time. 
Also note that, in the BS2v case, it is necessary to iterate five independent parameters -- $\eta_{1}$, $\eta_{2}$, $\xi_{1}$, $\xi_{2}$ and $\theta_{02}$, since the solution also contains the plane wave background represented by the branch cut $[0,i]$, and the positioning of soliton eigenvalues relative to it provides essentially different TWBs. 
Meanwhile, iteration over the plane wave's phase $\Theta$ is not necessary, since it only leads to the corresponding phase rotation of the entire solution, see Section~\ref{Sec:Theory-3}. 

When compiling the BS2v database, the aforementioned parameters are iterated as follows: $\eta_{1,2}$ -- from $0.1$ to $1$ with $0.02$ step (with $\eta_{2}\le\eta_{1}$), $|\xi_{1,2}|$ -- from $0.02$ to $1$ with $0.02$ step, and $\theta_{02}$ -- from $0$ to $2\pi$ with $0.02\pi$ step. 
This iteration corresponds to that for the SS2v database with two exceptions. 
First, when iterating $\xi_{1,2}$, the step is doubled: for the SS2v solutions, there is a relation $\xi_{1} = -\xi_{2}$, meaning that the difference $|\xi_{1}-\xi_{2}|$ in the SS2v case is iterated with same $0.02$ step as for the BS2v solutions. 
Second, for the BS2v case, the step in $\eta_{1,2}$ is doubled, since otherwise this database becomes too large. 
When selecting the BS2v local maximums, it is also required that their amplitude exceeds the amplitude of the plane wave background by at least two times. 
In total, $59\,486\,863$ suitable local maximums are collected that do not transform into each other under the symmetry transformations. 
The parameters of the corresponding BS2v solutions are transformed with Eqs.~(\ref{TWB-x0theta-evolution}) to make these maximums appear at the point $x,t = 0$, and the auxiliary wavefield~(\ref{wavefield-auxiliary}) around each local maximum is recorded within the $W_{\mathrm{f}}$ window. 
Preliminary simulations show that iteration over a denser grid does not lead to a significant improvement in the results. 




\subsection{The fitting procedure}
\label{Sec:NumMethods-4}

RWs are approximated by the discussed above model exact solutions from the condition that models must best fit the RWs in the immediate vicinity of their maximums. 
Then, each fit is evaluated against the RW in question within the comparison window $W_{0}$~(\ref{region-W0}). 
Both the selection of the fitting parameters and the evaluation of the accuracy of the fit are performed via the corresponding auxiliary wavefields~(\ref{wavefield-auxiliary}), in which the information about the maximum amplitude of the original RW, together with the phase and chirp at the maximum point, is removed. 
The advantage of this method is that it allows the direct use of canonical equations for the model solutions, for instance, Eqs.~(\ref{A-breather}) and~(\ref{KM-breather}) for the AB and KB models, assuming unit amplitude of the plane wave background. 
Then, the fitting solutions can be obtained in the original $(x,t)$-variables by performing a transformation inverse to~(\ref{wavefield-auxiliary}), so that the fits will have maximum at the same point $(x_{0}, t_{0})$ and with the same amplitude $A_{0}$, phase $\Phi_{0}$ and chirp $\vartheta$ as the considered RW.

The RB1, RB2 and RB3 fits are completely determined by the described procedure. 
For the AB, KB and SS2 fits, there remains one parameter that needs to be calculated: for the AB~(\ref{A-breather}) and KB~(\ref{KM-breather}), this is the imaginary part of breather eigenvalue $\eta_{1}$, while for the SS2~(\ref{2-SS}) this is the ratio $\eta_{2}/\eta_{1}$ of the imaginary parts of soliton eigenvalues. 
In principal, these parameters can be found from the equality to some other values characterizing the RW maximum, for instance, the second spatial derivative of the wave amplitude $|\psi|_{xx}$. 
However, not all combinations of $|\psi|$ and $|\psi|_{xx}$ at the maximum point are allowed for these exact solutions, which often makes such a fitting procedure impossible. 
For this reason, another method is used, in which the missing parameter is determined by minimizing (with the Newton's method) the difference between RWs and their fits in a small sub-window $(\chi,\tau)\in W_{\mathrm{f}} = [-\Delta\chi_{\mathrm{f}}, \Delta\chi_{\mathrm{f}}]\times[-\Delta\tau_{\mathrm{f}}, \Delta\tau_{\mathrm{f}}]$, see Fig.~\ref{fig:fig01}, around the maximum. 
The values $\Delta\chi_{\mathrm{f}}\approx 0.2$ and $\Delta\tau_{\mathrm{f}}\approx 0.26$ used correspond to the situation when, if a RW is a Peregrine breather, then at the points $(\pm\Delta\chi_{\mathrm{f}}, 0)$ and $(0, \pm\Delta\tau_{\mathrm{f}})$ its amplitude equals $90$\% of the maximum. 

Note that, for the AB and KB fits, this method almost always results in the imaginary parts of their eigenvalues asymptotically approaching unity, so that the corresponding breathers converge to the RB1 solution. 
For this reason, fitting with these two models will not be discussed further in the present paper.

The AB2 fit depends on two parameters, $\eta_{1,2}$, that need to be found during the fitting procedure. 
These parameters are found in the same way as discussed above using the Newton's method in two dimensions ($\eta_{1}$ and $\eta_{2}$). 
The parameters of the SS2v and BS2v fits are determined by minimizing the difference between RWs and these solutions in the same subwindow $W_{\mathrm{f}}$ by searching through the SS2v and BS2v databases, respectively. 

In the described above fitting procedure, the RB1, RB2 and RB3 models can be called zero-parameter fits, since they are completely determined by the condition that models must best fit the RWs in the immediate vicinity of their maximums. 
The SS2 model can be called one-parameter fit, as one internal parameter determining its shape needs to be found. 
Similarly, the AB2 model represents a two-parameter fit, while the SS2v and BS2v models depend on the same internal parameters that are iterated to construct the corresponding databases (i.e., these are three- and five-parameter fits, respectively). 
The number of internal parameters can be understood as model flexibility: the more of them, the greater the possibility of fitting the model to a given RW. 
From this point of view, the BS2v model should provide the best results, followed by the SS2v, then the AB2, then the SS2, and then the RB1, RB2 and RB3 models. 

Also note that there are alternative methods for fitting and subsequent comparing RWs with model solutions. 
For instance, one can fit a RW with an exact solution by fitting their expansions in Taylor series up to the second-order terms near the local maximum. 
In these expansions, there exist specific relations between the gradients and elements of the Hessian matrix. 
Accounting for them and using the representation of wavefield $\psi = r\,e^{i\phi}$ via its amplitude $r$ and phase $\phi$, it is easy to see that, for such a fitting procedure, one needs to know only the following eight parameters at the RW maximum point $(x_{0}, t_{0})$,
\begin{eqnarray}
	\{\,r,\, r_{xx},\, r_{xt},\, r_{tt},\, \phi,\, \phi_{x},\, \phi_{xt},\, \phi_{tt}\}|_{(x_{0},t_{0})}. 
	\label{altermative-fitting-string}
\end{eqnarray}
Indeed, at $(x_{0}, t_{0})$, the amplitude gradients are zero, $r_{x} = r_{t} = 0$, and from the 1D-NLSE it follows that 
\begin{eqnarray}
	\phi_{xx}|_{(x_{0},t_{0})} = 0, \quad\,\,\, \phi_{t}|_{(x_{0},t_{0})} = \frac{r_{xx}}{2r} - \frac{\phi_{x}^{2}}{2} + r^{2}. 
	\label{altermative-fitting-relations}
\end{eqnarray}
For this approach, (i) the calculation of the auxiliary wavefields~(\ref{wavefield-auxiliary}) is not required, and (ii) RWs can be compared against their fits on the original numerical grid that comes from the simulation of wavefield dynamics. 
It has been implemented for comparison, and the results obtained turned out to be the same.


\section{Rogue waves: frequency of occurrence and distributions by intensity, chirp and in space-time}
\label{Sec:Results1}


\subsection{Frequency of occurrence}
\label{Sec:Results1-1}

\begin{figure}[t]\centering
	\includegraphics[width=8.5cm]{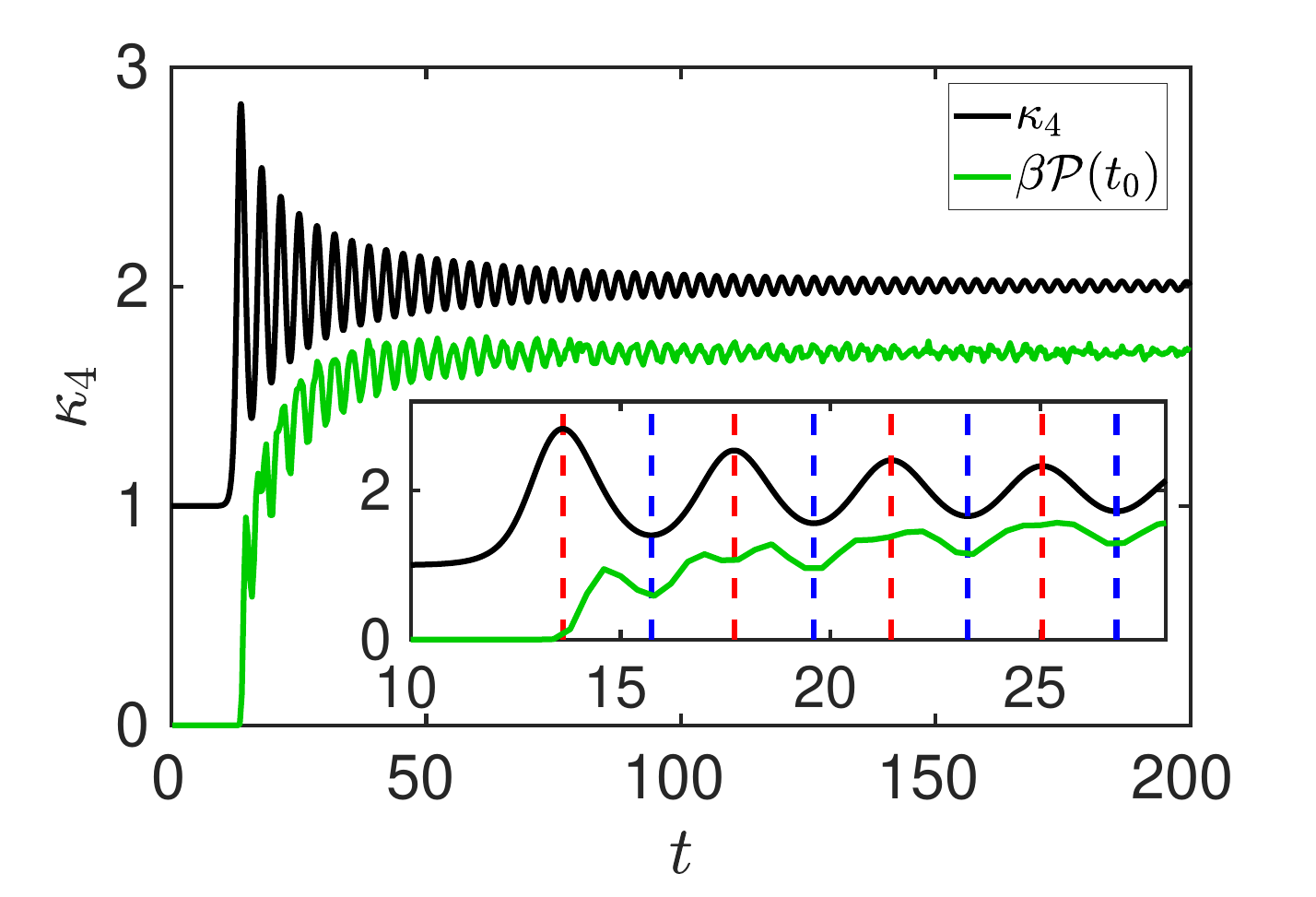}
	
	\caption{\small {\it (Color on-line)} 
	Evolution of the fourth-order moment of amplitude $\kappa_{4} = \langle\overline{|\psi|^{4}}\rangle/\langle\overline{|\psi|^{2}}\rangle^{2}$ (black) and distribution $\mathcal{P}(t_{0})$ of moments of time $t_{0}$ when RWs reach their maximum amplitude (green), scaled with a coefficient $\beta = 307$ for better visualization. 
	The inset shows a zoomed-in beginning of the nonlinear stage of MI (the transient regime), with the red dashed lines indicating the moments of the first four local maximums of $\kappa_{4}$ at $t = 13.6$, $17.7$, $21.4$ and $25$, and the blue dashed lines -- the moments of the first four local minimums at $t = 15.7$, $19.6$, $23.3$ and $26.8$.
	}
	\label{fig:fig02}
\end{figure}

Within the framework of the 1D-NLSE~(\ref{NLSE}), the MI of a plane wave of unit amplitude affects modulations with wavenumbers $k\in(-2, 2)$; the maximum increment $\gamma_{\max} = 1$ of MI is achieved at wavenumbers $k = \pm\sqrt{2}$. 
With initial conditions~(\ref{IC-PW}) and initial noise amplitude $a_{0} = 10^{-5}$, the MI reaches its nonlinear stage at $t\simeq 12$, as illustrated in Fig.~\ref{fig:fig02} with evolution of the fourth-order moment of amplitude $\kappa_{4} = \langle\overline{|\psi|^{4}}\rangle/\langle\overline{|\psi|^{2}}\rangle^{2}$. 
In~\cite{agafontsev2015integrable}, it has been discovered that the subsequent dynamics turns out to be oscillatory. 
In particular, the moments of amplitude perform quasi-harmonic oscillations with an amplitude decaying as $t^{-3/2}$, a frequency equal to twice the maximum growth rate of MI, $s = 2\gamma_{\max}$, and a nonlinear phase shift decaying as $t^{-1/2}$. 
Asymptotically, the system approaches its long-time statistically stationary state, in which the statistical functions, such as the moments of amplitude, Fourier spectrum, spatial correlation functions and probability density function (PDF) of intensity $|\psi|^{2}$, no longer change with time. 

Distribution $\mathcal{P}(t_{0})$ of moments of time $t_{0}$ when RWs reach their maximum amplitude is shown in Fig.~\ref{fig:fig02} with the green line. 
Here and below, $\mathcal{P}$ denotes the PDF of its arguments, the integral of which over their range of values is normalized to unity; in the figure, the PDF $\mathcal{P}(t_{0})$ is scaled for better joint visualization with the evolution of the fourth-order moment $\kappa_{4}(t)$. 
Unsurprisingly, the PDF $\mathcal{P}(t_{0})$, which can be called the frequency of RW occurrence, equals zero until the beginning of nonlinear stage of MI. 
Surprisingly, at the moment of time $t\approx 13.6$, when the fourth-order moment reaches its first (largest) local maximum $\kappa_{4}\approx 2.9$, the frequency $\mathcal{P}(t_{0})$ only begins to increase from the zero level. 
At later times, it increases in an oscillatory manner with a decaying amplitude of oscillations and asymptotically reaches its stationary level $P_{\infty}$ at long time. 
During this evolution, the local minimums of $\kappa_{4}(t)$ and $\mathcal{P}(t_{0})$ coincide, but the local maximums of $\mathcal{P}(t_{0})$ lag behind those of $\kappa_{4}(t)$. 
Moreover, in oscillations of the fourth-order moment $\kappa_{4}$ from the second to the fourth, one can see oscillations of $\mathcal{P}(t_{0})$ with doubled frequency, when between two neighboring local minimums of $\kappa_{4}$ there are two local maximums of $\mathcal{P}(t_{0})$ -- one before the local maximum of $\kappa_{4}$, and the second after. 

The frequency $\mathcal{P}(t_{0})$ reaches half of its asymptotic long-time value $P_{\infty}$ during the second oscillation of $\kappa_{4}$ and becomes close to $P_{\infty}$ by the fifth oscillation of $\kappa_{4}$ at $t_{0}\gtrsim 30$. 
Since in the transient regime $12\lesssim t\lesssim 30$ the frequency $\mathcal{P}(t_{0})$ changes rapidly and remains significantly smaller than its long-time value $P_{\infty}$, further this paper will consider only those RWs that emerge near the statistically stationary state of MI. 

Note that this stationary state is essentially nonlinear~\cite{agafontsev2015integrable}: its potential energy is twice larger than the kinetic one, $\langle|H_{nl}|\rangle/\langle H_{l}\rangle \approx 2$, so that the nonlinear effects are leading in the wave dynamics. 
Also, remarkably, the PDF of intensity in this state coincides with the exponential distribution, $\exp(-|\psi|^{2})$, which characterizes a random superposition of linear waves with unit average intensity $\overline{|\psi|^{2}} = 1$.

\begin{figure}[t]\centering
	\includegraphics[width=8.5cm]{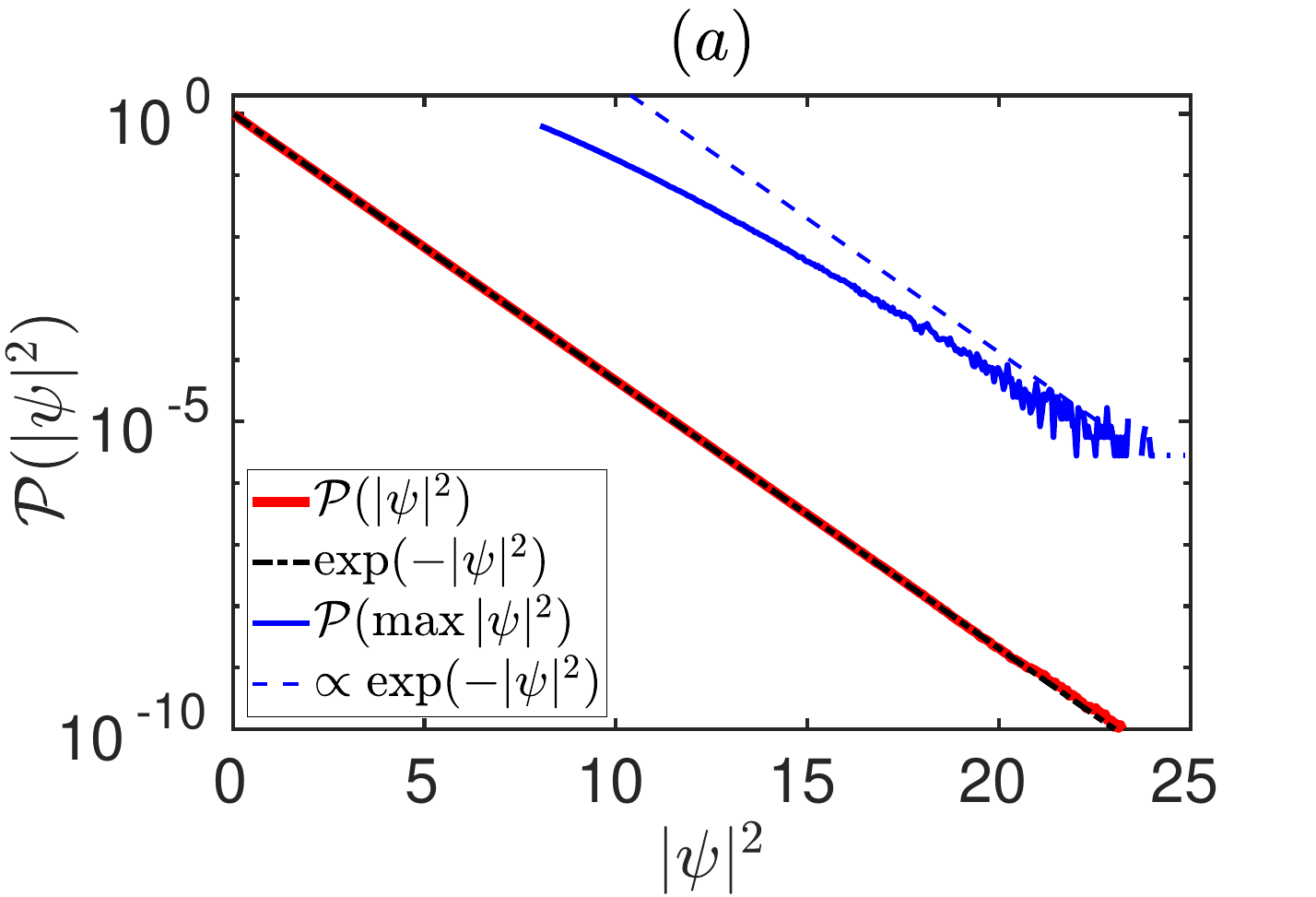}\\
	\includegraphics[width=8.5cm]{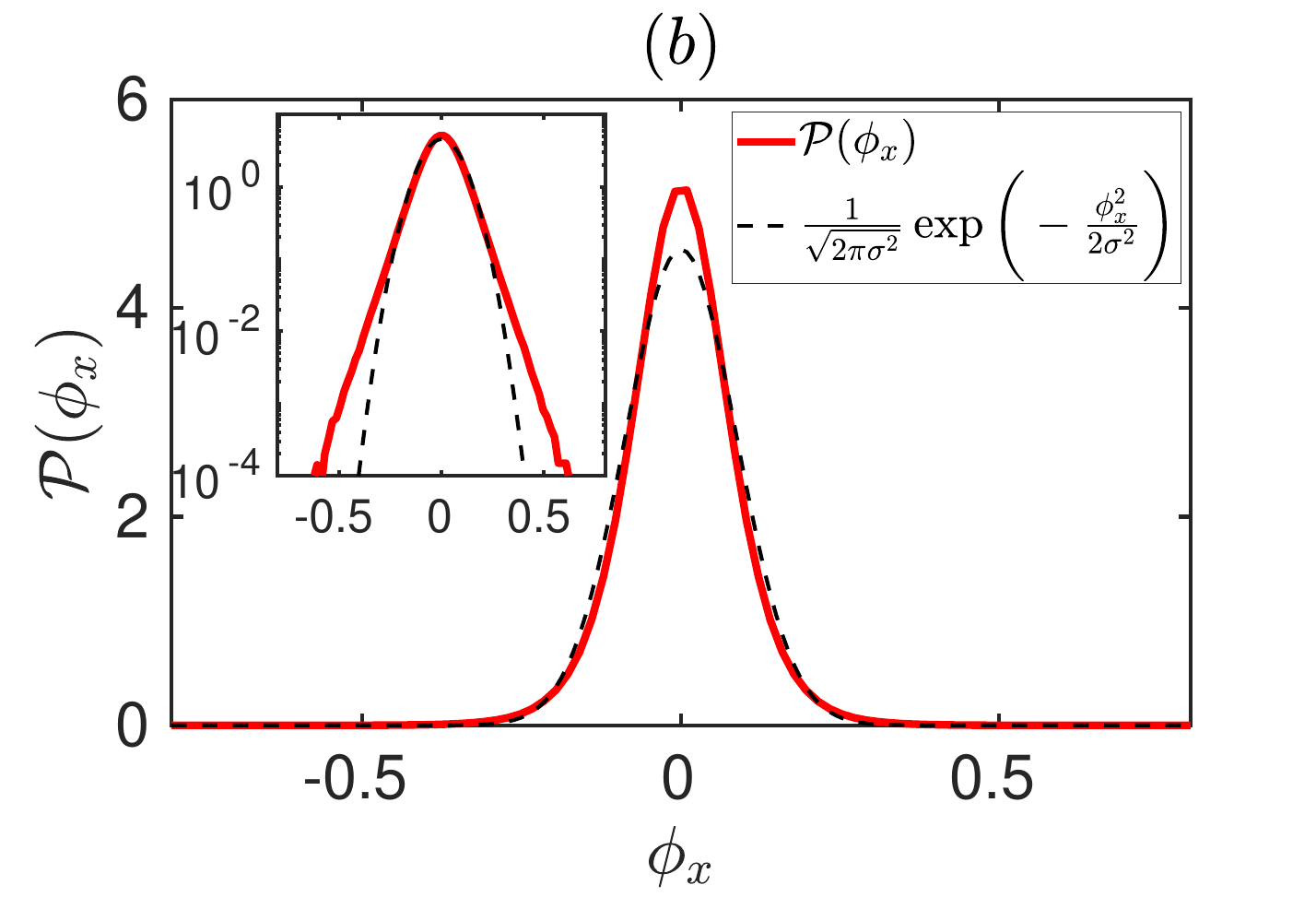}
	
	\caption{\small {\it (Color on-line)} 
	(a) PDFs of intensity for the whole wavefield $\mathcal{P}(|\psi|^{2})$ (red) and for the maximums of collected RWs $\mathcal{P}(\max|\psi|^{2})$ (blue), and (b) PDF of chirp $\phi_{x} = \partial\,\mathrm{arg}\,\psi/\partial x$ at the RW maximums, for RWs close to the statistically stationary state $t\in [80, 200]$ of MI. 
	In panel (a), the black dash-dot line shows the exponential distribution $\exp(-|\psi|^{2})$, while the blue dashed line indicates exponential dependency $\propto\exp(-|\psi|^{2})$. 
	In panel (b), the inset shows the same PDF in semi-logarithmic scale, while the black dashed line marks the Gaussian distribution of the same variance.
	}
	\label{fig:fig03}
\end{figure}

\begin{figure}[t]\centering
	\includegraphics[width=8.5cm]{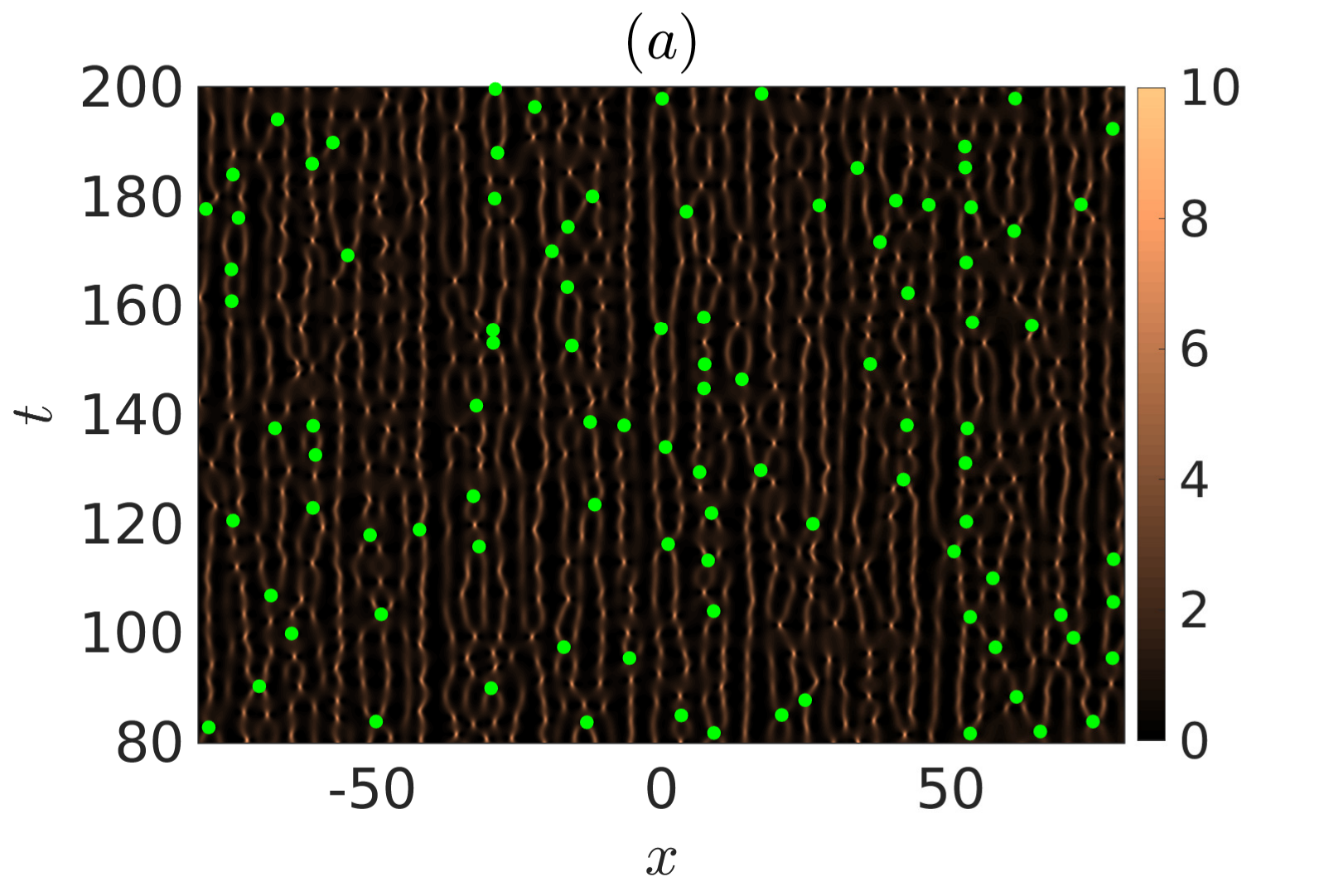}\\
	\includegraphics[width=8.5cm]{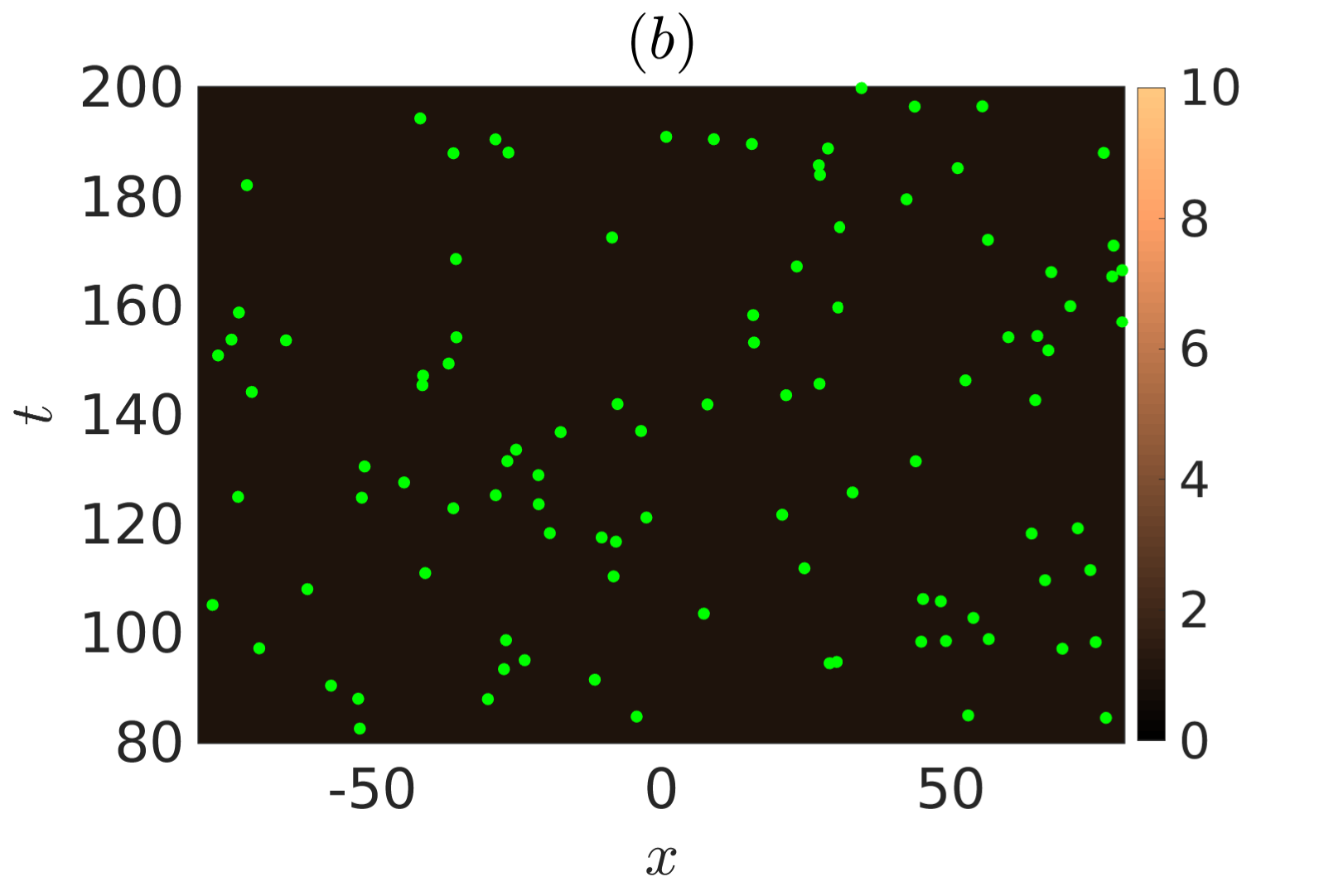}
	
	\caption{\small {\it (Color on-line)} 
	(a) Space-time evolution near the statistically stationary state of MI, $t\in[80,200]$, from one realization of initial conditions~(\ref{IC-PW}). 
	Color shows the intensity $|\psi|^{2}$, while green dots indicate the positions of RW maximums. 
	(b) Random uniformly distributed positions (green dots) of the same number as the number of RWs detected in the experiment from panel (a). 
	The color background is added for ease of comparison with panel (a) and corresponds to unit intensity. 
	}
	\label{fig:fig04}
\end{figure}

\begin{figure}[t]\centering
	\includegraphics[width=8.5cm]{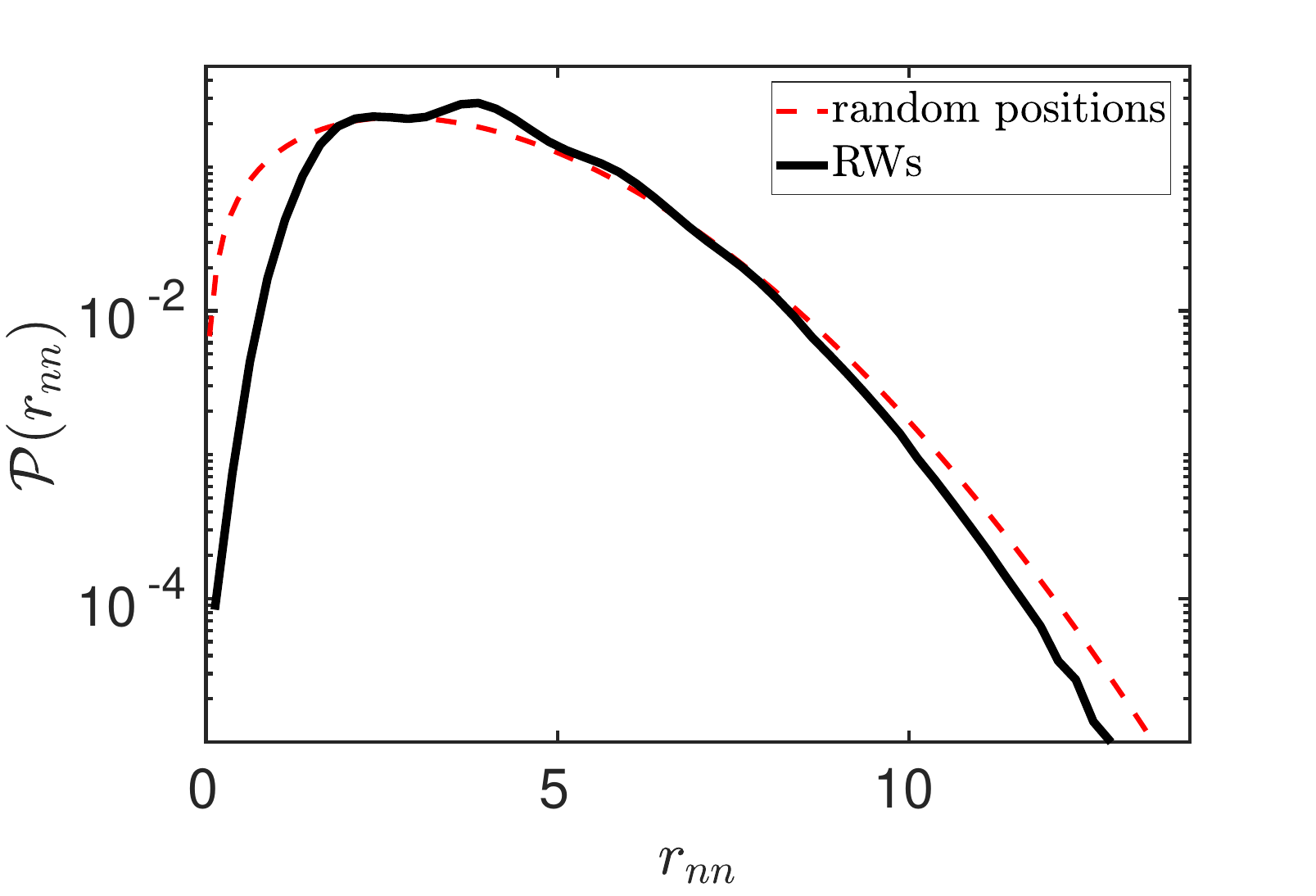}
	
	\caption{\small {\it (Color on-line)} 
	Ensemble-averaged PDFs of space-time distance to the nearest neighbor $r_{\mathrm{nn}}$ for RW maximums (thick black) and random uniformly distributed positions~(\ref{PDF-nn-random-positions}) with density $\mu = 2.12\times 10^{-2}$ per unit area (dashed red). 
	When calculating $r_{\mathrm{nn}}$, the space-time distance between two points is measured according to Eq.~(\ref{distance-space_time}) and the periodic boundary conditions are forced in the time interval $t\in[80,200]$ to model points outside of this interval. 
	}
	\label{fig:fig05}
\end{figure}


\subsection{Distributions by intensity, chirp and in space-time}
\label{Sec:Results1-2}

In the time interval $t\in [80, 200]$, which can be considered as sufficiently close to the stationary state, within the box $x\in[-L/2, L/2)$, $L=128\sqrt{2}\pi$, and using the ensemble of $20\,000$ simulations from initial conditions~(\ref{IC-PW}), a total of $7\,184\,000$ RWs have been collected. 
Hence, one simulation produced in average $(359.2 \pm 16.7)$ RWs, so that one RW event took place per $(190.4\pm 8.9)$ units of spatio-temporal area; here and below, the first number in round brackets means the mean and the second means one standard deviation. 
The maximum amplitude and intensity of RWs equal $\max|\psi| = (3.06\pm 0.21)$ and $\max|\psi|^{2} = (9.4\pm 1.4)$, respectively. 

Figure~\ref{fig:fig03}(a) shows the distribution of RWs by intensity at the maximum point, $\mathcal{P}(\max|\psi|^{2})$, together with the PDF of intensity for the entire wavefield $\mathcal{P}(|\psi|^{2})$ averaged over the ensemble and time interval $t\in [80, 200]$. 
Note that since RWs with $\max|\psi|^{2} < 8$ do not exist, the PDF for local maximums is shifted upwards relative to the PDF of the entire wavefield. 
As demonstrated in Fig.~\ref{fig:fig03}(a), the PDF for local maximums $\mathcal{P}(\max|\psi|^{2})$ is slightly bent downwards and decays slower than exponentially ($\propto e^{-\max|\psi|^{2}}$). 
No areas with a noticeable change in behavior are visible over the entire measured interval $8 \le \max|\psi|^{2} \lesssim 25$. 
This suggests that the mechanism of RW formation should be universal for RWs of significantly different amplitudes, e.g., for RWs with $\max|\psi| \simeq 3$  and $\simeq 5$. 
The PDF for the entire wavefield $\mathcal{P}(|\psi|^{2})$ coincides with the exponential distribution $\exp(-|\psi|^{2})$, confirming the result first found in~\cite{agafontsev2015integrable}. 

At the points of RW maximums, the spatial phase slope (chirp) $\phi_{x} = \partial\,\mathrm{arg}\,\psi/\partial x$ is narrowly distributed around zero, see Fig.~\ref{fig:fig03}(b), with $\mathrm{std}(\phi_{x})\approx 0.09$. 
The tails of its PDF at $|\phi_{x}|\gtrsim 0.25$ lie significantly higher than the Gaussian distribution of the same variance, so that, for a few RWs, chirp turns out to be rather significant, $|\phi_{x}|\simeq 0.5$. 
This parameter is important, since when the fits with exact solutions are transformed to the original $(x,t)$-variables using the transformation inverse to~(\ref{wavefield-auxiliary}), it affects their eigenvalues and phases, see Appendix~\ref{Sec:App1}.

Figure~\ref{fig:fig04}(a) shows the typical space-time evolution of intensity $|\psi|^{2}$, as well as the positions of RW maximums, near the statistically stationary state of MI for a simulation from one realization of initial conditions. 
Fig.~\ref{fig:fig04}(b) illustrates random uniformly distributed points of the same number as the number of RWs detected in the experiment shown in Fig.~\ref{fig:fig04}(a). 
The two figures look very similar. 
For random positions one can notice several groups, in which a number of points practically merge into one point, and no such groups are visible for the positions of RW maximums. 
Hence, RWs do not appear to be located too close to each other. 

To verify the latter hypothesis, one can calculate a space-time distance $r_{\mathrm{nn}}$ to the nearest neighbor for each RW, find the ensemble-averaged distribution of these distances $\mathcal{P}(r_{\mathrm{nn}})$, and compare the result with that calculated using random uniformly distributed positions. 
For such a comparison, one needs to define a space-time distance between two RW maximums located at the points $(x_{i}, t_{i})$ and $(x_{j}, t_{j})$, which can be done as follows,
\begin{eqnarray}
	r_{ij} = \sqrt{\frac{(x_{i} - x_{j})^{2}}{\Delta x_{\mathrm{RB1}}^{2}} + \frac{(t_{i} - t_{j})^{2}}{\Delta t_{\mathrm{RB1}}^{2}}}.
	\label{distance-space_time}
\end{eqnarray}
Here $\Delta x_{\mathrm{RB1}} = \sqrt{3} \approx 1.7$ and $\Delta t_{\mathrm{RB1}} = \sqrt{5.4} \approx 2.3$ are the characteristic sizes of the RB1 solution~(\ref{P-breather}) in space and time, respectively, which are used in Eq.~(\ref{distance-space_time}) to sum dimensionless quantities under the square root and make the distances $r_{ij}$ and $r_{\mathrm{nn}}$ dimensionless as well. 
More specifically, $\Delta x_{\mathrm{RB1}}$ is distance between the two zeros of the RB1 in the $x$-direction, while $\Delta t_{\mathrm{RB1}}$ is full width at half maximum (FWHM) of the RB1 in the $t$-direction. 
Note that, if $\Delta x_{\mathrm{RB1}}$ were defined as a FWHM too, then the value obtained would be too small and the time difference between RWs would have little effect on $r_{\mathrm{nn}}$.

Figure~\ref{fig:fig05} shows two PDFs of distance to the nearest neighbor: one for the RW maximums (thick black), and the other is the known result~\cite{chiu2013stochastic},
\begin{eqnarray}
	\mathcal{P}_{\mathrm{rnd}}(r_{\mathrm{nn}}) = 2\pi\mu\,r_{\mathrm{nn}}\,\exp(-\pi\mu r_{\mathrm{nn}}^{2}),
	\label{PDF-nn-random-positions}
\end{eqnarray}
for random uniformly distributed positions in a two-dimensional space (dashed red), where $\mu$ is the density of points per unit area. 
With $359.2$ RWs per one simulation, and for the space-time distance defined as in~(\ref{distance-space_time}), the density equals $\mu\approx 2.12\times 10^{-2}$. 
Relative to the PDF~(\ref{PDF-nn-random-positions}), the PDF for RWs demonstrates a much faster decay when the distance $r_{\mathrm{nn}}$ decreases below $2$, a small bump at $r_{\mathrm{nn}}\simeq 4$, and a slightly faster decaying tail at large $r_{\mathrm{nn}}$. 
Hence, RWs indeed do not emerge too close to each other. 
The mean distance to the nearest neighbor equals $\langle r_{\mathrm{nn}}^{\mathrm{RW}}\rangle \approx 3.8$ for RWs and $\langle r_{\mathrm{nn}}^{\mathrm{rnd}}\rangle = 0.5\mu^{-1/2} \approx 3.4$ for random positions, so that the Clark -- Evans aggregation index $R = \langle r_{\mathrm{nn}}^{\mathrm{RW}}\rangle/\langle r_{\mathrm{nn}}^{\mathrm{rnd}}\rangle \approx 1.1 > 1$ indicates a slight repulsion between RWs. 


\begin{figure*}[t]\centering
	\includegraphics[width=0.32\linewidth]{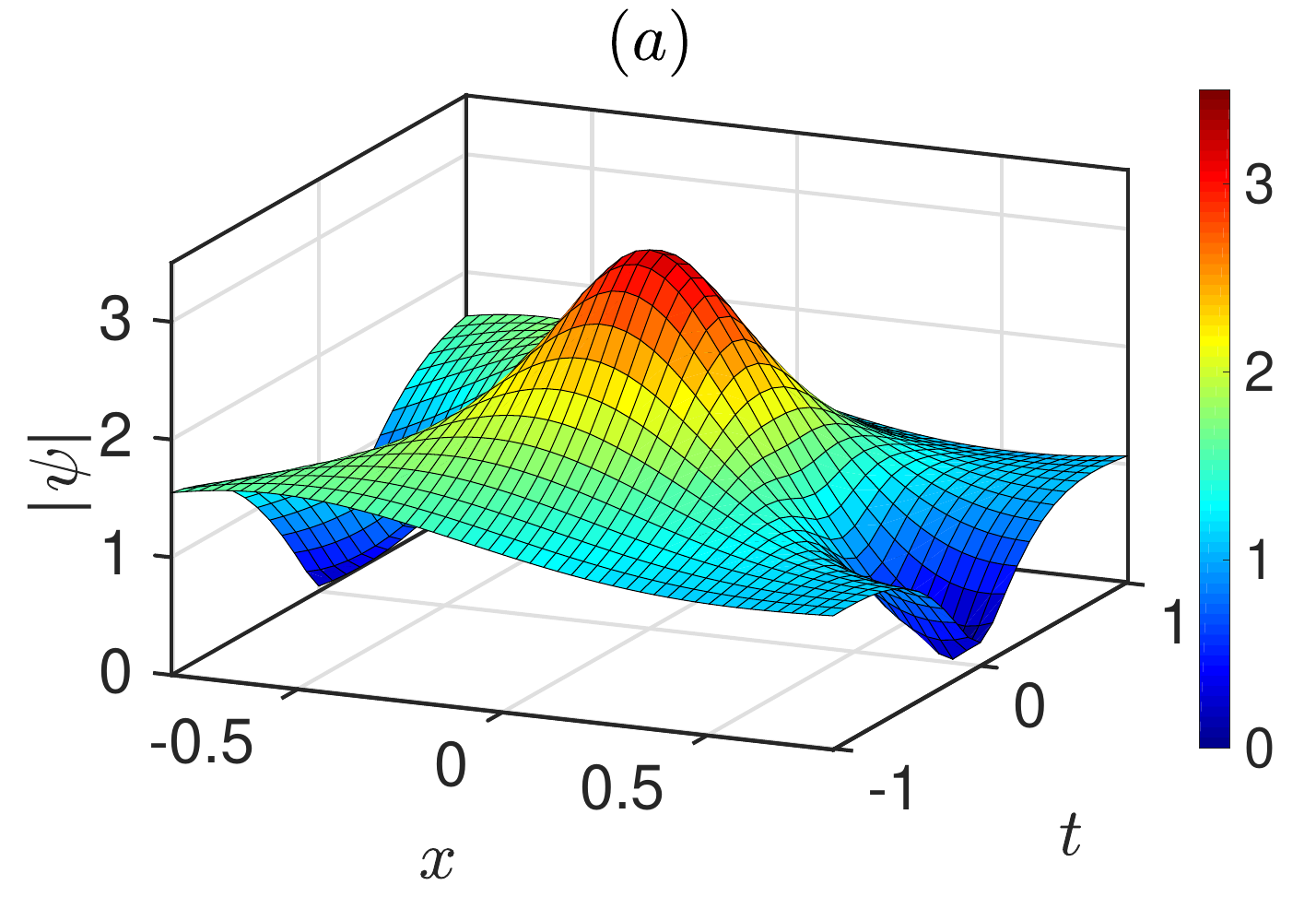}
	\includegraphics[width=0.32\linewidth]{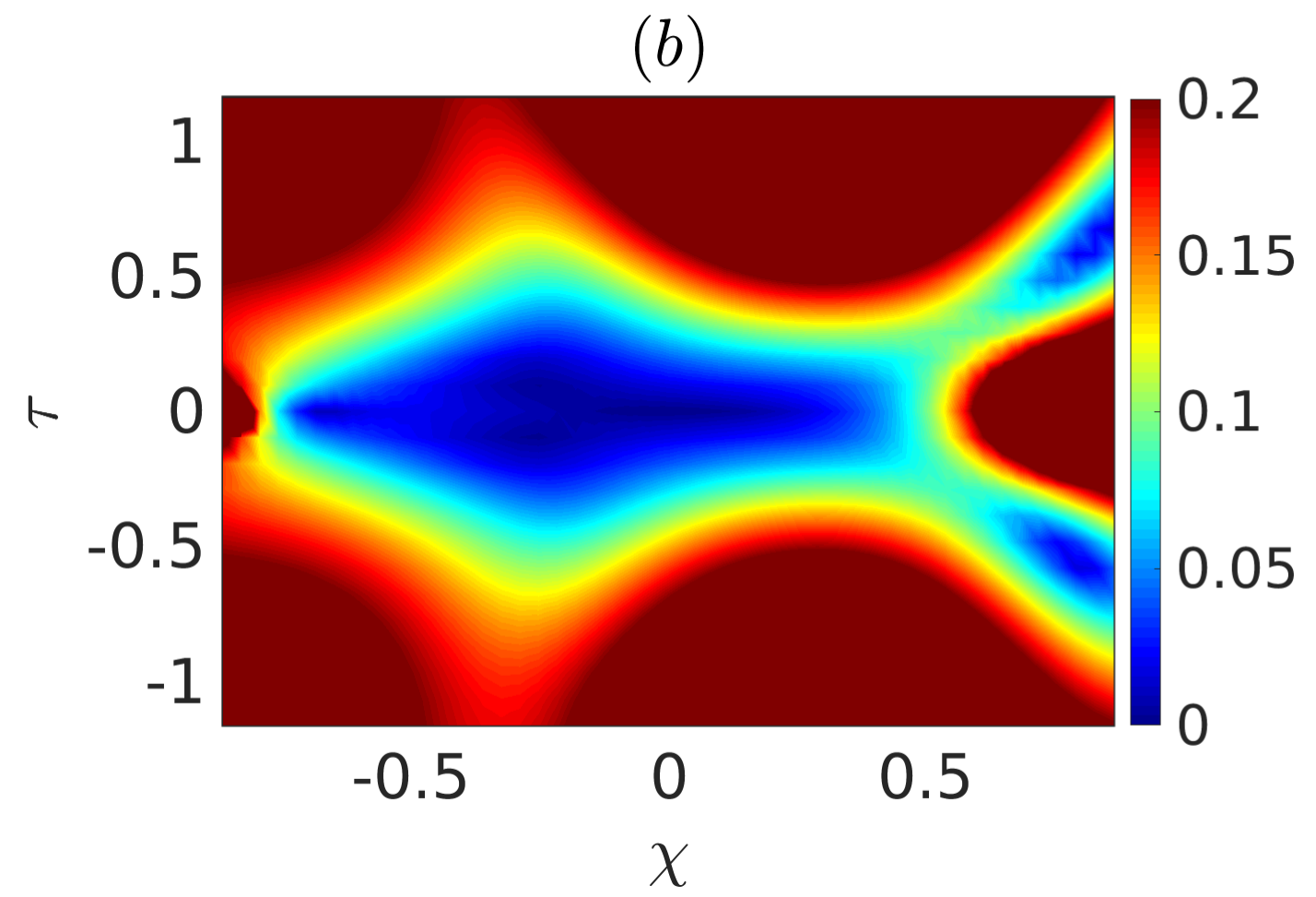}
	\includegraphics[width=0.32\linewidth]{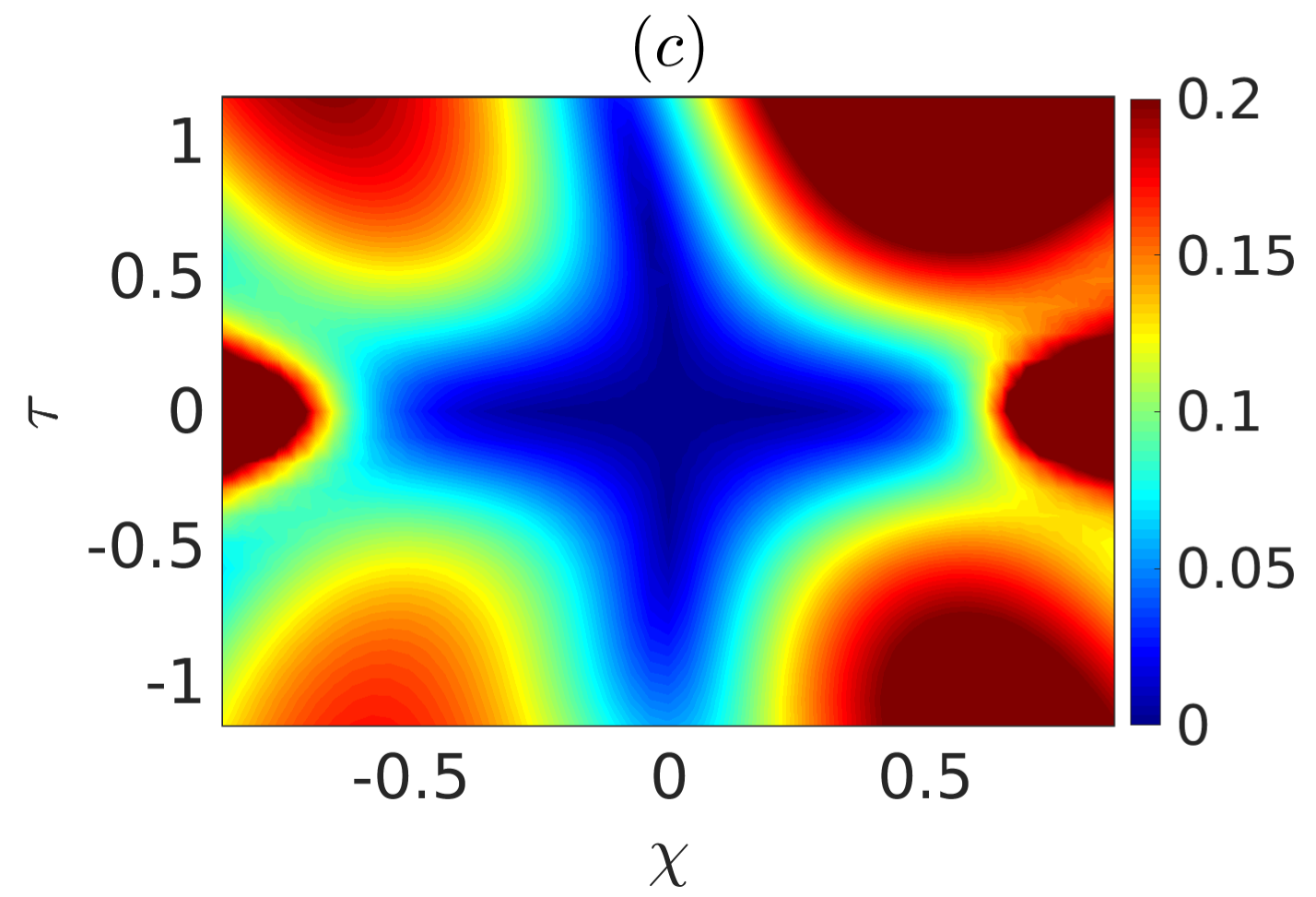}\\
	\includegraphics[width=0.32\linewidth]{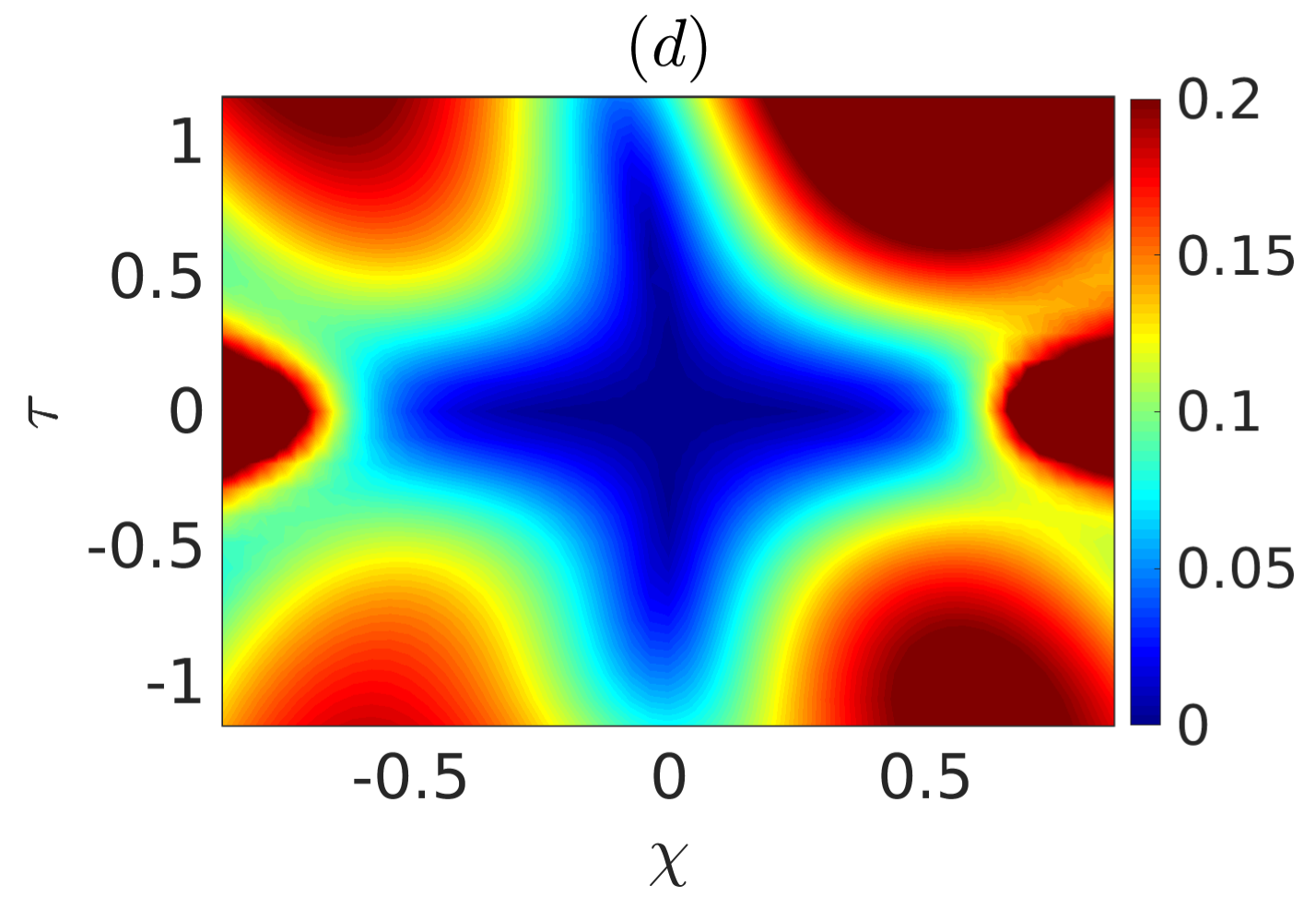}
	\includegraphics[width=0.32\linewidth]{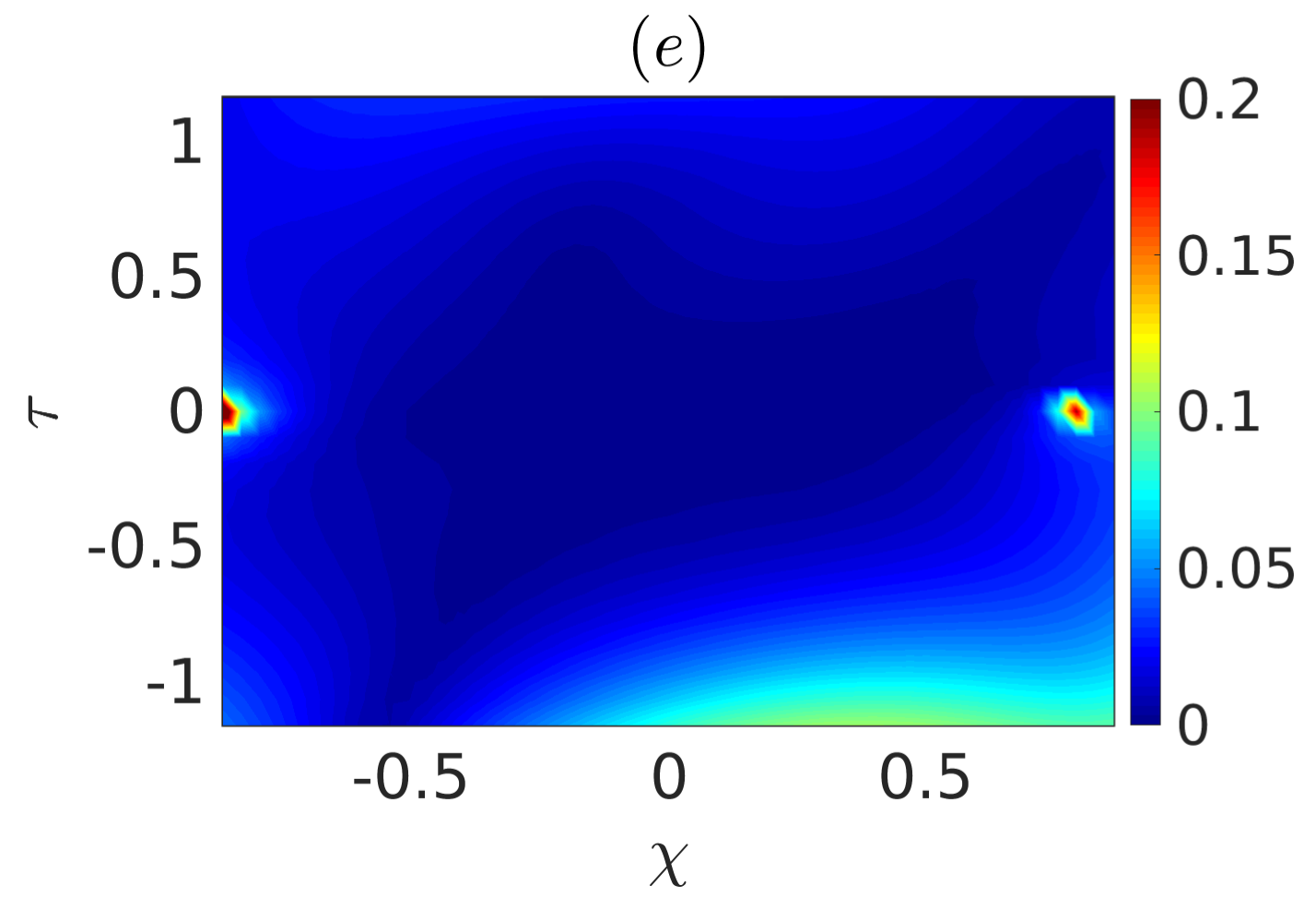}
	\includegraphics[width=0.32\linewidth]{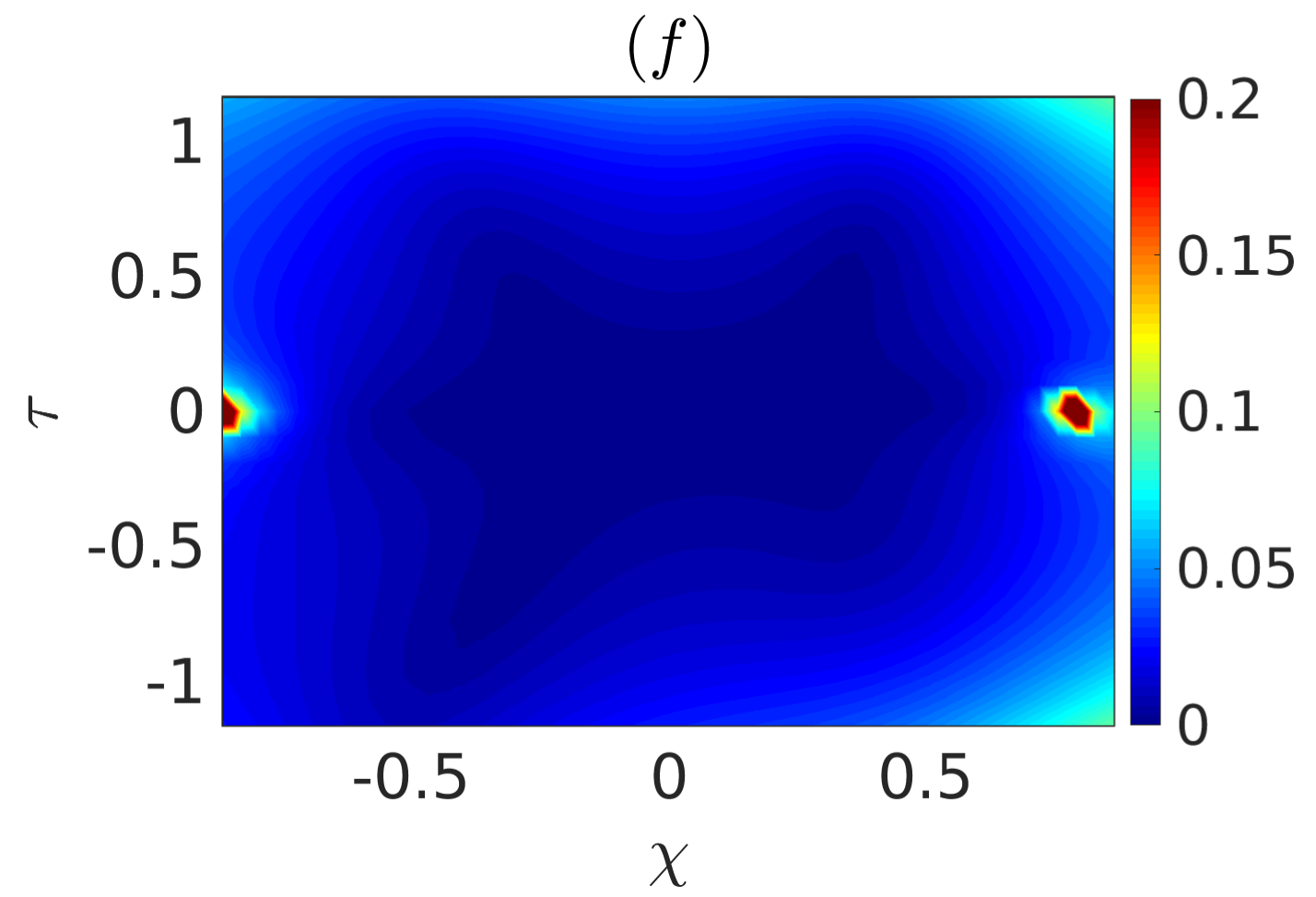}
	
	\caption{\small {\it (Color on-line)} 
	(a) Space-time representation of amplitude $|\psi(x,t)|$ near one of the collected RW events, and (b-f) relative deviations~(\ref{deviation-relative}) in the $(\chi, \tau)$-space between the auxiliary wavefields~(\ref{wavefield-auxiliary}) of this RW and its fits with exact solutions: (b) - RB1, (c) - SS2, (d) - AB2, (e) - SS2v, and (f) - BS2v. 
	In panel (a), the RW is shifted to the coordinate origin for better visualization. 
	In panels (b-f), relative deviations greater than $0.2$ are shown with constant deep red color. 
	The areas in the $(x,t)$-space in panel (a) and in the $(\chi, \tau)$-space in panels (b-f) are in one-to-one correspondence with each other. 
	The parameters of the fits with exact solutions are presented in Table~\ref{tab:M1}. 
	}
	\label{fig:fig06}
\end{figure*}

\begin{figure}[t]\centering
	\includegraphics[width=8.5cm]{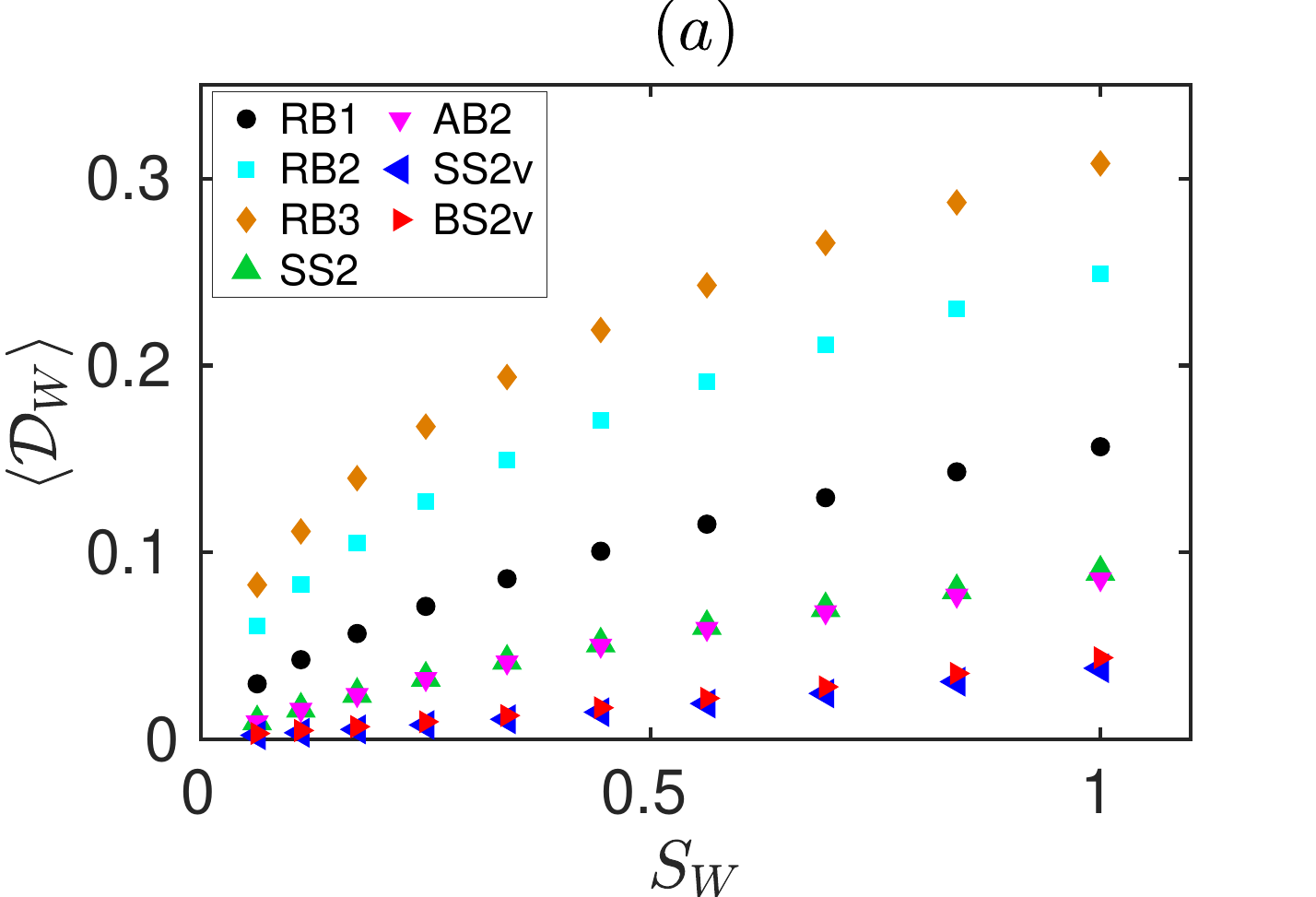}\\
	\includegraphics[width=8.5cm]{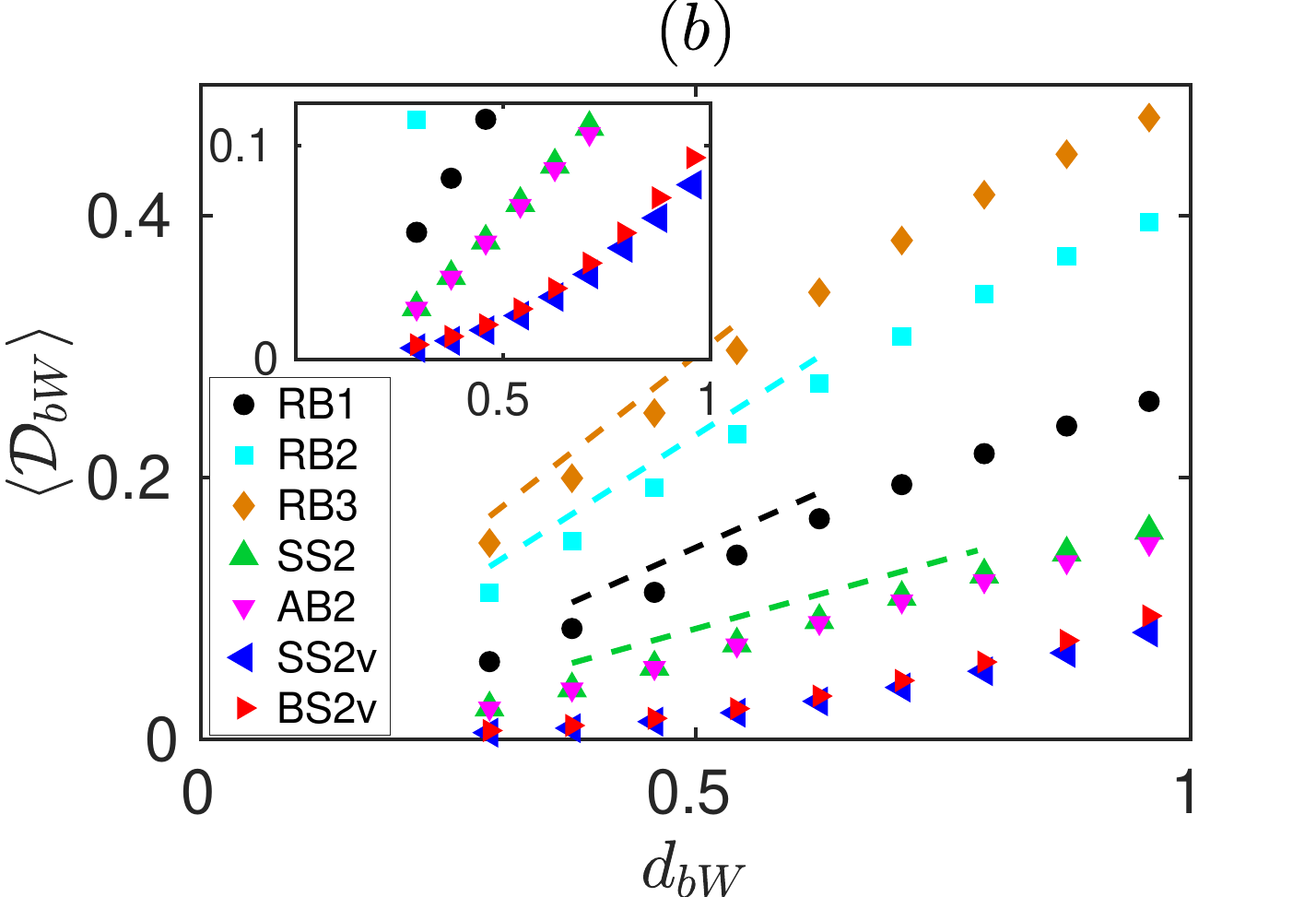}
	
	\caption{\small {\it (Color on-line)}
	Integral deviations~(\ref{deviation-intergral}) between RWs and their fits with exact solutions, averaged over all collected RWs: (a) calculated over the $10$ windows $W_{m}$~(\ref{region-Wm}), $m=0, ..., 9$, and (b) calculated over the $9$ bands between the windows $bW_{m} = W_{m}\backslash W_{m+1}$, $m = 0, ..., 8$. 
	In panel (a), $S_{W}$ stands for the area of window $W_{m}$ divided by the area of the largest window $W_{0}$. 
	In panel (b), $d_{bW}$ denotes the mean diagonal of the band divided by the diagonal of window $W_{0}$ (i.e., it is a dimensionless distance from the RW maximum), while the dashed lines indicate regions of linear growth with $d_{bW}$ and the inset shows zoomed-in SS2v and BS2v dependencies. 
	}
	\label{fig:fig07}
\end{figure}

\begin{figure}[t]\centering
	\includegraphics[width=8.5cm]{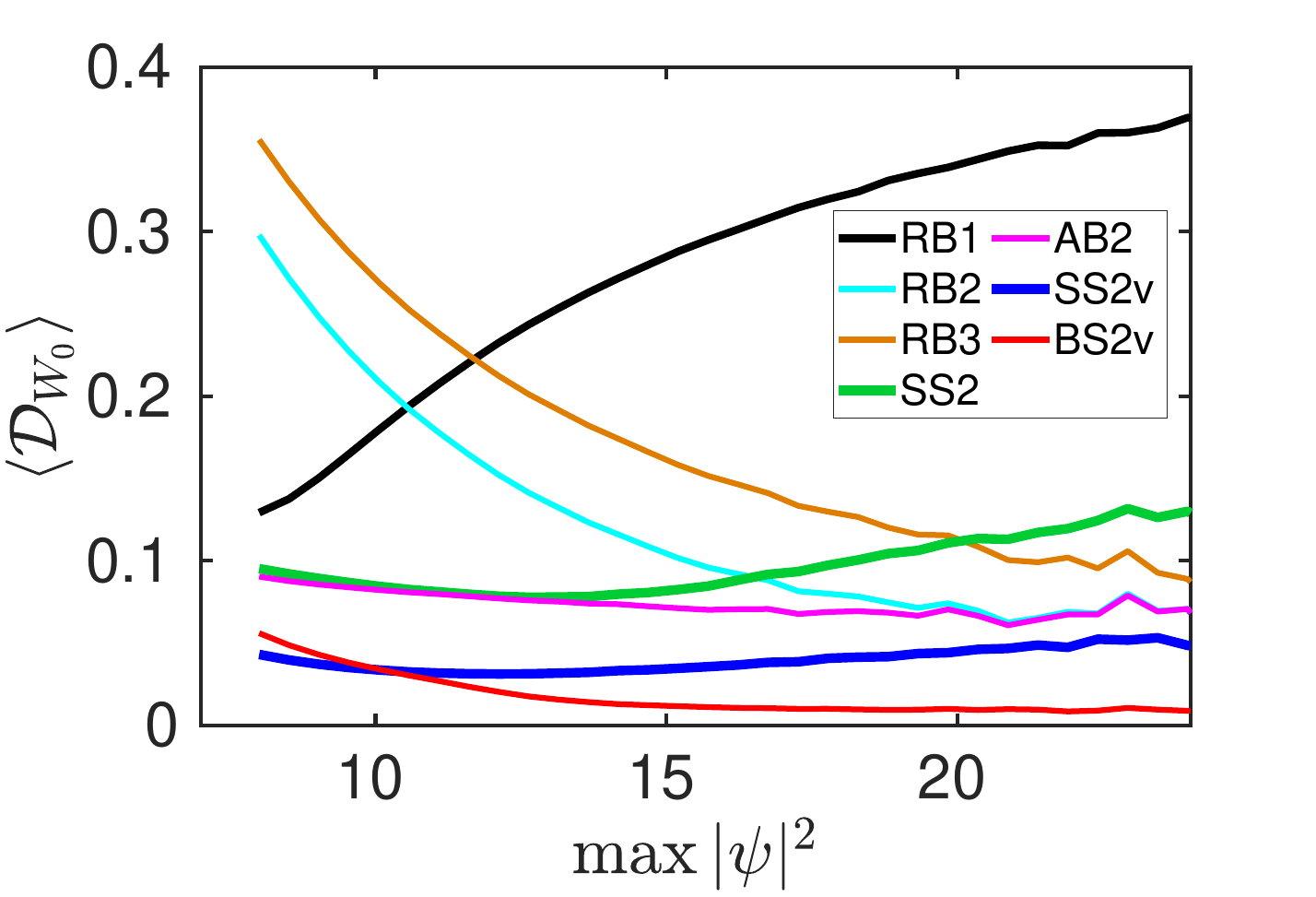}
	
	\caption{\small {\it (Color on-line)}
	Integral deviation $\mathcal{D}_{W_{0}}$~(\ref{deviation-intergral}) between RWs and their fits with exact solutions, averaged over RWs with specific maximum intensity $\max|\psi|^{2}$.
	}
	\label{fig:fig08}
\end{figure}


Note that the values $\langle r_{\mathrm{nn}}^{\mathrm{RW}}\rangle$ and $\langle r_{\mathrm{nn}}^{\mathrm{rnd}}\rangle$ look surprisingly small due to the quasi-random or random distribution of points: if the points were distributed orderly on a uniform grid, then the distance between nearest neighbors would be $\sqrt{2}\mu^{-1/2}$ (one point sits on an area of size $\mu^{-1}$), which is by $\sqrt{8}$ times larger than $\langle r_{\mathrm{nn}}^{\mathrm{rnd}}\rangle = 0.5\mu^{-1/2}$. 
Also, when calculating the PDF of distance to the nearest neighbor for RWs, the periodic boundary conditions are forced in the time interval $[80,200]$ to model RWs outside of this interval. 
Without this artificial condition, the result~(\ref{PDF-nn-random-positions}) would not be valid for random uniformly distributed positions and the PDFs in Fig.~\ref{fig:fig05} (both numerically computed in this case) would have a higher tail at $r_{\mathrm{nn}}\gtrsim 7$. 

In comparison with a linear system discussed in Appendix~\ref{Sec:App2}, the 1D-NLSE near the statistically stationary state of MI generates a much larger number of RWs: by $3.7$ times larger if the linear system has the same Fourier spectrum as near the stationary state of MI, and by $8.2$ times larger if the linear system has Gaussian spectrum of the same characteristic width. 
However, in average, one RW affects by the same times smaller spatiotemporal area, so that the resulting distributions of wavefield intensity practically coincide for these systems. 
In the linear case, RWs appear as interactions of high-intensity long-living structures with each other and with a rapidly changing shorter-wave oscillations, so that they accumulate into clusters of closely spaced RWs located on these high-intensity structures. 
The PDF of maximum intensity for these RWs coincides with the exponential distribution $e^{8-\max|\psi|^{2}}$, while the PDF of their chirp $\phi_{x}$ matches with a Gaussian distribution. 


\section{Rogue waves: comparison with exact solutions and distributions of fitting parameters}
\label{Sec:Results2}


\subsection{Comparison with exact solutions}
\label{Sec:Results2-1}

Figure~\ref{fig:fig06}(a) shows the space-time representation of amplitude $|\psi(x,t)|$ for one of the collected RW events with $\max|\psi|\approx 3.2$. 
The relative deviations 
\begin{eqnarray}
	\frac{|\Psi(\chi, \tau) - \Psi_{\mathrm{s}}(\chi, \tau)|}{|\Psi(\chi, \tau)|},
	\label{deviation-relative}
\end{eqnarray}
in the $(\chi, \tau)$-space between the auxiliary wavefield $\Psi$~(\ref{wavefield-auxiliary}) of this RW and the auxiliary wavefields $\Psi_{\mathrm{s}}$ of its fits with exact solutions are demonstrated in Fig.~\ref{fig:fig06}(b-f) for the RB1, SS2, AB2, SS2v and BS2v solutions. 
The parameters of the fits, together with the integral deviations $\mathcal{D}_{W_{0}}$~(\ref{deviation-intergral}) calculated over the largest comparison window $W_{0}$~(\ref{region-W0}) for them, are provided in Table~\ref{tab:M1}. 
The BS2v and SS2v solutions turn out to be the best dynamical models for the presented RW; their integral deviations $\mathcal{D}_{W_{0}}$ are smaller than for the RB1 by $10.3$ and $9.1$ times, respectively. 
The next best models are the SS2 and AB2, with their integral deviations smaller by around $40$\% than that for the RB1. 
The RB2 and RB3 models demonstrate significantly worse deviations than the RB1, see Table~\ref{tab:M1}, and for this reason they are not shown in Fig.~\ref{fig:fig06}. 

\begin{table}[t]
\caption{
Parameters of the fits with exact solutions to the RW presented in Fig.~\ref{fig:fig06}. 
In the table, the given parameters reproduce the best-fitting solutions $\psi_{\mathrm{s}}$ with maximums located at $x,t=0$, zero phase $\mathrm{arg}\,\psi_{\mathrm{s}}(0,0) = 0$ and a specific value of the chirp $\phi_{x}(0,0)$, $\phi_{x} = \partial\,\mathrm{arg}\,\psi_{\mathrm{s}}/\partial x$; these solutions need to be re-scaled with factors $\alpha$, see Eq.~(\ref{transformations}), to achieve the same maximum amplitude $\max|\psi|\approx 3.2$ as has the RW. 
The BS2v solution needs to be additionally transformed to the zero chirp, $\phi_{x}=0$, while the parameters of the SS2v model have already been transformed accordingly. 
The quantity $\mathcal{D}_{W_{0}}$ represents the integral deviation~(\ref{deviation-intergral}) between the RW and the corresponding fits calculated over the largest window $W_{0}$~(\ref{region-W0}). 
The rows are sorted in ascending order in $\mathcal{D}_{W_{0}}$.
}
\begin{tabular}{|c|c|c|c|c|}
\hline
  model & parameters & $\alpha$ & $\phi_{x}(0,0)$ & $\mathcal{D}_{W_{0}}$ \\ \hline
\multirow{4}{*}{BS2v} & $\eta_{1} = 0.6$, $\eta_{2} = 0.3$, & \multirow{4}{*}{$1.22$} & \multirow{4}{*}{$0.012$} & \multirow{4}{*}{$0.017$} \\
                  &  $\xi_{1} = 0.02$, $\xi_{2} = -0.04$, &                   &                   &                   \\
                  &  $x_{1} = 5.3$, $x_{2} = -8.7$,     &                   &                   &                   \\ 
                  &  $\theta_{1} = 4.3$, $\theta_{2} = 1.8$, $\Theta = 0.12$  &                   &                   &                   \\ \hline
\multirow{4}{*}{SS2v} & $\eta_{1} = 1$, $\eta_{2} = 0.72$, & \multirow{4}{*}{$0.99$} & \multirow{4}{*}{$0$} & \multirow{4}{*}{$0.019$} \\
                  &  $\xi_{1} = -0.008$, $\xi_{2} = 0.012$, &                   &                   &                   \\
                  &  $x_{1} = 0.22$, $x_{2} = 0.38$,     &                   &                   &                   \\ 
                  &  $\theta_{1} = 0.008$, $\theta_{2} = 6.27$  &                   &                   &                   \\ \hline
  SS2 & $\eta_{1} = 1.07$, $\eta_{2} = 0.54$  & $1$ & $0$ & $0.11$ \\ \hline
  AB2 & $\eta_{1} = 0.82$, $\eta_{2} = 0.15$  & $1.09$ & $0$ & $0.11$ \\ \hline
  RB1 & -- & $1.07$ & $0$ & $0.17$ \\ \hline
  RB2 & -- & $0.64$ & $0$ & $0.22$ \\ \hline
  RB3 & -- & $0.46$ & $0$ & $0.28$ \\ \hline
\end{tabular}
\label{tab:M1}
\end{table}

Integral deviations $\langle\mathcal{D}_{W_{m}}\rangle$~(\ref{deviation-intergral}), averaged over all collected RWs and calculated over the $10$ windows $W_{m}$~(\ref{region-Wm}), $m = 0, ..., 9$, are demonstrated in Fig.~\ref{fig:fig07}(a). 
In average, the SS2v and BS2v solutions represent the best models to describe RW dynamics, with mean integral deviations over the largest window equal $\langle\mathcal{D}_{W_{0}}\rangle \approx 0.038$ and $0.044$, respectively. 
Remarkably, the two solutions show very similar deviations for all $10$ windows $W_{m}$; compared to the RB1 case, these deviations are smaller by around $4$ (for $W_{0}$) to $10$ (for $W_{9}$) times. 
The next two best models are the AB2 and SS2 solutions, with mean deviation over the largest window $\langle\mathcal{D}_{W_{0}}\rangle \approx 0.09$ for both cases. 
For these two solutions, the deviations are also very similar for all windows $W_{m}$ and are smaller than deviations for the RB1 by $1.8$ (for $W_{0}$) to $3.2$ (for $W_{9}$) times. 
The RB2 and RB3 turn out to be, in average, significantly worse models than the RB1, with $\langle\mathcal{D}_{W_{0}}\rangle$ reaching $0.25$ and $0.31$, respectively, vs. $0.16$ for the RB1 model. 

Mean integral deviations $\langle\mathcal{D}_{bW_{m}}\rangle$, calculated over the bands $bW_{m} = W_{m}\backslash W_{m+1}$, $m = 0, ..., 8$, between the windows, are presented in Fig.~\ref{fig:fig07}(b).
They repeat the same pattern as discussed for deviations calculated over the windows: the SS2v and BS2v solutions represent the best performing models, followed by the AB2 and SS2 solutions, and then by the RB1. 
Interestingly, for all models except the SS2v and BS2v, one can observe regions (marked in Fig.~\ref{fig:fig07}(b) by the dashed lines), where the mean deviation $\langle\mathcal{D}_{bW}\rangle$ grows linearly with increasing dimensionless distance $d_{bW}$ from the RW maximum. 
Meanwhile, deviations for the SS2v and BS2v models have not yet reached such a region and grow with $d_{bW}$ faster than linearly. 
Thus, one can suggest that (i) at small $d_{bW}$, the deviations between RWs and their fits with exact solutions are small and increase with $d_{bW}$ faster than linearly, (ii) at larger $d_{bW}$, this behavior is replaced with a linear growth with $d_{bW}$, and (iii) at even larger distances, the deviations saturate due to the influence of the background wavefield. 
For models that are more accurate, changes in behavior with $d_{bW}$ take place at larger distances from the RW maximum, and, in this sense, the SS2v and BS2v models are qualitatively more efficient. 
For the largest band $bW_{0} = W_{0}\backslash W_{1}$, the mean deviations for the SS2v and BS2v models equal $\langle\mathcal{D}_{bW_{0}}\rangle\approx 0.082$ and $0.094$, respectively, so that, even at such a significant distance, these models reproduce RW dynamics fairly accurately.

While the integral deviations for the SS2v and BS2v models, averaged over all collected RWs, are very similar, for individual RWs they behave differently. 
In particular, the deviation $\mathcal{D}_{W_{0}}$ for the SS2v model, averaged over the $10$\% best SS2v fits, equals $\langle\mathcal{D}_{W_{0}}\rangle_{0.1b}\approx 0.01$, while, if calculated over the $10$\% worst fits, it is $\langle\mathcal{D}_{W_{0}}\rangle_{0.1w}\approx 0.1$. 
The corresponding values for the BS2v model equal $0.0075$ and $0.1$, respectively. 
Hence, the best BS2v fits reproduce RWs better than the best SS2v fits. 

\begin{figure*}[t]\centering
	\includegraphics[width=0.32\linewidth]{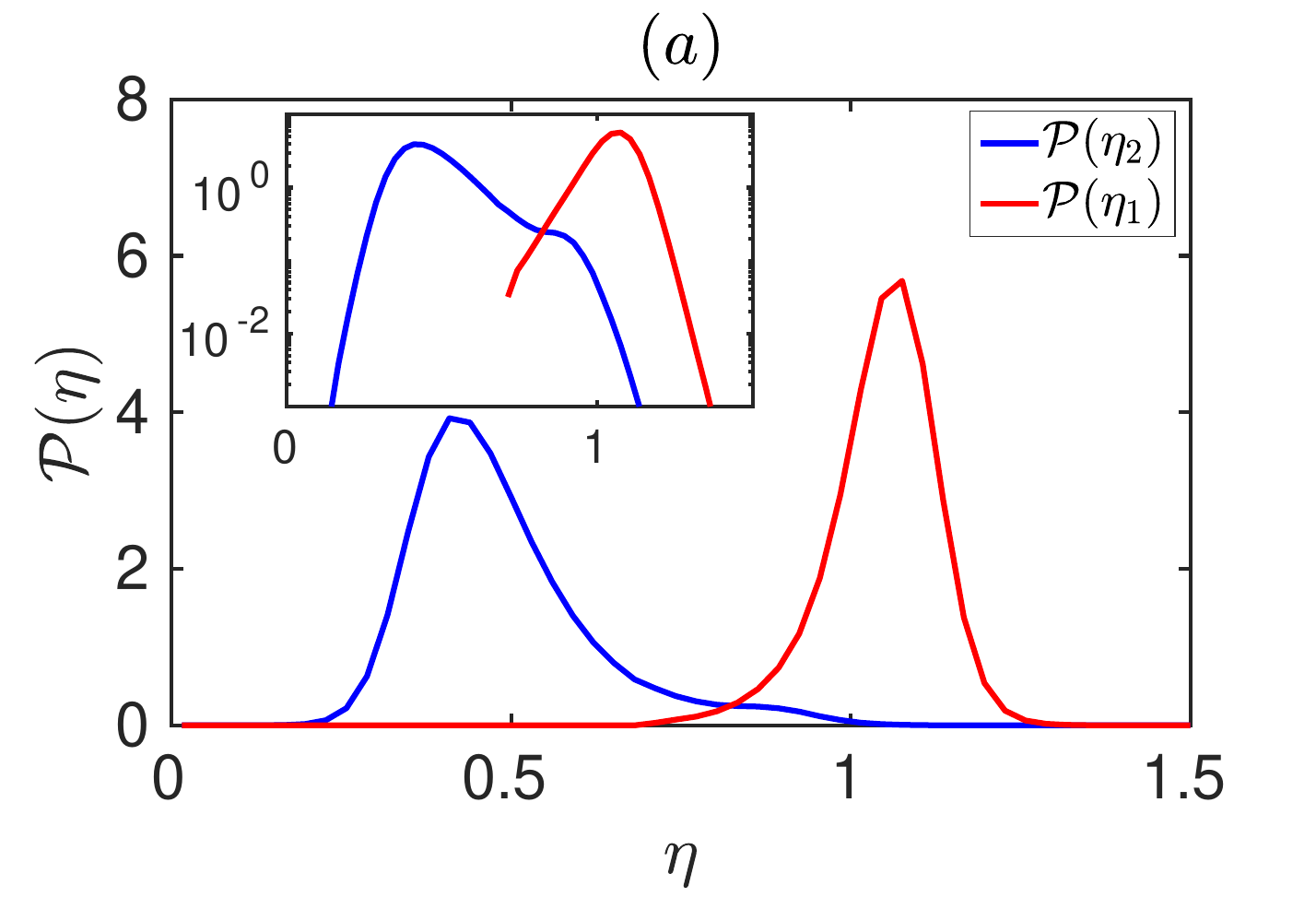}
	\includegraphics[width=0.32\linewidth]{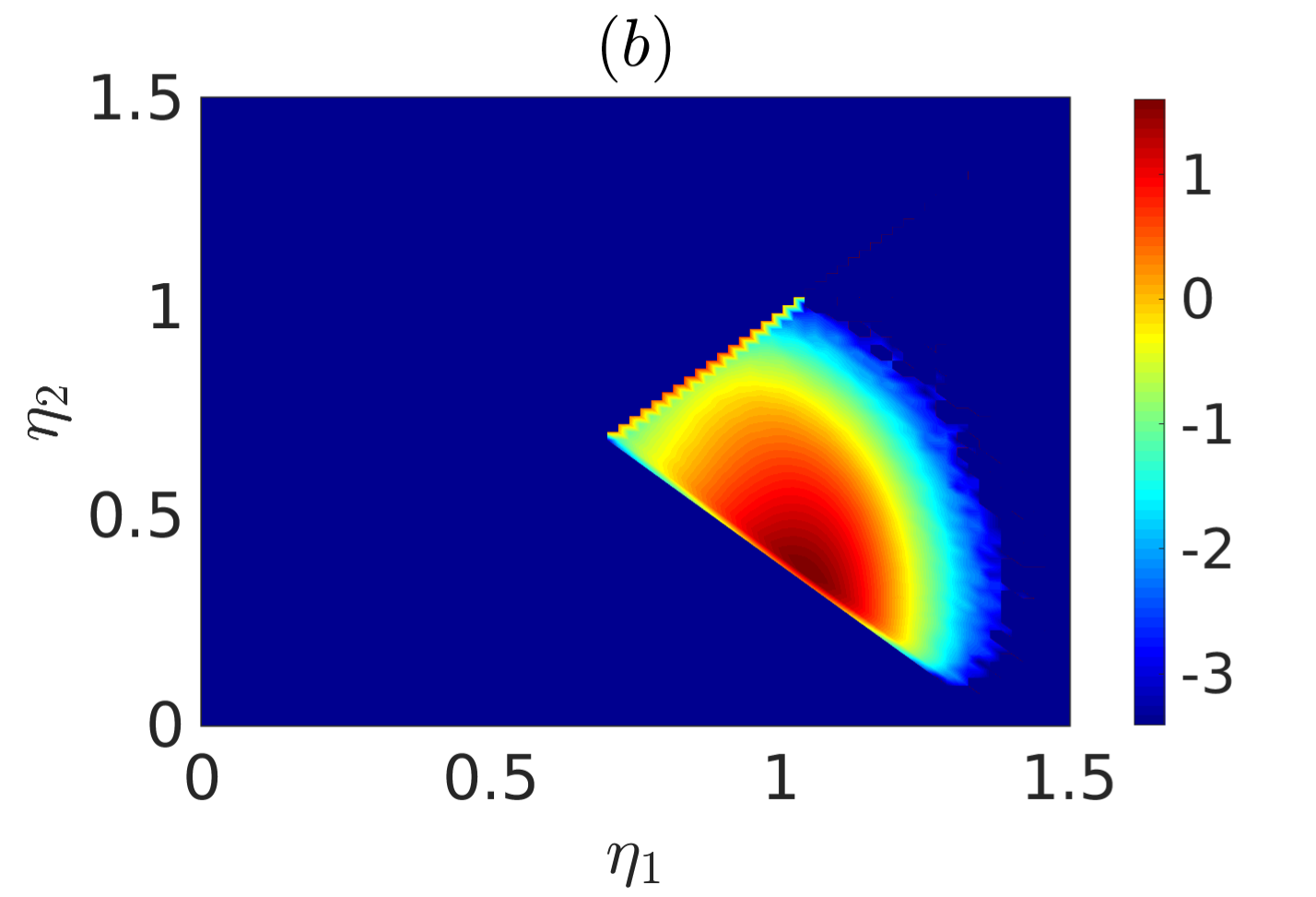}
	\includegraphics[width=0.32\linewidth]{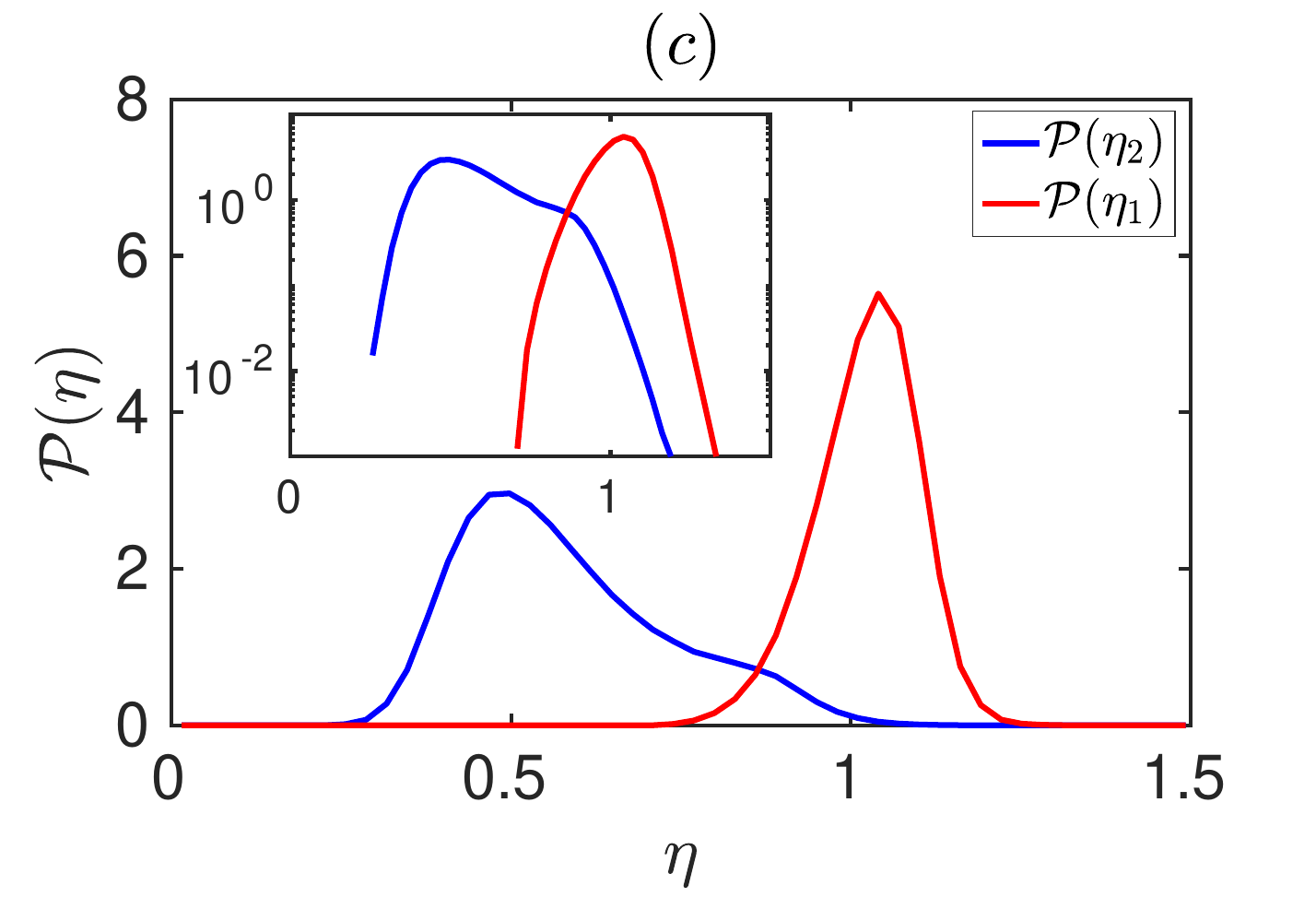}\\
	\includegraphics[width=0.32\linewidth]{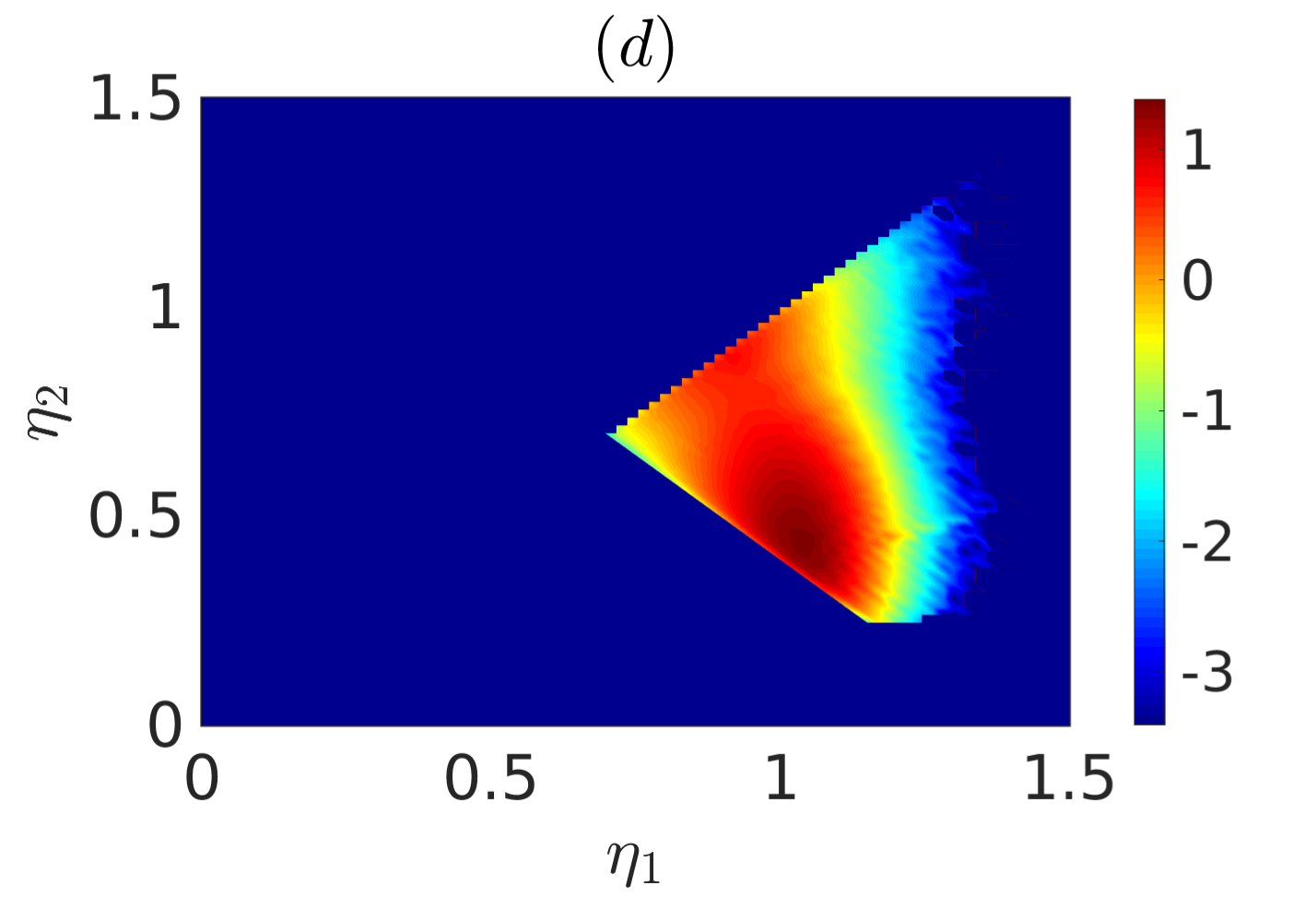}
	\includegraphics[width=0.32\linewidth]{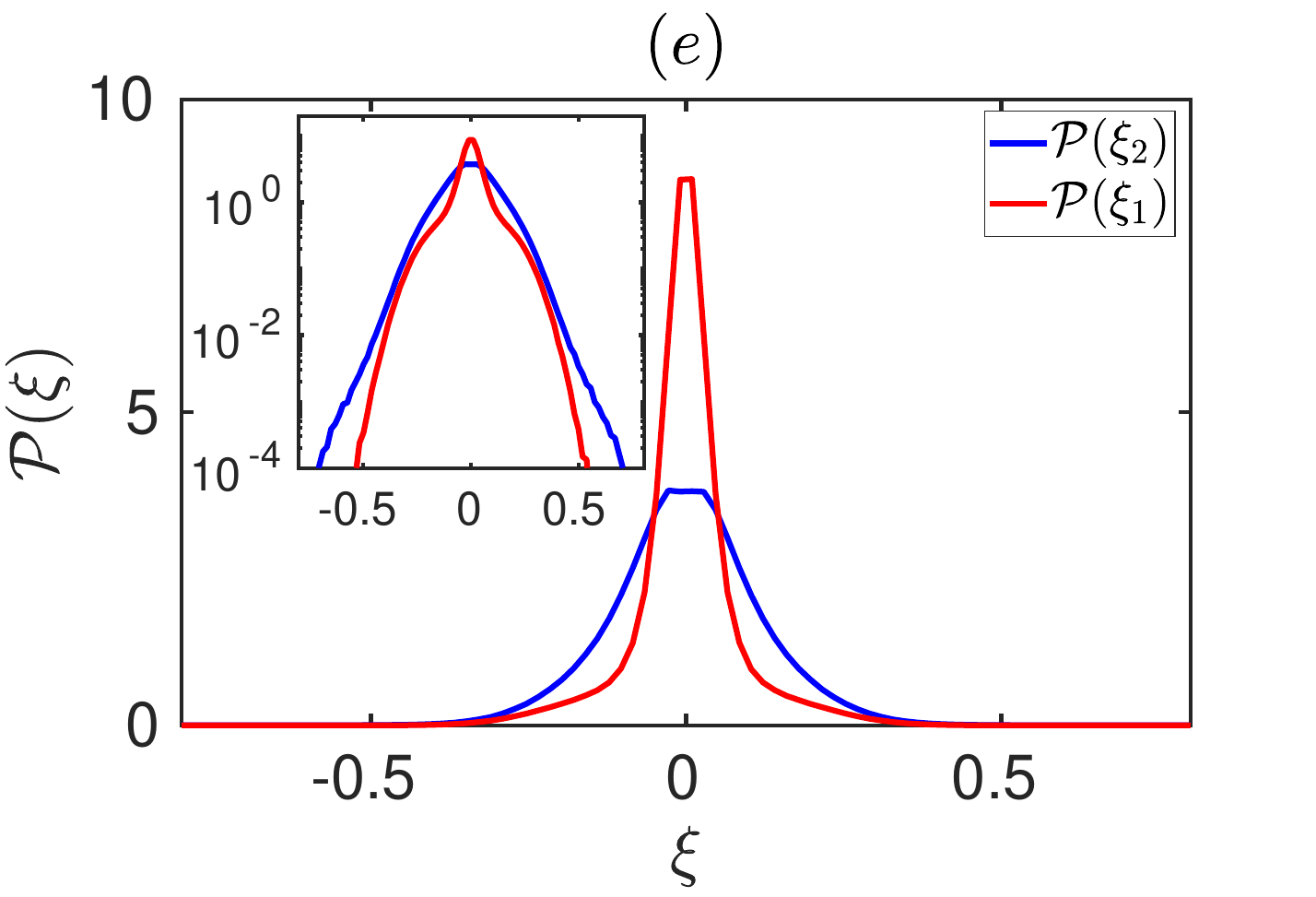}
	\includegraphics[width=0.32\linewidth]{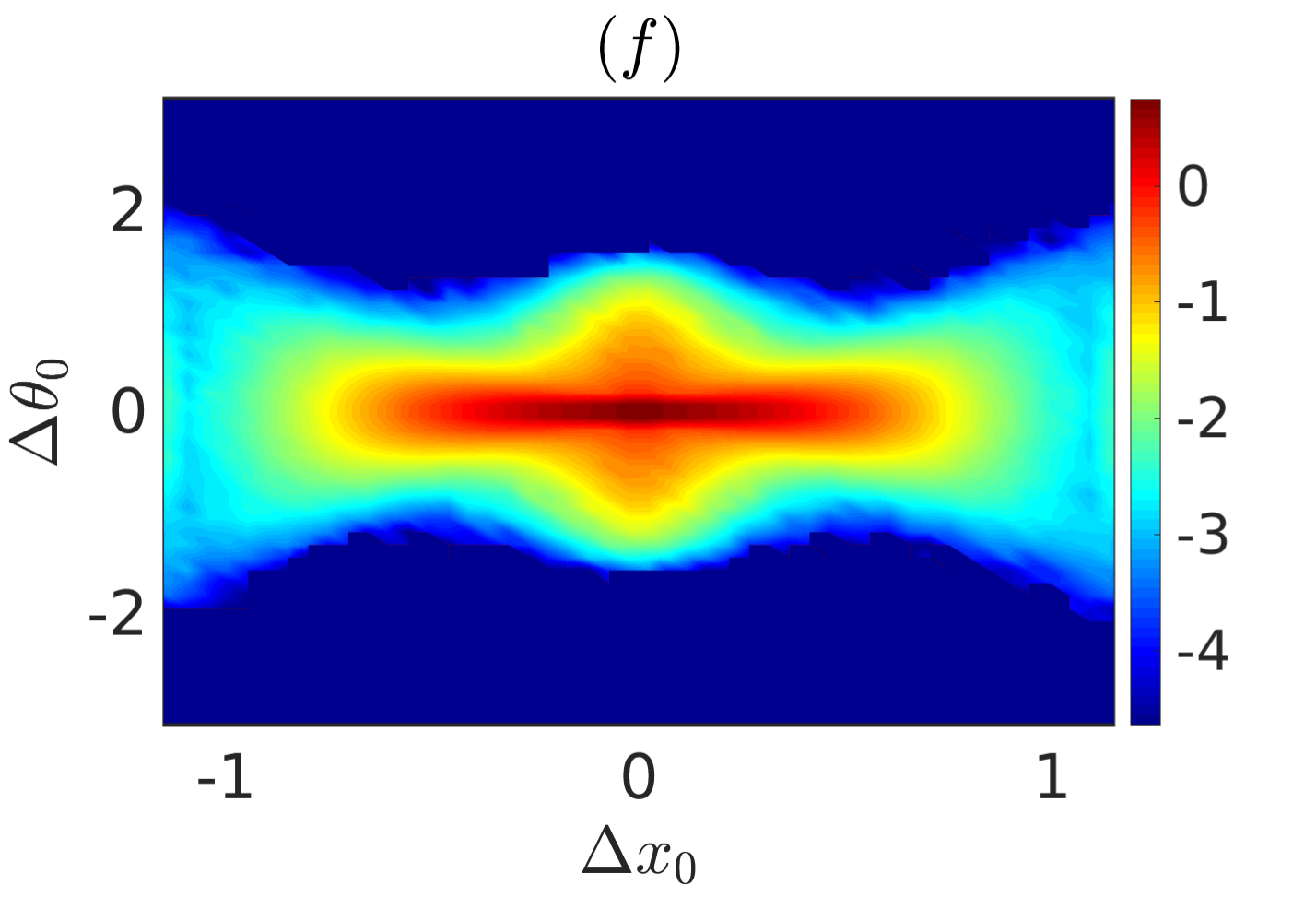}
	
	\caption{\small {\it (Color on-line)} 
	(a,b) Distributions of parameters $\eta_{1}$ and $\eta_{2}$ of the SS2 fits: (a) PDFs of the imaginary parts of soliton eigenvalues $\mathcal{P}(\eta_{2})$ (blue) and $\mathcal{P}(\eta_{1})$ (red), and (b) logarithm of their joint PDF, $\log_{10}\mathcal{P}(\eta_{1},\eta_{2})$. 
	(c-f) Distributions of parameters $\eta_{1,2}$, $\xi_{1,2}$, $\Delta x_{0} = x_{01}-x_{02}$ and $\Delta\theta_{0} = \theta_{01}-\theta_{02}$ of the SS2v fits: (c-d) same distributions as in panels (a-b), but for the SS2v case, (e) PDFs of the real parts of soliton eigenvalues $\mathcal{P}(\xi_{2})$ (blue) and $\mathcal{P}(\xi_{1})$ (red), and (f) logarithm of the joint PDF of differences in soliton positions and phases, $\log_{10}\mathcal{P}(\Delta x_{0},\Delta\theta_{0})$. 
	The insets in panels (a), (c) and (e) show the same distributions in semi-logarithmic scale. 
	}
	\label{fig:fig09}
\end{figure*}

\begin{figure*}[t]\centering
	\includegraphics[width=0.32\linewidth]{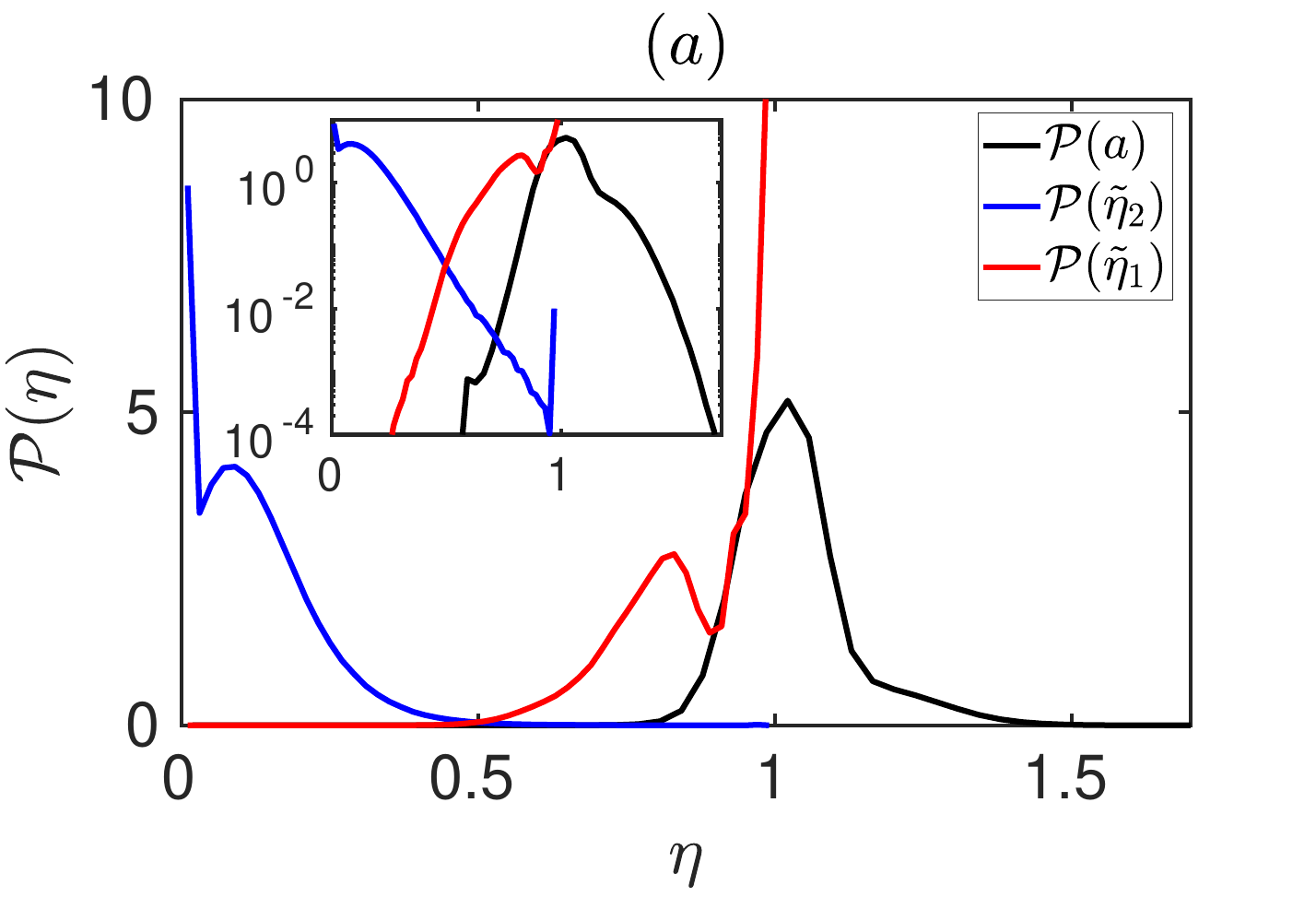}
	\includegraphics[width=0.32\linewidth]{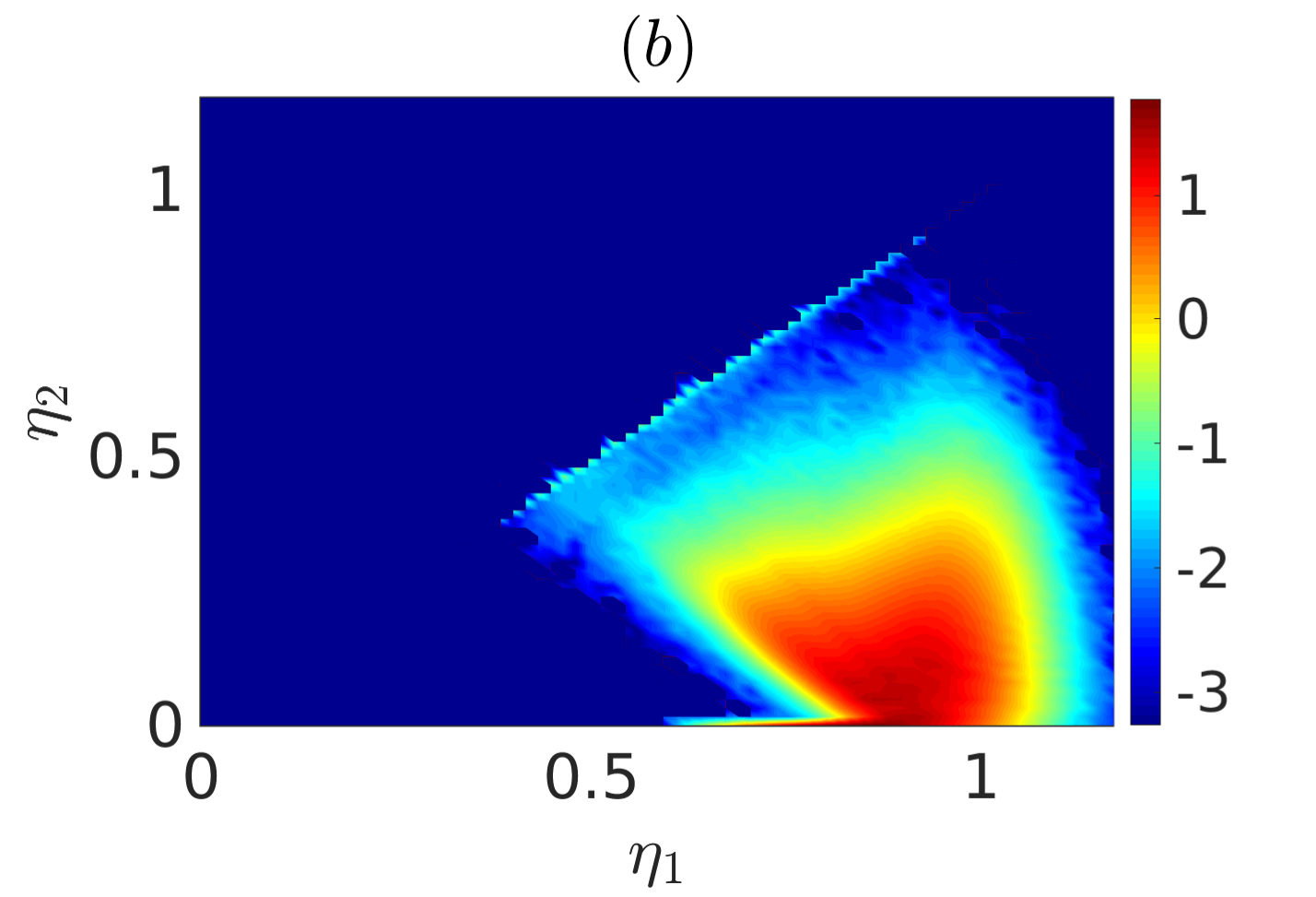}
	\includegraphics[width=0.32\linewidth]{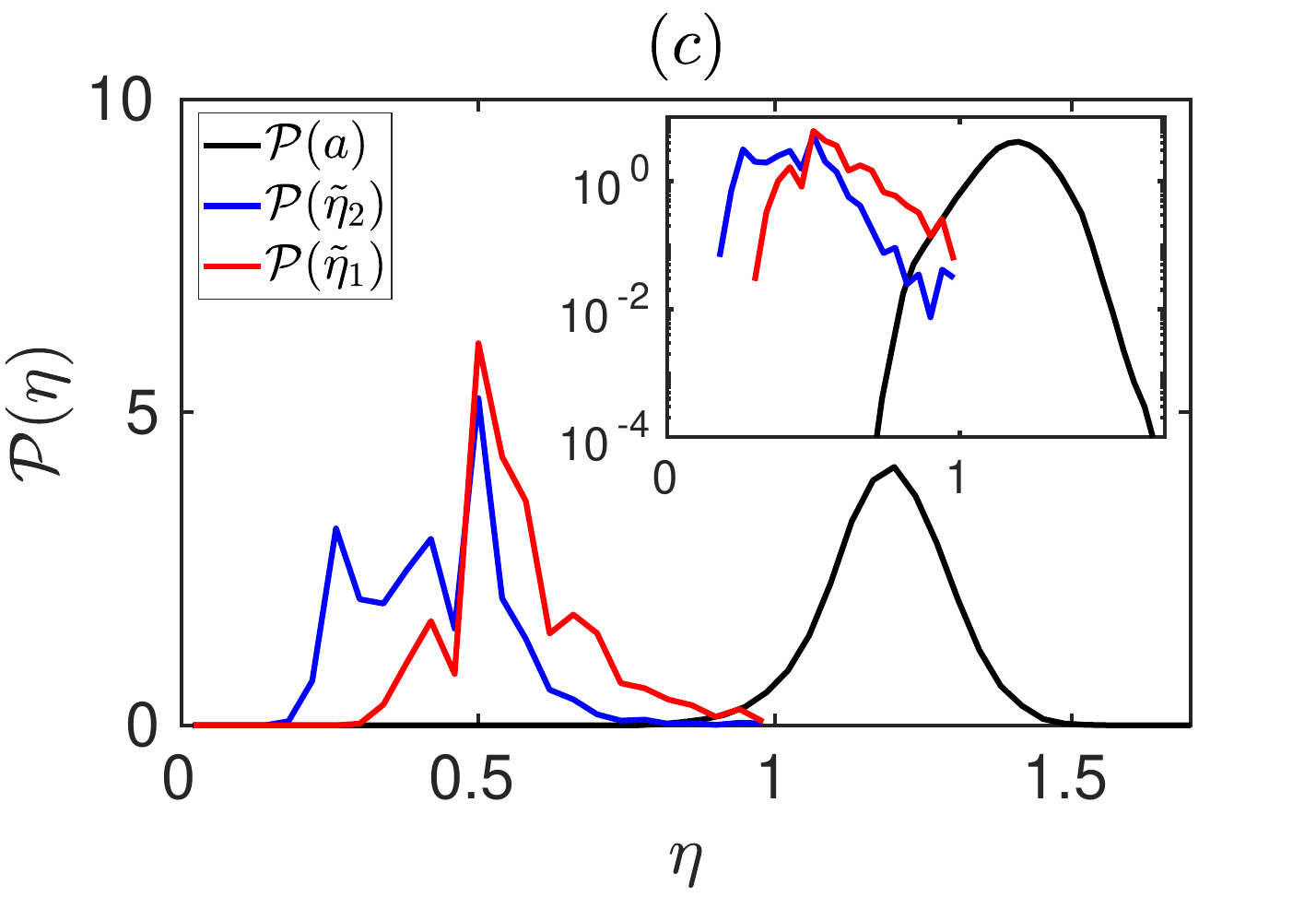}\\
	\includegraphics[width=0.32\linewidth]{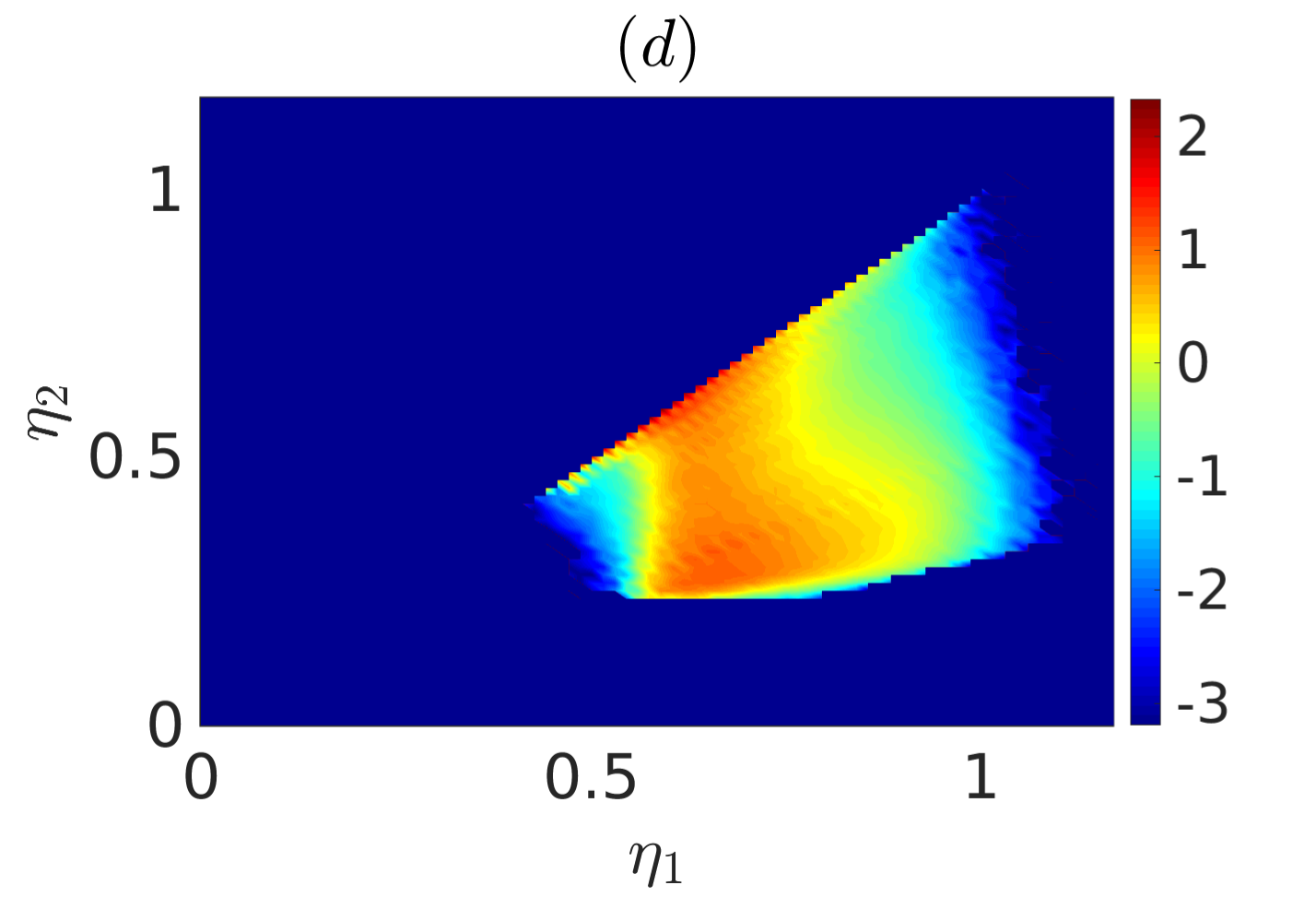}
	\includegraphics[width=0.32\linewidth]{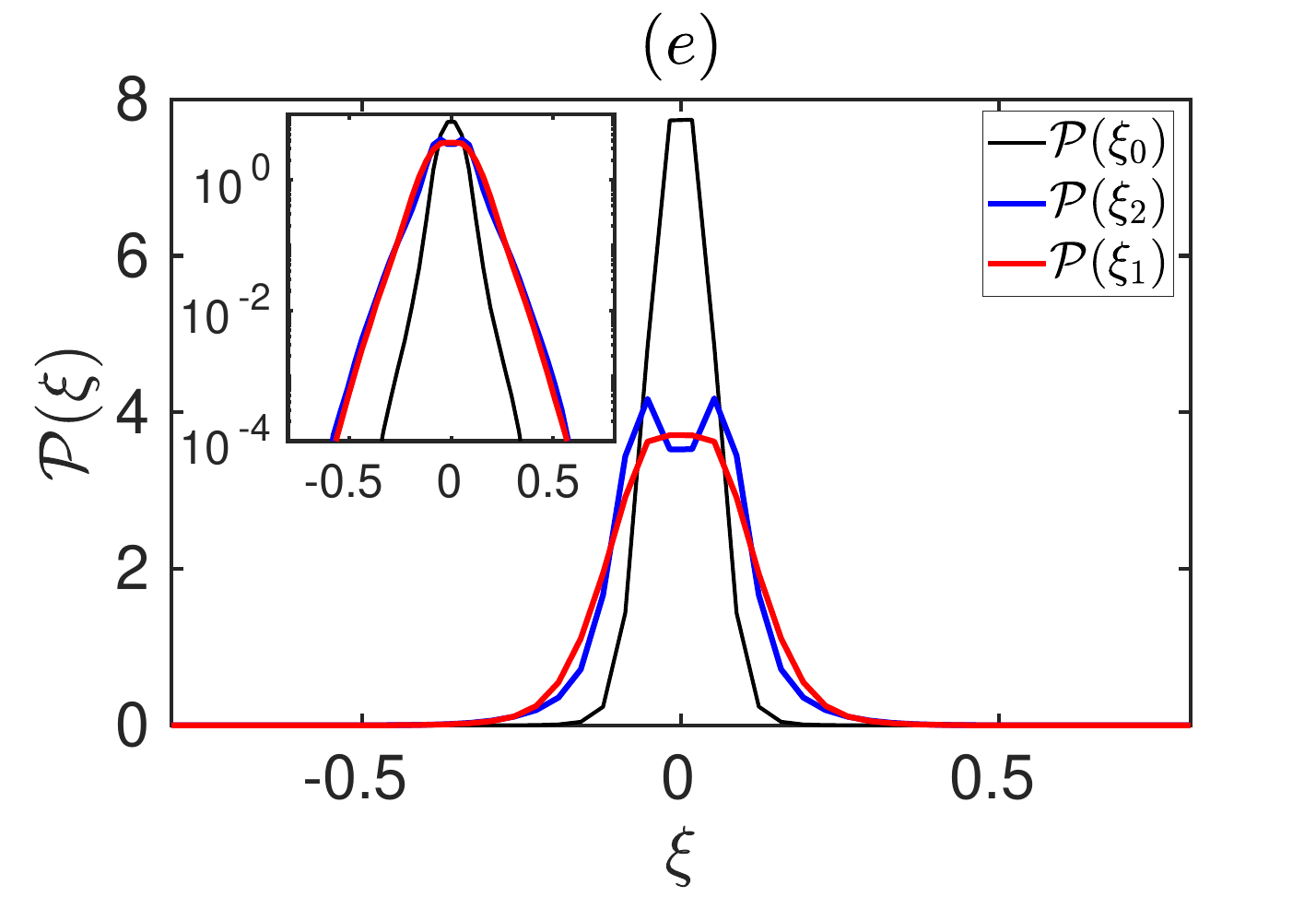}
	\includegraphics[width=0.32\linewidth]{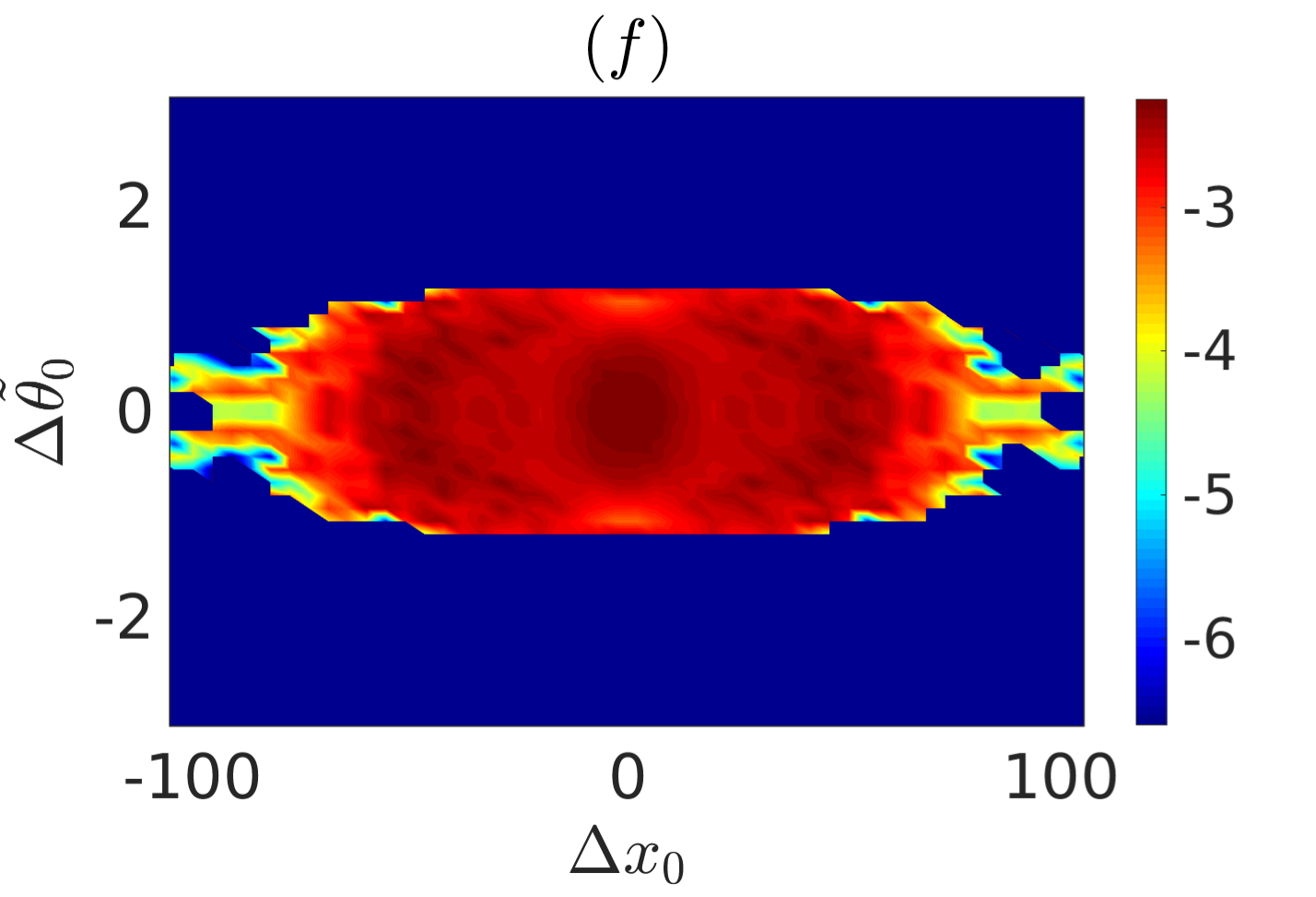}
	
	\caption{\small {\it (Color on-line)} 
	(a,b) Distributions of parameters $\eta_{1}$ and $\eta_{2}$ of the AB2 fits: (a) PDFs $\mathcal{P}(\tilde{\eta}_{2})$ (blue) and $\mathcal{P}(\tilde{\eta}_{1})$ (red) of the imaginary parts of breather eigenvalues normalized by the amplitude $a$ of the plane wave background, $\tilde{\eta}_{1,2} = \eta_{1,2}/a$, and also the PDF $\mathcal{P}(a)$ (black) of the plane wave amplitude, and (b) logarithm of the joint PDF of (non-normalized) $\eta_{1}$ and $\eta_{2}$, $\log_{10}\mathcal{P}(\eta_{1},\eta_{2})$. 
	(c-f) Distributions of parameters $\eta_{1,2}$, $\xi_{0,1,2}$, $\Delta x_{0} = x_{01}-x_{02}$ and $\Delta\tilde{\theta}_{0} = \tilde{\theta}_{01}-\tilde{\theta}_{02}$, see Eq.~(\ref{phases-BS2v-modified}), of the BS2v fits: (c-d) same distributions as in panels (a-b), but for the BS2v case, (e) PDFs of the real parts of breather eigenvalues $\mathcal{P}(\xi_{2})$ (blue) and $\mathcal{P}(\xi_{1})$ (red), and also the PDF $\mathcal{P}(\xi_{0})$ (black) of the intersection $\xi_{0}$ between the (shifted) branch cut and the real axis, and (f) logarithm of the joint PDF of differences in breather positions and modified phases, $\log_{10}\mathcal{P}(\Delta x_{0},\Delta\tilde{\theta}_{0})$. 
	The insets in panels (a), (c) and (e) show the same distributions in semi-logarithmic scale. 
	}
	\label{fig:fig10}
\end{figure*}

As shown in Fig.~\ref{fig:fig08}, the deviations between RWs and their fits also behave differently for RWs of different maximum intensity $\max|\psi|^{2}$. 
In particular, for the RB1 model, the mean deviation is minimal, $\langle\mathcal{D}_{W_{0}}\rangle \approx 0.13$, for the smallest RWs and increases with increasing $\max|\psi|^{2}$ up to $0.37$ for the largest RWs. 
At $\max|\psi|^{2}\gtrsim 12$, this model turns out to be (in average) the least accurate. 
On the contrary, for the RB2 and RB3 models, the mean deviation is maximal, $\langle\mathcal{D}_{W_{0}}\rangle\approx 0.3$ and $0.36$, for the smallest RWs and decreases with increasing $\max|\psi|^{2}$ up to $\langle\mathcal{D}_{W_{0}}\rangle\approx 0.07$ and $0.09$ for the largest RWs, respectively. 
For large intensities, the RB2 model, together with the AB2 model, ranks third in accuracy after the BS2v and SS2v models, confirming the results published in~\cite{agafontsev2021rogue}. 

For the SS2 and SS2v models, the mean deviation slowly decreases from the values of $\langle\mathcal{D}_{W_{0}}\rangle\approx 0.095$ and $0.043$ for the smallest RWs until intensities $\max|\psi|^{2}\approx 13$ and $12$, where it reaches minimums, $\langle\mathcal{D}_{W_{0}}\rangle\approx 0.078$ and $0.031$, respectively. 
Then, the mean deviation slowly grows with $\max|\psi|^{2}$ up to the values of $\langle\mathcal{D}_{W_{0}}\rangle\approx 0.13$ and $0.06$ for the largest RWs. 
For the AB2 and BS2v models, the mean deviation slowly decreases from $\langle\mathcal{D}_{W_{0}}\rangle\approx 0.1$ and $0.056$ for the smallest RWs to $0.07$ and $0.01$ for the largest RWs, respectively. 
At $\max|\psi|^{2}\gtrsim 10$, the BS2v model becomes (in average) the most accurate; at $\max|\psi|^{2}\gtrsim 15$, its mean deviation $\langle\mathcal{D}_{W_{0}}\rangle$ stays around $0.01$, which is almost twice smaller than the deviations for the SS2v and BS2v fits shown in Fig.~\ref{fig:fig06}(e,f). 


\subsection{Distribution of parameters for the SS2, SS2v, AB2 and BS2v models}
\label{Sec:Results2-2}

This subsection studies the distributions of parameters of the SS2, SS2v, AB2 and BS2v fits written in the original $(x,t)$ variables. 
These fits are obtained by (i) applying the transformation inverse to~(\ref{wavefield-auxiliary}) to the auxiliary wavefield of the fits and (ii) transforming the parameters of these solutions accordingly as discussed in Appendix~\ref{Sec:App1}. 
In the $(x,t)$-space, the corresponding solutions have maximums at the same points and with the same amplitude, phase and chirp as the collected RWs. 

Figure~\ref{fig:fig09}(a,b) shows distributions of imaginary parts of soliton eigenvalues for the SS2 fits of RWs collected near the statistically stationary state of MI. 
The larger soliton has eigenvalue narrowly distributed approximately between $0.9$ and $1.2$, with mean $\mathbb{E}(\eta_{1})\approx 1.05$ and standard deviation $\mathrm{std}(\eta_{1})\approx 0.08$. 
The eigenvalue of the smaller soliton has a wider distribution approximately between $0.3$ and $0.8$, with $\mathbb{E}(\eta_{2})\approx 0.49$ and $\mathrm{std}(\eta_{1})\approx 0.13$. 
The joint distribution $\mathcal{P}(\eta_{1},\eta_{2})$ in Fig.~\ref{fig:fig09}(b) has range of values bounded by two lines: $\eta_{2}\le\eta_{1}$, due to the condition that $\eta_{1}$ refers to the larger soliton, and $\max|\psi^{\mathrm{SS2}}| = 2(\eta_{1} + \eta_{2})\ge \sqrt{8}$, as the maximum amplitude of the SS2 solution~(\ref{2-SS}) must exceed the minimum RW amplitude. 
This joint PDF has one clearly visible region of high values around a maximum at the point $\eta_{1}\approx 1.05$, $\eta_{1}\approx 0.37$, which lies near the second line (i.e., corresponds to the smallest RWs).

The imaginary parts of soliton eigenvalues for the SS2v fits are distributed similarly, see Fig.~\ref{fig:fig09}(c-d), with a few differences. 
First, the PDF for the larger soliton is shifted towards slightly smaller values, with $\mathbb{E}(\eta_{1})\approx 1.02$ and $\mathrm{std}(\eta_{1})\approx 0.08$, while the PDF for the smaller soliton is moderately shifted to larger values, $\mathbb{E}(\eta_{2})\approx 0.58$ and $\mathrm{std}(\eta_{2})\approx 0.15$. 
Note that the maximum amplitude of the SS2v solution is bounded from above as $\max|\psi^{\mathrm{SS2v}}|\le 2(\eta_{1} + \eta_{2})$, see e.g.~\cite{bertola2017maximal}, and since the equality in this relation is not always realized, the sum $(\eta_{1} + \eta_{2})$ for the SS2v fits must, in average, exceed that for the SS2 fits. 
Secondly, the joint PDF $\mathcal{P}(\eta_{1},\eta_{2})$ in Fig.~\ref{fig:fig09}(d) has a second region of elevated values near $\eta_{1}\simeq 0.9$, $\eta_{2}\simeq 0.9$, linked to the first region that lies around the maximum at the point $\eta_{1}\approx 1.03$, $\eta_{1}\approx 0.45$. 

The real parts of soliton eigenvalues $\xi_{1,2}$ are narrowly distributed around zero, see Fig.~\ref{fig:fig09}(e), with $\mathrm{std}(\xi_{1})\approx 0.08$, $\mathrm{std}(\xi_{2})\approx 0.12$ and $\mathrm{std}(\xi_{2}-\xi_{1})\approx 0.17$. 
The tails of the corresponding PDFs, especially for the smaller soliton, are exponential-like, and, for a few RWs, the real parts of the eigenvalues turn out to be quite significant, $|\xi_{1,2}|\simeq 0.5$. 
The joint distribution $\mathcal{P}(\Delta x_{0}, \Delta\theta_{0})$ of differences $\Delta x_{0} = x_{01}-x_{02}$ and $\Delta\theta_{0} = \theta_{01}-\theta_{02}$ in soliton positions $x_{0n}$ and phases $\theta_{0n}$ in Fig.~\ref{fig:fig09}(f) has one central region of high values at $|\Delta x_{0}|\lesssim 0.3$ and $|\Delta\theta_{0}|\lesssim 0.2$, so that for most RWs these differences are rather small. 

Figure~\ref{fig:fig10}(a,b) shows distributions of parameters for the AB2 fits. 
The amplitude $a$ of the plane wave background is distributed narrowly around unity, with  $\mathbb{E}(a)\approx 1.03$ and $\mathrm{std}(a)\approx 0.1$; however, its distribution has a ``fat tail'' at $a\gtrsim 1.2$. 
Distributions of the imaginary parts of breather eigenvalues reveal that the AB2 model works in three different regimes: 
\begin{itemize}
	\item when the smaller breather represents a small correction, $\eta_{2}/a\ll 1$, so that the AB2 solution becomes close to the AB. If also $\eta_{1}/a \approx 1$, then the AB2 solution becomes close to the RB1;
	\item when $\eta_{2}/a \approx \eta_{1}/a \approx 1$, see the inset in Fig.~\ref{fig:fig10}(a), so that the AB2 solution becomes close to the RB2;
	\item a ``genuine'' AB2 model with $\eta_{1}/a$ widely distributed around $0.8$ and $\eta_{2}/a\gtrsim 0.1$. 
\end{itemize}
In particular, this explains why in Fig.~\ref{fig:fig08} the curves corresponding to the RB2 and AB2 models practically merge at $\max|\psi|^{2}\gtrsim 20$: at that large intensities, the AB2 model transforms into the RB2 model. 
The joint PDF $\mathcal{P}(\eta_{1}, \eta_{2})$ in Fig.~\ref{fig:fig10}(b) takes high values near $\eta_{1}\simeq 0.9$ and $\eta_{2}\lesssim 0.1$, so that, for the majority of RWs, the smaller breather represents a small correction. 
Interestingly, this small correction turns out to be enough to significantly increase the accuracy of the AB2 fits compared to the RB1 and AB models, see Figs.~(\ref{fig:fig07})-(\ref{fig:fig08}) (recall that the AB model shows results practically indistinguishable from the RB1 model, see Section~\ref{Sec:NumMethods-3}). 

The imaginary parts of breather eigenvalues for the BS2v model demonstrate a different behavior. 
In particular, $\eta_{1}/a$ is distributed approximately between $0.4$ and $0.7$ with a sharp maximum of its PDF at $\eta_{1}/a = 0.5$, see Fig.~\ref{fig:fig10}(c), while $\eta_{2}/a$ is distributed more evenly and approximately between $0.2$ and $0.6$. 
The joint PDF $\mathcal{P}(\eta_{1}, \eta_{2})$ in Fig.~\ref{fig:fig10}(d) has two regions of high values: (i) near the line $\eta_{1} = \eta_{2}$ at around $\eta_{1,2}\simeq 0.6$, and (ii) around $\eta_{1}\simeq 0.6$ and $\eta_{2}\simeq 0.3$. 
Compared to the AB2 model, the amplitude $a$ of the plane wave background is shifted to higher values, with $\mathbb{E}(a)\approx 1.19$ and $\mathrm{std}(a)\approx 0.1$. 

The real parts of breather eigenvalues $\xi_{1,2}$ are narrowly distributed around zero, see Fig.~\ref{fig:fig10}(e), with $\mathrm{std}(\xi_{1})\approx 0.1$, $\mathrm{std}(\xi_{2})\approx 0.09$ and $\mathrm{std}(\xi_{2}-\xi_{1})\approx 0.16$. 
The corresponding PDFs turn out to be nearly identical; their tails are exponential-like, and, for a few fits, the real parts are rather significant, $|\xi_{1,2}|\simeq 0.5$. 
Note that, in the $(x,t)$-variables, these fits have the branch cut shifted along the real axis according to the second expression in Eq.~(\ref{transformations-NBS}). 
The intersection point $\xi_{0}$ between this shifted branch cut and the real axis has almost twice narrower distribution around zero, see Fig.~\ref{fig:fig10}(e), with $\mathrm{std}(\xi_{0})\approx 0.06$. 
The PDF $\mathcal{P}(\xi_{0})$ has exponential-like tails, and the BS2v fits for a few RWs have significantly shifted branch cuts, $|\xi_{0}|\simeq 0.3$. 
Also note that the angular shape of the PDFs shown in Fig.~\ref{fig:fig10}(e) comes from limitations of the BS2v database, in which the minimal real part of breather eigenvalues equals $0.02$. 

The joint distribution $\mathcal{P}(\Delta x_{0}, \Delta\tilde{\theta}_{0})$ of differences $\Delta x_{0} = x_{01}-x_{02}$ and $\Delta\tilde{\theta}_{0} = \tilde{\theta}_{01}-\tilde{\theta}_{02}$ in breather positions $x_{0n}$, $n = 1,2$, and modified phases
\begin{eqnarray}
	\tilde{\theta}_{0n} = \theta_{0n} + 2\,\mathrm{Re}[\zeta_{n}]\,x_{0n},
	\label{phases-BS2v-modified}
\end{eqnarray}
demonstrated in Fig.~\ref{fig:fig10}(f), has a central region of high values at $|\Delta x_{0}|\lesssim 10$ and $|\Delta\tilde{\theta}_{0}|\lesssim 0.4$, and also several smaller regions of high values approximately within $30\lesssim|\Delta x_{0}|\lesssim 60$ and $|\Delta\tilde{\theta}_{0}|\lesssim 1$. 
At $|\Delta x_{0}|\gtrsim 80$ or $|\Delta\tilde{\theta}_{0}|\gtrsim 1.2$, this PDF equals zero. 
Note that the amplitude of norming constants~(\ref{C_param_TWB}) for TWBs is defined by the breather positions $x_{0n}$, 
$$
	|C_{n}^{\mathrm{TWB}}| = \exp\big(-2\,\mathrm{Im}[\zeta_{n}]\,x_{0n}\big),
$$
and, since $|\mathrm{Im}[\zeta_{n}]|\ll 1$ for $|\xi_{n}|\ll 1$, the coordinates $x_{0n}$ turn out to be large (with the characteristic widths $|\mathrm{Im}[\zeta_{n}]|^{-1}$ of breathers being large as well). 
The phase of the norming constants depends on both breather positions $x_{0n}$ and phases $\theta_{0n}$,
$$
	\mathrm{arg}\,C_{n}^{\mathrm{TWB}} = \pi + \Theta + \theta_{0n} + 2\,\mathrm{Re}[\zeta_{n}]\,x_{0n}.
$$
Since $\mathrm{Re}[\zeta_{n}]\simeq 1$, the term $2\,\mathrm{Re}[\zeta_{n}]\,x_{0n}$ turns out to be large and leads to a situation when the correlation is visible between the modified phases~(\ref{phases-BS2v-modified}), but is not visible between the original phases $\theta_{0n}$. 

The results discussed above demonstrate that, while the SS2 and SS2v fits represent rather similar solutions, compare Fig.~\ref{fig:fig09}(a,b) with Fig.~\ref{fig:fig09}(c,d), the AB2 and BS2v fits are significantly different. 
In particular, the joint PDFs $\mathcal{P}(\eta_{1},\eta_{2})$ in Fig.~\ref{fig:fig10}(b,d) occupy different regions, and the distributions $\mathcal{P}(\eta_{1,2}/a)$ in Fig.~\ref{fig:fig10}(a,c) are very different as well. 
Since the real parts of soliton or breather eigenvalues in the SS2v and BS2v solutions are small, see Figs.~\ref{fig:fig09}-\ref{fig:fig10}(e), the key difference between the SS2 and AB2 models from the one hand and the SS2v and BS2v models from the other hand is that, in the former, the interacting nonlinear structures have the same positions, $x_{01}=x_{02}$, while, in the latter, these positions are different. 
This effect should be much more pronounced when comparing the breather models: since the real parts of their eigenvalues are small, $|\xi_{1,2}|\ll 1$, the characteristic widths of breathers are large, $|\mathrm{Im}[\zeta_{1,2}]|^{-1}\gg 1$, so that a collision of such breathers may provide much more opportunities for the emergence of RWs. 
In other words, the AB2 model is a significantly worse simplification of the BS2v model than the SS2 is of the SS2v model.

\section{Discussions}
\label{Sec:Discussions}

It is known that a plane wave of unit amplitude perturbed by a small noise can be represented both as a breather gas~\cite{soto2016integrable,akhmediev2016breather,grinevich2018finite,grinevich2018exact,grinevich2019finite} and as a soliton gas~\cite{gelash2019bound,gelash2021solitonic}. 
In both of these approaches, the eigenvalues of, respectively, solitons or breathers are concentrated in the complex $\lambda$-plane very close to or at the $[0,i]$ segment of the imaginary axis. 
The best in accuracy SS2v and BS2v models represent ad-hoc approximations involving only two nonlinear modes (solitons or breathers), which, near the RW maximums, replace the whole wavefield composed of a large number of nonlinear modes. 
Meanwhile, as demonstrated above, the parameters of the approximating nonlinear modes turn out to be surprisingly similar to the parameters of the IST spectrum of the whole wavefield, although they do not coincide exactly. 
In particular, the eigenvalues of the SS2v and BS2v models lie close to the $[0,i]$ segment, while the plane wave amplitude in the BS2v model turns out to be close to unity; the deviations usually do not exceed $\simeq 0.3$ for both the eigenvalues and the plane wave amplitude. 

These observations raise a question whether one can put forward a hypothesis that RWs emerge due to synchronization of a few (for instance, two) nonlinear modes in presence of many other nonlinear modes that may lead to significant distortions. 
Note that, under certain conditions, adding just one nonlinear mode to a wavefield that already contains hundreds of them can completely change this wavefield in the physical space across its entire characteristic width~\cite{agafontsev2024multisoliton}. 
Hence, such a hypothesis represents a very naive assumption that hundreds of other nonlinear modes of the wavefield do not destroy the RW formed by only a few nonlinear modes, but only affect their synchronization conditions~\cite{gelash2021solitonic}, distort the RW in the physical space and also distort the parameters of the two effective nonlinear modes of the SS2v and BS2v fits approximating this RW. 

To evaluate this hypothesis properly, one needs to analyze the IST spectrum of the whole wavefield and pinpoint within it the two synchronized modes of the SS2v and BS2v solutions, which lies beyond the scope of the present paper. 
However, the data collected shows that the deviations of nonlinear modes of these models from the IST spectrum of the whole wavefield, on the one hand, and the deviations of these models from RWs in the physical space, on the other hand, are correlated positively. 
In other words, it is possible to assume that the larger the distortions from the other nonlinear modes present in the wavefield, the worse the performance of the fits with the two models and, simultaneously, the larger the deviations of their two nonlinear modes from the IST spectrum of the whole wavefield. 

In particular, the integral deviation $\mathcal{D}_{W_{0}}$ for the SS2v model positively correlates with the square of the arithmetic mean of real parts of soliton eigenvalues, $[(\xi_{1}+\xi_{2})/2]^{2}$, with Pearson's correlation coefficient $r_{\mathrm{P}}\approx 0.42$. 
For the BS2v model, $\mathcal{D}_{W_{0}}$ positively correlates with the square of the shift of the branch cut along the real axis, $\xi_{0}^{2}$, with $r_{\mathrm{P}}\approx 0.35$. 
For each model, the corresponding quantity can be considered as distortions due to the other nonlinear modes present in the wavefield. 

For the SS2v model, the construct $(\eta_{1}-1)^{2}$ correlates with $\mathcal{D}_{W_{0}}$ with $r_{\mathrm{P}}\approx 0.34$. 
In the soliton gas approximation of the plane wave solution~\cite{gelash2019bound,gelash2021solitonic}, the density of solitonic states diverges when the imaginary part $\eta$ approaches unity, and it can be assumed that, out of the two interacting modes, the one that corresponds to the larger soliton should have the imaginary part of its eigenvalue close to unity. 
Then, the combination $(\eta_{1}-1)^{2}$ can be considered as a distortion due to the other nonlinear modes. 
The LASSO regression method gives an optimal combination of $\eta_{1}$ and $\eta_{1}^{2}$ in the form $(\eta_{1}-0.91)^{2}$ with $r_{\mathrm{P}}\approx 0.49$, which can be interpreted that $\eta_{1}$ should actually be close to $0.91$. 
For the BS2v model, the square of deviation of the plane wave amplitude $a$ from unity, $(a-1)^{2}$, which can also be considered as a distortion, correlates with $\mathcal{D}_{W_{0}}$ with $r_{\mathrm{P}}\approx 0.77$.

Note, however, that while the discussed above correlation coefficients are positive, they are not particularly large, and one can construct other quantities from parameters of these two models which will correlate with $\mathcal{D}_{W_{0}}$ better and will not have any physical meaning. 
For instance, for the SS2v model, the construct $|\eta_{1}|^{n}$ correlates with $\mathcal{D}_{W_{0}}$ the better the larger the exponent $n$. 

There can be suggested an alternative hypothesis that, for RWs to appear, all nonlinear modes in the system must be synchronized, and no modes in the vicinity of a RW maximum can be considered as a small correction. 
However, in this case it is unclear why the nonlinear modes of the SS2v and BS2v fits turn out to be so close to the IST spectrum of the whole wavefield. 


\section{Conclusions}
\label{Sec:Conclusions}

This paper presents a systematic numerical examination of RWs that emerge in the nonlinear stage of the noise-induced MI of a plane wave. 
It is known~\cite{agafontsev2015integrable,kraych2019statistical} that at this stage the statistical functions, such as the moments of amplitude, Fourier spectrum, spatial correlation functions and PDF of intensity, perform damped oscillations with time around their asymptotic values. 
This study shows that the frequency of RW occurrence begins to increase from the zero level simultaneously with the fourth-order moment $\kappa_{4}$ reaching its first (largest) local maximum ($\kappa_{4}\approx 2.9$), and then grows in an oscillatory manner approaching its asymptotic value at long time. 
Thus, the first few local maximums of the fourth-order moment correspond, in fact, to a significantly less frequent occurrence of RWs compared to the statistically stationary state of MI (when $\kappa_{4}\approx 2$), presenting a counterexample to the common assertion that values of $\kappa_{4}$ greater than $2$ correspond to more frequent RWs. 

Then, the study focuses on RWs that appear near the statistically stationary state of MI. 
It turns out that the 1D-NLSE generates a much larger number of such RWs than a comparable linear system: by $3.7$ times larger if the linear system has the same Fourier spectrum as near the stationary state of MI, and by $8.2$ times larger if the linear system has Gaussian spectrum of the same characteristic width. 
However, in average, one RW affects by the same times smaller spatiotemporal area, so that the resulting distributions of wavefield intensity practically coincide for these systems. 
The distribution of the 1D-NLSE RWs by maximum intensity represents a function which decays slower than exponentially with increasing intensity and does not exhibit regions with a noticeable change in behavior. 
This suggests that the mechanism of RW formation should be universal for RWs of significantly different amplitudes, e.g., for RWs with $\max|\psi|\simeq 3$ and $\simeq 5$. 
At the RW maximums, the spatial phase slope (chirp) $\partial\,\mathrm{arg}\,\psi/\partial x$ is narrowly distributed around zero. 
In space-time, the points of RW maximums are distributed almost indistinguishably ``by eye'' from random uniformly distributed positions. 
The distribution of distance to the nearest neighbor for RWs is very similar to that for random positions, except that RWs do not emerge too close to each other (usually no closer than $2$ characteristic sizes of the Peregrine breather). 

Then, more than seven million of collected RWs near the statistically stationary state of MI are compared with nine exact solutions of the 1D-NLSE, some of which have been previously suggested as models for RWs. 
The nine exact solution include the rational breathers of the first (RB1, also known as the Peregrine breather), second (RB2) and third (RB3) orders, the Akhmediev (AB) and Kuznetsov (KB) breathers, a bound state of two solitons in which they have identical positions (SS2), a simplified superposition of two Akhmediev breathers (AB2), a general collision of two solitons (SS2v) and a general collision of two breathers (BS2v). 
Note that the last two models are difficult to work with since the positions of their local maximums in space-time are unknown; for this reason, the procedure of fitting the RWs with them is carried out using pre-compiled databases of such solutions. 
Also note that each RW is fitted with exact solutions in a small neighborhood around its maximum, so that the corresponding fits have maximums at the same point and with the same amplitude, phase and chirp as the considered RW (in particular, within this fitting procedure, all exact solutions are scaled in amplitude).

Among the nine models considered, the RB1 model shows, in average, a fairly mediocre accuracy in reproducing the wavefield of RWs. 
Moreover, its accuracy deteriorates with increasing RW amplitude, so that, at $\max|\psi|\gtrsim 3.5$, this model becomes the least accurate. 
Its undeniable advantage is that it is the simplest and, within the framework of the implemented fitting procedure, it does not depend on any internal parameters. 
The AB and KB models depend on a single internal parameter (the ratio $\eta_{1}/a$ of the imaginary part of breather eigenvalue to the plane wave amplitude), but the fitting with them for almost all RWs results in breathers that converge to the RB1 model. 
The RB2 and RB3 models do not depend on any internal parameters and, in average, demonstrate the worst accuracy in reproducing RWs. 
However, their accuracy improves with increasing RW amplitude, and, for RWs with $\max|\psi|\gtrsim 4$, they become comparable with the remaining four models. 

The SS2 model represents a simplified interaction of two solitons when their velocities, positions and, at the moment of reaching the maximum, phases are identical. 
This model is rather simple (similar in complexity with the AB and KB solutions) and depends on only one internal parameter -- the ratio $\eta_{2}/\eta_{1}$ of the imaginary parts of soliton eigenvalues. 
Nevertheless, in average, it shows a good accuracy in reproducing RWs, which is significantly better than that of the RB1 model and does not change much with increasing RW amplitude. 

The AB2 model is a simplified situation when two ABs emerge at the same time and with the same spatial phase shift (i.e., at the same place). 
It depends on two internal parameters -- the ratios $\eta_{1,2}/a$ of the imaginary parts of breather eigenvalues to the plane wave amplitude, however, the explicit relation for it is very cumbersome. 
In average across all RWs, it shows practically the same accuracy to that of the SS2 model, but for sufficiently large RWs, $\max|\psi|\gtrsim 4$, it becomes significantly more accurate. 
However, the distribution of internal parameters of the AB2 model reveals that, depending on RW, it switches, in fact, between three models -- the AB (including the RB1), RB2 and a genuine AB2. 

The SS2v model represents a general collision of two solitons when their velocities are different and no other constraints apply. 
This model depends on three internal parameters and shows, in average, the best accuracy in reproducing RWs (around $4$ times better than that for the RB1), which practically does not change with increasing RW amplitude. 
Similarly, the BS2v model is a general collision of two Tajiri-Watanabe breathers with different velocities, which depends on five internal parameters. 
Due to limitations in computational power, in the present paper this model is further limited by consideration of breathers with eigenvalues on the opposite sides of the branch cut. 
In average across all RWs, its accuracy is slightly worse than for the SS2v model, but significantly improves with increasing RW amplitude, and for RWs with $\max|\psi|\gtrsim 3.5$ it becomes by far the most accurate.

A plane wave of unit amplitude perturbed by a small noise can be represented both as a soliton gas and as a breather gas, in which the eigenvalues of, respectively, solitons or breathers are concentrated in the complex $\lambda$-plane very close to or at the $[0,i]$ segment of the imaginary axis. 
The most accurate SS2v and BS2v models represent ad-hoc approximations involving only two nonlinear modes (solitons or breathers), which, near the RW maximums, replace the whole wavefield composed of a large number of nonlinear modes. 
Interestingly, the parameters of these two approximating nonlinear modes turn out to be surprisingly similar to the IST spectrum of the whole wavefield, which makes it possible to put forward a hypothesis that RWs may emerge due to synchronization of a few nonlinear modes in presence of many other modes that may lead to significant distortions. 
This hypothesis is in line with the observation that the larger the deviations of the SS2v and BS2v models from RWs in the physical space, the larger the deviations of nonlinear modes of these models from the IST spectrum of the whole wavefield (assuming that both kinds of deviations are caused by distortions from the other nonlinear modes). 
An alternative hypothesis is that, for RWs to appear, all nonlinear modes in the system must be synchronized, and no modes in a vicinity of a RW maximum can be considered as a small correction. 


\begin{center}
\textbf{Potential for improvement}
\end{center}

The accuracy of the SS2v and BS2v models can be somewhat improved by compiling a denser databases of such solutions. 
The BS2v model can be improved by allowing breathers to have eigenvalues with the imaginary part greater than the plane wave amplitude, $\eta_{1,2}/a > 1$, and by considering breathers whose eigenvalues lie on the same side of the branch cut. 
Databases of the SS2v and BS2v solutions require a lot of memory, and searching through them during the fitting procedure is computationally intensive. 
An alternative approach may be possible in training a neural network on a database and then using it to obtain the optimal fitting parameters. 

The fitting procedure can also be modified. 
For instance, one can perform fitting in a sufficiently large spatiotemporal window (of order of characteristic sizes of the RB1 solution), waiving the requirement that the fits with exact solutions must have maximums at the same point and with the same amplitude, phase and chirp, as the considered RW. 
This procedure will add five additional internal parameters to each model -- the position and time of the maximum, as well as its amplitude, phase and chirp, which will significantly increase the computational demands. 
The resulting fits should better match RWs in the physical space, but will depend significantly on the fitting window, and, in particular, might be significantly influenced by the wavefield at its edges.


\begin{center}
\textbf{Declaration of competing interests}
\end{center}

The author declares he has no known competing financial interests or personal relationships that could have appeared to influence the work reported in this paper.


\begin{center}
\textbf{Acknowledgements}
\end{center}

The author thanks Andrey Gelash and Evgenii Kuznetsov for very valuable discussions. 
This work has been supported by the RSF grant 25-72-31023.


\appendix

\section{Transformations of multi-soliton and multi-breather solutions}
\label{Sec:App1}

If two multi-soliton solutions $\psi$ and $\tilde{\psi}$ are related by transformation
\begin{equation}
	\tilde{\psi}(x,t) = \alpha\,e^{i\big(\alpha v x - \frac{\alpha^{2}v^{2}t}{2} + \Theta_{0}\big)}\psi\bigg(\alpha\, x - \alpha^{2}vt,\, \alpha^{2}t\bigg),
	\label{transformations-app}
\end{equation}
then the eigenvalues, positions and phases of their solitons satisfy
\begin{eqnarray}
    && \tilde{\lambda}_{n} = \alpha\bigg(\lambda_{n} - \frac{v}{2}\bigg), \quad \tilde{x}_{0n} = \frac{x_{0n}}{\alpha}, \nonumber\\
    && \tilde{\theta}_{0n} = \theta_{0n} + \Theta_{0} + v\, x_{0n}. 
	\label{transformations-NSS}
\end{eqnarray}
These relations are obtained by splitting the transformation~(\ref{transformations-app}) into the rescaling, Galilean and gauge parts, and then finding the relations between the $\mathbf{\tilde{q}}_{n}$-vectors of the modified and original solutions; see Eqs.~(\ref{NSsolution})-(\ref{tildeq-N-soliton}) and also the last sentence in Section~\ref{Sec:Theory-1}. 
In Eq.~(\ref{transformations-NSS}), the relation for the eigenvalues is derived by equating the $(x,t)$-dependencies of the $\mathbf{\tilde{q}}_{n}$-vectors, and the relations for the positions and phases are obtained from the observation that for the rescaling and Galilean transformations the norming constants remain the same, while for the gauge transformation they are multiplied by $e^{i\Theta_{0}}$. 

For the multi-breather case, the corresponding relations read as
\begin{eqnarray}
    && \tilde{a} = \alpha\,a, \quad \tilde{\lambda}_{n} = \alpha\bigg(\lambda_{n} - \frac{v}{2}\bigg), \quad \tilde{x}_{0n} = \frac{x_{0n}}{\alpha}, \nonumber\\
    && \tilde{\theta}_{0n} = \theta_{0n}, \quad \tilde{\Theta} = \Theta + \Theta_{0},
	\label{transformations-NBS}
\end{eqnarray}
where $a$ and $\tilde{a}$ are the amplitudes, while $\Theta$ and $\tilde{\Theta}$ are the phases of the corresponding plane waves. 
These relations are obtained in the same way with the exception of the Galilean transformation, for which Eqs.~(\ref{NBsolution})-(\ref{qn_breathers}) change significantly and cannot be reduced to their original form simply by redefining breather parameters. 
This happens because the dressing procedure starts from a modified wave function $\mathbf{\Phi}^{\mathrm{B}}_{0}$ having a branch cut shifted along the real axis, which leads to a different formulation of Eqs.~(\ref{NBsolution})-(\ref{qn_breathers}); it is assumed that the norming constants are not changed during this procedure. 

For the mirror reflection in space, the parameters of multi-soliton and multi-breather solutions transform as
\begin{eqnarray}
	&& \tilde{\psi}(x,t) = \psi(-x,t): \quad\quad \tilde{\lambda}_{n} = -\lambda_{n}^{*}, \nonumber\\
	&& \tilde{x}_{0n} = -x_{0n}, \quad \tilde{\theta}_{0n} = \theta_{0n}, \quad \tilde{\Theta} = \Theta,
	\label{transformations-mirror-x}
\end{eqnarray}
where the last equality is written for the multi-breather case. 
For the mirror reflection in time, the corresponding relations read as
\begin{eqnarray}
	&& \tilde{\psi}(x,t) = \psi^{*}(x,-t): \quad\quad \tilde{\lambda}_{n} = -\lambda_{n}^{*}, \nonumber\\
	&& \tilde{x}_{0n} = x_{0n}, \quad \tilde{\theta}_{0n} = -\theta_{0n}, \quad \tilde{\Theta} = -\Theta.
	\label{transformations-mirror-x}
\end{eqnarray}


\begin{figure}[t]\centering
	\includegraphics[width=8.5cm]{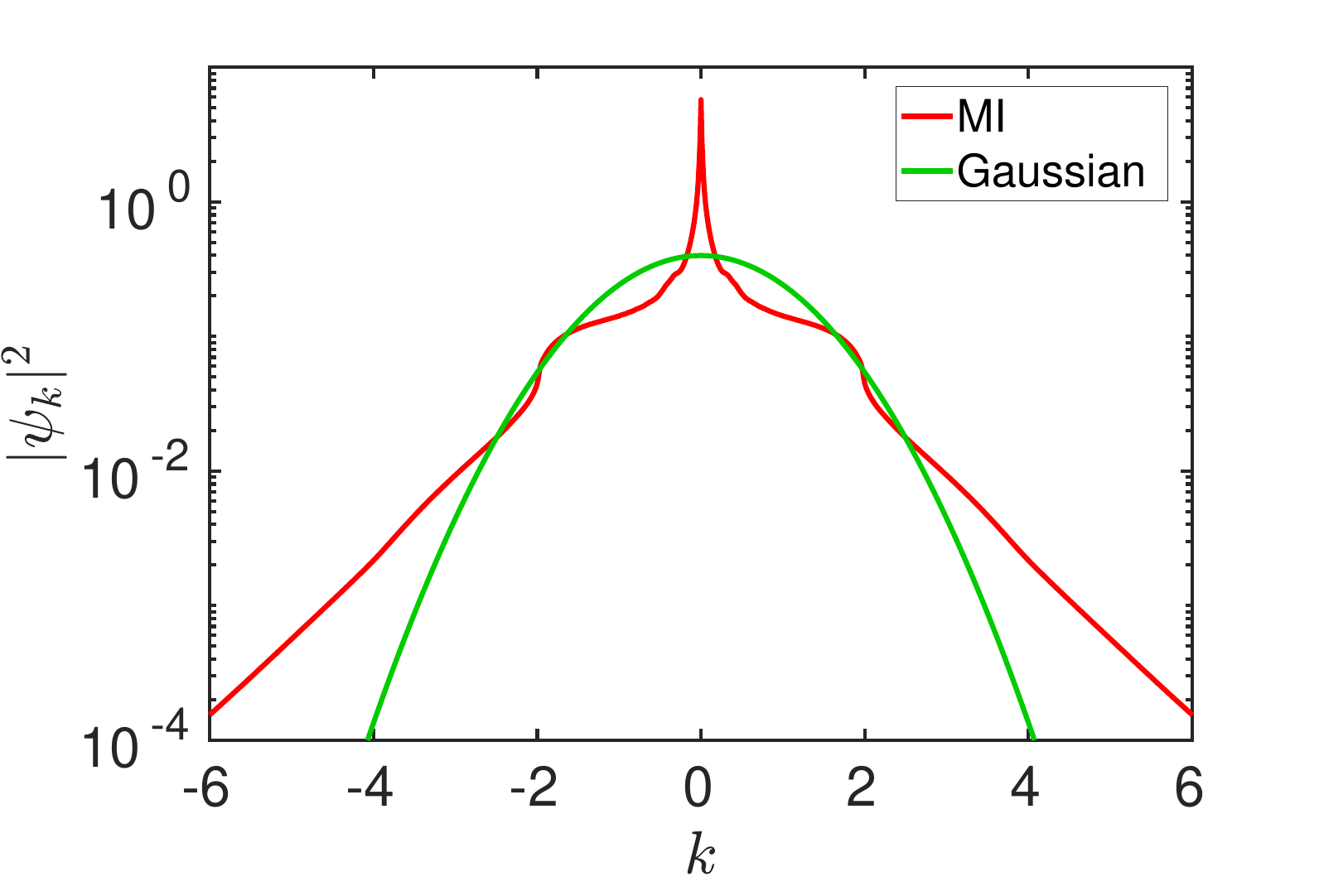}
	
	\caption{\small {\it (Color on-line)}
	Fourier spectrum $S_{k}$, see Eq.~(\ref{Sk}), near the statistically stationary state of MI (red), averaged over both the time interval $t\in[80,200]$ and the ensemble of initial conditions discussed in the main part of this paper. 
	The green line shows the Gaussian spectrum $S_{k}\propto e^{-2 k^{2}/\delta k^{2}}$ with $\delta k = 2$, which has the same characteristic spectral width $\sum_{k}k^{2}|\psi_{k}|^{2}/\sum_{k}|\psi_{k}|^{2} = 1$.
	}
	\label{fig:fig11}
\end{figure}

\begin{figure}[t]\centering
	\includegraphics[width=8.5cm]{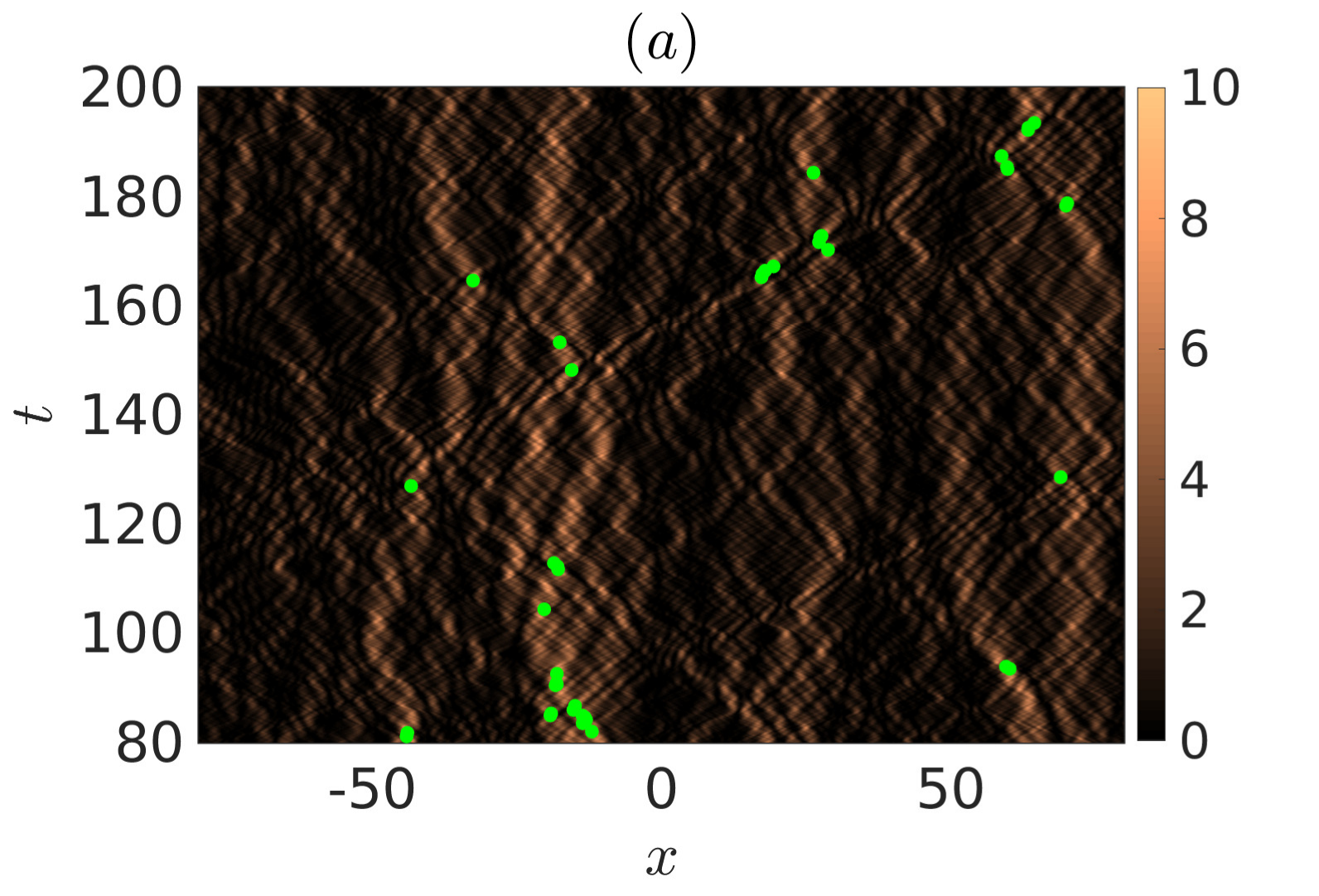}\\
	\includegraphics[width=8.5cm]{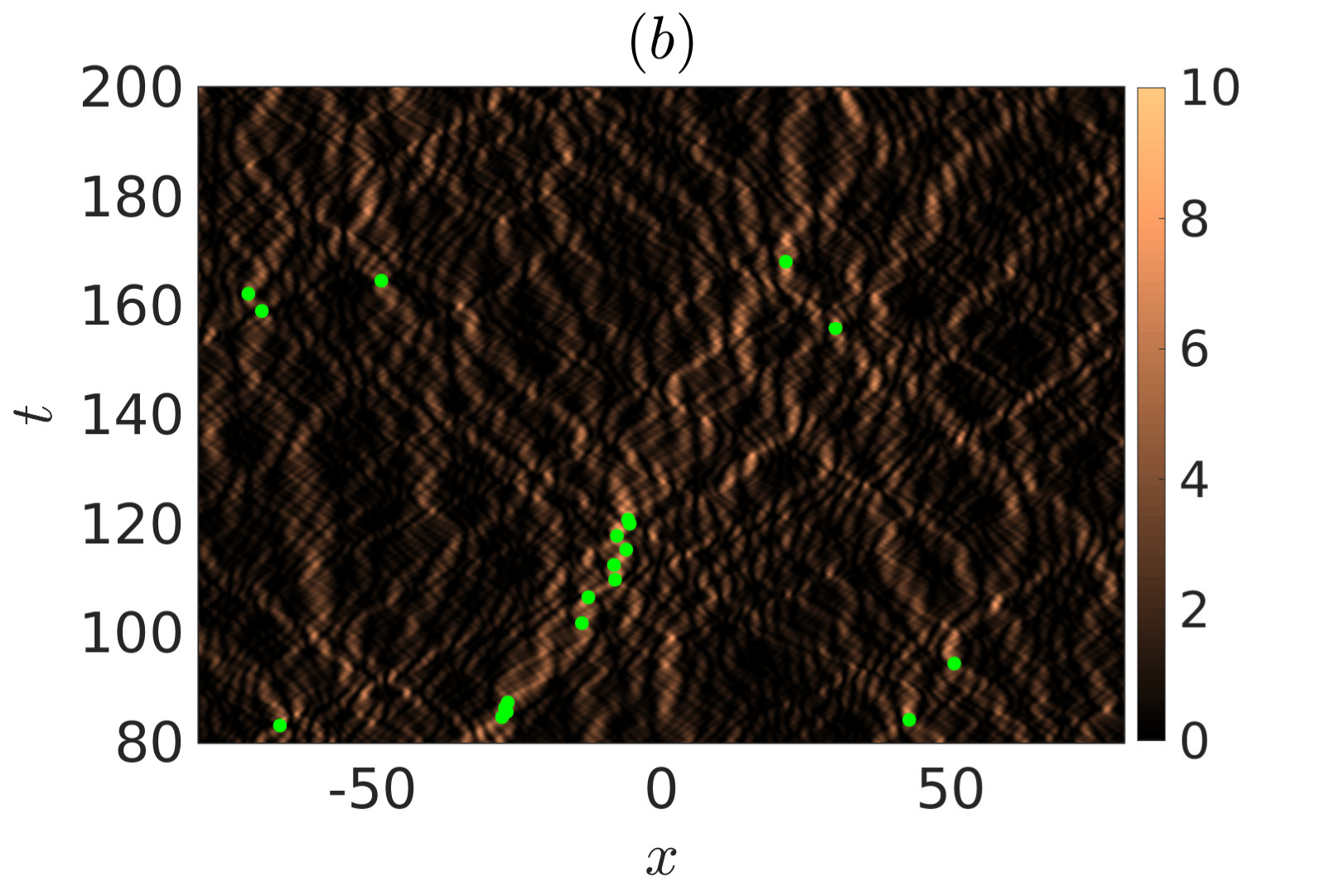}
	
	\caption{\small {\it (Color on-line)} 
	Space-time evolution in the framework of Eq.~(\ref{NLSE-linear}) from one realization of initial conditions~(\ref{IC-linear}): (a) for the MI spectrum and (b) for the Gaussian spectrum. 
	Color shows the intensity $|\psi|^{2}$, and green dots indicate the positions of RW maximums. 
	While the scales are the same as in Fig.~\ref{fig:fig04}, both panels represent enlargements of regions where relatively many RWs are detected; in the adjacent regions of space-time there are significantly fewer RWs.
	}
	\label{fig:fig12}
\end{figure}

\begin{figure}[t]\centering
	\includegraphics[width=8.5cm]{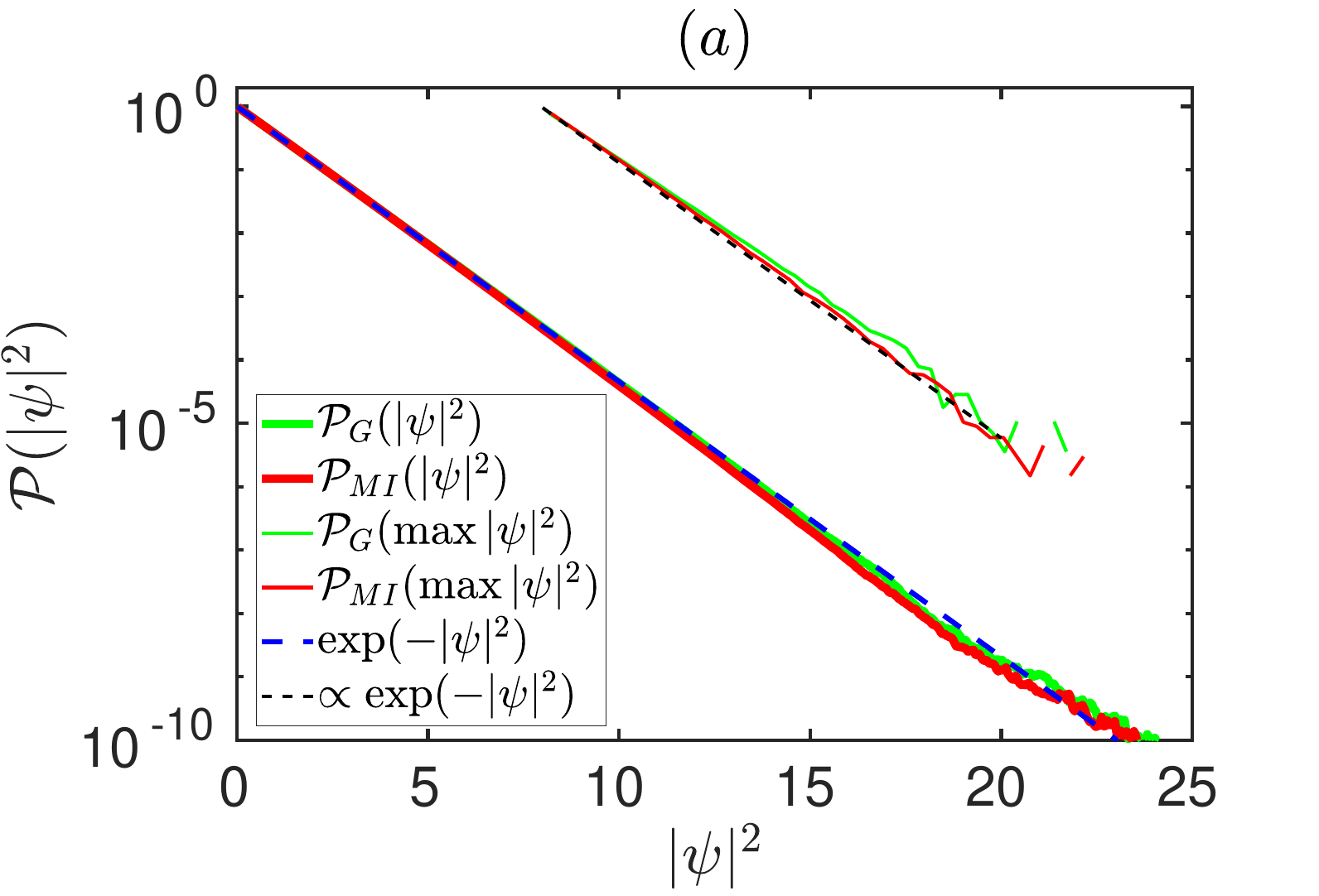}\\
	\includegraphics[width=8.5cm]{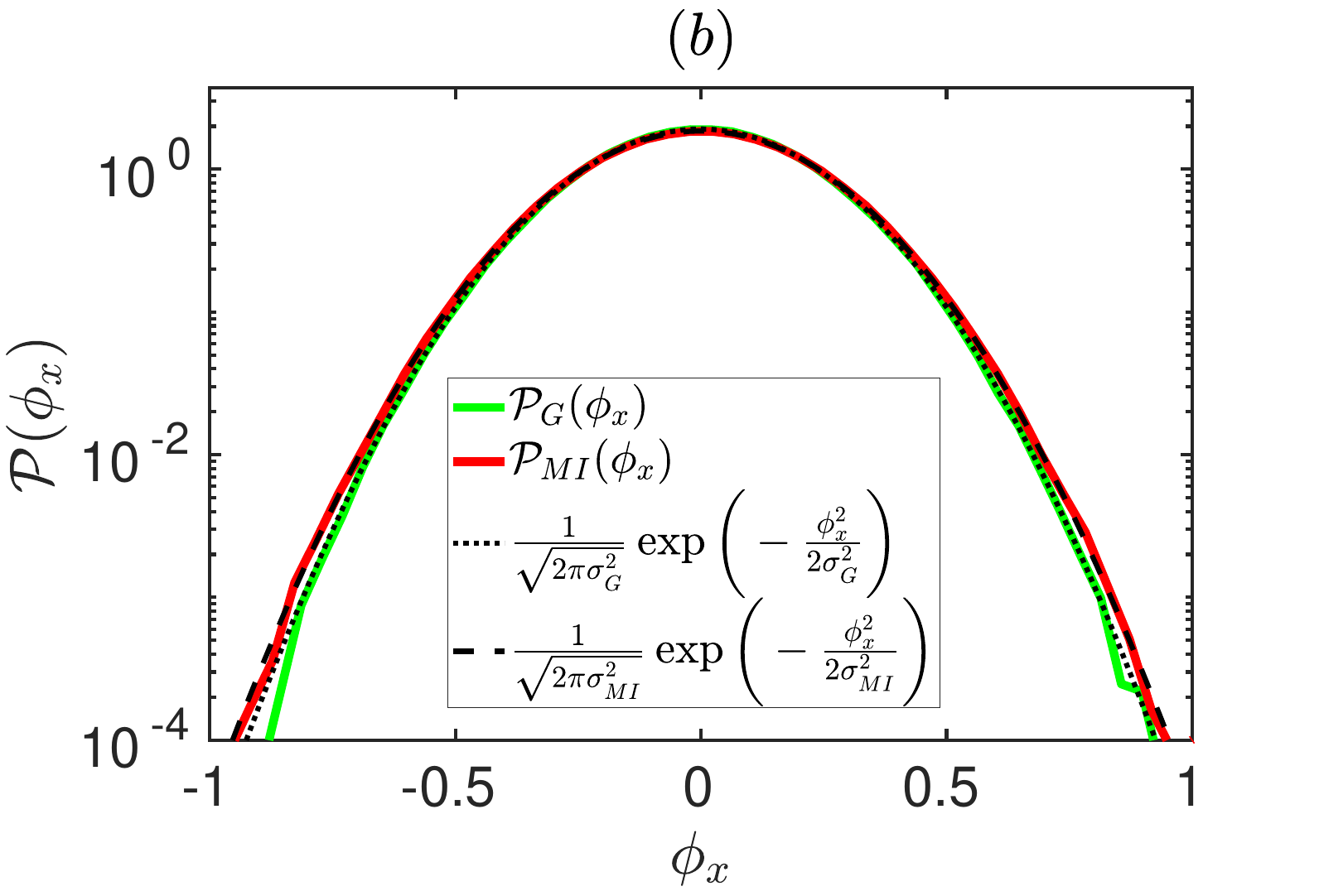}
	
	\caption{\small {\it (Color on-line)} 
	(a) PDFs of intensity for the whole wavefield $\mathcal{P}(|\psi|^{2})$ (thick lines) and for the maximums of collected RWs $\mathcal{P}(\max|\psi|^{2})$ (thin lines), and (b) PDF of chirp $\phi_{x}=\partial\,\mathrm{arg}\psi/\partial x$ at the RW maximums. 
	Green lines and the subscript $\mathrm{G}$ correspond to the Gaussian spectrum, while red lines and the subscript $\mathrm{MI}$ correspond to the MI spectrum. 
	In panel (a), dashed lines show the exponential PDFs $e^{-|\psi|^{2}}$ (thick blue) and $e^{8-\max|\psi|^{2}}$ (thin black). 
	In panel (b), black lines indicate Gaussian distributions of the same variances. 
	}
	\label{fig:fig13}
\end{figure}


\section{Rogue waves in a linear system}
\label{Sec:App2}

This Appendix discusses RWs that emerge in a linear system, 
\begin{equation}
	i\psi_t + \frac{1}{2}\psi_{xx} = 0,
	\label{NLSE-linear}
\end{equation}
which is obtained from the 1D-NLSE~(\ref{NLSE}) by discarding the nonlinear term. 
When studying such RWs, it is essential to correctly set up the initial spectrum, as it is conserved during the evolution. 
For instance, for a twice narrower spectrum, the structures in the physical space will be twice larger, while in time they will evolve four times slower, so that the system will produce $8$ times less RWs per unit of spatiotemporal area.

In this Appendix, two initial spectra are considered. 
The first one is the Fourier spectrum 
\begin{equation}
	S_{k} = \langle|\psi_{k}|^{2}\rangle/\Delta k,
	\label{Sk}
\end{equation}
near the statistically stationary state of MI, see Fig.~\ref{fig:fig11}, averaged over both the time interval $t\in[80,200]$ and the ensemble of initial conditions discussed in the main part of this paper. 
Here $\psi_{k}$ is the Fourier-transformed wavefield and $\Delta k$ is the distance between neighbor wavenumbers. 
However, this spectrum has a divergency~\cite{agafontsev2015integrable} by power law near the zeroth harmonic, which may lead to a non-standard behavior. 
For this reason, a Gaussian spectrum $S_{k}\propto e^{-2 k^{2}/\delta k^{2}}$, $\delta k = 2$, having the same characteristic width $\sum_{k}k^{2}|\psi_{k}|^{2}/\sum_{k}|\psi_{k}|^{2} = 1$ is also considered. 
Both spectra have unit average intensity, $\overline{|\psi|^{2}} = 1$.

Since Eq.~(\ref{NLSE-linear}) is linear, it is integrated numerically in one time step as $\psi(x,t) = \mathcal{F}\big[e^{-\frac{i k^{2}t}{2}}\mathcal{F}[\psi(x,0)]\big]$, where $\mathcal{F}$ is the Fourier transform. 
The initial wavefield is defined as a partially coherent wave
\begin{equation}
	\psi|_{t=0} = \sum_{m} \big[S_{k}\Delta k\big]^{1/2} e^{ik_{m}x + i\phi_{m}},
	\label{IC-linear}
\end{equation}
with the discussed above spectrum $S_{k}$ and random phases $\phi_{m}$ for each $k_{m}$ and each realization of initial conditions. 
For direct comparison with the main part of the paper, the experiments are performed for both spectra in the periodic box of size $L = 128\sqrt{2}\pi$ and RWs are detected within the time interval $t\in[80,200]$; each ensemble contains $20\,000$ simulations that differ from each other only in the random realization of the initial phases $\phi_{m}$.

\renewcommand{\arraystretch}{1.2}

\begin{table}[t]
\caption{
Comparison of RWs that emerge (i) within the 1D-NLSE~(\ref{NLSE}) near the statistically stationary state $t\in[80,200]$ of MI (second column), (ii) within the linear system~(\ref{NLSE-linear}) from the MI spectrum (third column), and (iii) within the linear system~(\ref{NLSE-linear}) from the Gaussian spectrum (fourth column). 
In the table, the second row shows how many units of space-time area it takes per $1$ RW event, $\mu$ is the density of RWs per unit of the RB1-rescaled area, see Section~\ref{Sec:Results1-2}, $\langle r_{\mathrm{nn}}^{\mathrm{RW}}\rangle$ is the mean spatio-temporal distance to the nearest neighbor RW, $R$ is the Clark -- Evans aggregation index, and $\phi_{x}$ is the spatial phase slope (chirp) at the points of RW maximums. 
As in the main part of this paper, the first number in round brackets means the mean and the second means one standard deviation. 
}
\begin{center}
 \begin{tabular}{| c | c | c | c |}
  \hline
  Parameter 			& 1D-NLSE 			& linear, MI 		& linear, Gaussian 	\\ \hline
  area per 1 RW 		& $(190.4 \pm 8.9)$ & $(860 \pm 370)$ 	& $(1650 \pm 380)$ 	\\ \hline
  $\mu\times 10^{3}$	& $(21.2 \pm 1.0)$ 	& $(5.7 \pm 2.9)$ 	& $(2.6 \pm 0.6)$ 	\\ \hline
  $\langle r_{\mathrm{nn}}^{\mathrm{RW}}\rangle$ 		& $3.8$ 	& $1.6$ 	& $5.3$ 	\\ \hline
  $R$ 				& $1.1$ 				& $0.25$ 			& $0.54$ 			\\ \hline
  $\max|\psi|$ 		& $(3.06 \pm 0.21)$ 	& $(3.00 \pm 0.16)$ & $(3.01 \pm 0.17)$ \\ \hline
  $\max|\psi|^{2}$ 	& $(9.4 \pm 1.4)$ 		& $(9 \pm 1)$ 		& $(9.1 \pm 1.1)$ 	\\ \hline
  $\phi_{x}$ 		& $(0.00 \pm 0.09)$ 	& $(0.00 \pm 0.22)$ & $(0.00 \pm 0.21)$ \\ \hline
 \end{tabular}
\end{center}
 \label{tab:M2}
\end{table}

Figure~\ref{fig:fig12}(a,b) shows the typical space-time evolution of intensity $|\psi|^{2}$ together with the positions of RW maximums for simulations from one realization of initial conditions~(\ref{IC-linear}) for both spectra $S_{k}$. 
Note that, while the scales in both panels of the figure match those used in the similar Fig.~\ref{fig:fig04}(a) for the 1D-NLSE case, both panels represent enlargements of regions where relatively many RWs are detected; in the adjacent regions of space-time there are significantly fewer RWs. 
One can see that the number of RWs for both variants of the spectrum is much smaller than for the 1D-NLSE, however, in space-time, the structures of high intensity are much larger. 

The latter structures are formed as random synchronizations of harmonics. 
In particular, if for a certain position in space and time the phases of harmonics within the interval of wavenumbers $k_{0}\pm\delta k_{0}$ are synchronized, then the corresponding structure in the physical space has characteristic spatial width $\sim \delta k_{0}^{-1}$, moves with the group velocity $v_{\mathrm{g}} = k_{0}$, and disperse within characteristic time $\sim \delta k_{0}^{-2}$. 
As follows from Fig.~\ref{fig:fig12}(a,b), RWs are formed as superpositions of such high-intensity long-living structures with each other and with a rapidly changing shorter-wave oscillations, so that the distance between neighbor RWs can be quite small. 
For the MI spectrum, which is concentrated in harmonics with small wavenumbers, the high-intensity structures move with smaller velocities and appear more vertical in Fig.~\ref{fig:fig12}(a). 

Across the entire simulation ensembles, a total of $1\,925\,116$ and $872\,608$ RWs have been detected for the MI and Gaussian spectra, respectively; these numbers are smaller than for the 1D-NLSE by $3.7$ and $8.2$ times. 
The mean values and standard deviations of various quantities for these RWs are given in Table~\ref{tab:M2}. 
For the second and third rows of this table, the large standard deviations originate due to a large variation in the number of RWs detected in different simulations; to decrease this variation, one needs a much larger observation window both in space and in time. 
The mean distance to the nearest neighbor RW turns out to be small for both spectra (for the MI spectrum it is $2.4$ times smaller than for the 1D-NLSE), and, given a much smaller density $\mu$ of RWs per unit area, this leads to the Clark -- Evans aggregation index $R$ significantly smaller than unity, indicating clusterization of RWs in space and time. 

Figure~\ref{fig:fig13}(a) shows the distribution of RWs by intensity at the maximum point, $\mathcal{P}(\max|\psi|^{2})$, together with the PDF of intensity for the entire wavefield $\mathcal{P}(|\psi|^{2})$ averaged over the ensemble and time interval $t\in [80, 200]$. 
The PDF for local maximums $\mathcal{P}(\max|\psi|^{2})$ practically coincides with the exponential distribution $e^{8-\max|\psi|^{2}}$ for both spectra, with larger deviations visible for the Gaussian spectrum; these larger deviations may be related to the smaller number of RWs detected in this experiment. 
The PDF for the entire wavefield $\mathcal{P}(|\psi|^{2})$ coincides with the exponential PDF $\exp(-|\psi|^{2})$ for the Gaussian spectrum, while for the MI spectrum it turns out to be slightly smaller at $|\psi|^{2}\simeq 18$. 
The latter may be related to the divergency of the MI spectrum near the zeroth harmonic: in this case, the central limit theorem cannot be applied, and the PDF of intensity does not have to be exponential. 
Alternatively, this may be an effect of the finite sizes of the observation window in space and time. 
At the points of RW maximums, the spatial phase slope (chirp) $\phi_{x}=\partial\,\mathrm{arg}\psi/\partial x$ is Gaussian-distributed around zero, see Fig.~\ref{fig:fig13}(b), with standard deviations $0.22$ and $0.21$ for the MI and Gaussian spectra, respectively. 

The time-averaged PDF of intensity for the entire wavefield $\mathcal{P}(|\psi|^{2})$ equals, effectively, a fraction of spatiotemporal area with the given intensity. 
For the statistically stationary state of MI, as well as for the linear system~(\ref{NLSE-linear}), this PDF is exponential, $\exp(-|\psi|^{2})$, or very close to it. 
However, the number of detected RWs for the nonlinear system is by several times larger. 
This means that one RW in the 1D-NLSE case affects, in average, by the same times smaller spatiotemporal area. 


%


\begin{thebibliography}{50}%
\makeatletter
\providecommand \@ifxundefined [1]{%
 \@ifx{#1\undefined}
}%
\providecommand \@ifnum [1]{%
 \ifnum #1\expandafter \@firstoftwo
 \else \expandafter \@secondoftwo
 \fi
}%
\providecommand \@ifx [1]{%
 \ifx #1\expandafter \@firstoftwo
 \else \expandafter \@secondoftwo
 \fi
}%
\providecommand \natexlab [1]{#1}%
\providecommand \enquote  [1]{``#1''}%
\providecommand \bibnamefont  [1]{#1}%
\providecommand \bibfnamefont [1]{#1}%
\providecommand \citenamefont [1]{#1}%
\providecommand \href@noop [0]{\@secondoftwo}%
\providecommand \href [0]{\begingroup \@sanitize@url \@href}%
\providecommand \@href[1]{\@@startlink{#1}\@@href}%
\providecommand \@@href[1]{\endgroup#1\@@endlink}%
\providecommand \@sanitize@url [0]{\catcode `\\12\catcode `\$12\catcode
  `\&12\catcode `\#12\catcode `\^12\catcode `\_12\catcode `\%12\relax}%
\providecommand \@@startlink[1]{}%
\providecommand \@@endlink[0]{}%
\providecommand \url  [0]{\begingroup\@sanitize@url \@url }%
\providecommand \@url [1]{\endgroup\@href {#1}{\urlprefix }}%
\providecommand \urlprefix  [0]{URL }%
\providecommand \Eprint [0]{\href }%
\providecommand \doibase [0]{https://doi.org/}%
\providecommand \selectlanguage [0]{\@gobble}%
\providecommand \bibinfo  [0]{\@secondoftwo}%
\providecommand \bibfield  [0]{\@secondoftwo}%
\providecommand \translation [1]{[#1]}%
\providecommand \BibitemOpen [0]{}%
\providecommand \bibitemStop [0]{}%
\providecommand \bibitemNoStop [0]{.\EOS\space}%
\providecommand \EOS [0]{\spacefactor3000\relax}%
\providecommand \BibitemShut  [1]{\csname bibitem#1\endcsname}%
\let\auto@bib@innerbib\@empty
\bibitem [{\citenamefont {Kharif}\ and\ \citenamefont
  {Pelinovsky}(2003)}]{kharif2003physical}%
  \BibitemOpen
  \bibfield  {author} {\bibinfo {author} {\bibfnamefont {C.}~\bibnamefont
  {Kharif}}\ and\ \bibinfo {author} {\bibfnamefont {E.}~\bibnamefont
  {Pelinovsky}},\ }\bibfield  {title} {\bibinfo {title} {{Physical mechanisms
  of the rogue wave phenomenon}},\ }\href@noop {} {\bibfield  {journal}
  {\bibinfo  {journal} {{Eur. J. Mech.-B/Fluids}}\ }\textbf {\bibinfo {volume}
  {22}},\ \bibinfo {pages} {603} (\bibinfo {year} {2003})}\BibitemShut
  {NoStop}%
\bibitem [{\citenamefont {Dysthe}\ \emph {et~al.}(2008)\citenamefont {Dysthe},
  \citenamefont {Krogstad},\ and\ \citenamefont {Muller}}]{dysthe2008oceanic}%
  \BibitemOpen
  \bibfield  {author} {\bibinfo {author} {\bibfnamefont {K.}~\bibnamefont
  {Dysthe}}, \bibinfo {author} {\bibfnamefont {H.~E.}\ \bibnamefont
  {Krogstad}},\ and\ \bibinfo {author} {\bibfnamefont {P.}~\bibnamefont
  {Muller}},\ }\bibfield  {title} {\bibinfo {title} {{Oceanic rogue waves}},\
  }\href@noop {} {\bibfield  {journal} {\bibinfo  {journal} {{Annu. Rev. Fluid
  Mech.}}\ }\textbf {\bibinfo {volume} {40}},\ \bibinfo {pages} {287} (\bibinfo
  {year} {2008})}\BibitemShut {NoStop}%
\bibitem [{\citenamefont {Onorato}\ \emph {et~al.}(2013)\citenamefont
  {Onorato}, \citenamefont {Residori}, \citenamefont {Bortolozzo},
  \citenamefont {Montina},\ and\ \citenamefont {Arecchi}}]{onorato2013rogue}%
  \BibitemOpen
  \bibfield  {author} {\bibinfo {author} {\bibfnamefont {M.}~\bibnamefont
  {Onorato}}, \bibinfo {author} {\bibfnamefont {S.}~\bibnamefont {Residori}},
  \bibinfo {author} {\bibfnamefont {U.}~\bibnamefont {Bortolozzo}}, \bibinfo
  {author} {\bibfnamefont {A.}~\bibnamefont {Montina}},\ and\ \bibinfo {author}
  {\bibfnamefont {F.~T.}\ \bibnamefont {Arecchi}},\ }\bibfield  {title}
  {\bibinfo {title} {{Rogue waves and their generating mechanisms in different
  physical contexts}},\ }\href@noop {} {\bibfield  {journal} {\bibinfo
  {journal} {{Phys. Rep.}}\ }\textbf {\bibinfo {volume} {528}},\ \bibinfo
  {pages} {47} (\bibinfo {year} {2013})}\BibitemShut {NoStop}%
\bibitem [{\citenamefont {Dudley}\ \emph {et~al.}(2019)\citenamefont {Dudley},
  \citenamefont {Genty}, \citenamefont {Mussot}, \citenamefont {Chabchoub},\
  and\ \citenamefont {Dias}}]{dudley2019rogue}%
  \BibitemOpen
  \bibfield  {author} {\bibinfo {author} {\bibfnamefont {J.~M.}\ \bibnamefont
  {Dudley}}, \bibinfo {author} {\bibfnamefont {G.}~\bibnamefont {Genty}},
  \bibinfo {author} {\bibfnamefont {A.}~\bibnamefont {Mussot}}, \bibinfo
  {author} {\bibfnamefont {A.}~\bibnamefont {Chabchoub}},\ and\ \bibinfo
  {author} {\bibfnamefont {F.}~\bibnamefont {Dias}},\ }\bibfield  {title}
  {\bibinfo {title} {{Rogue waves and analogies in optics and oceanography}},\
  }\href@noop {} {\bibfield  {journal} {\bibinfo  {journal} {Nat. Rev. Phys.}\
  }\textbf {\bibinfo {volume} {1}},\ \bibinfo {pages} {675} (\bibinfo {year}
  {2019})}\BibitemShut {NoStop}%
\bibitem [{\citenamefont {Dysthe}\ and\ \citenamefont
  {Trulsen}(1999)}]{dysthe1999note}%
  \BibitemOpen
  \bibfield  {author} {\bibinfo {author} {\bibfnamefont {K.~B.}\ \bibnamefont
  {Dysthe}}\ and\ \bibinfo {author} {\bibfnamefont {K.}~\bibnamefont
  {Trulsen}},\ }\bibfield  {title} {\bibinfo {title} {{Note on breather type
  solutions of the NLS as models for freak-waves}},\ }\href@noop {} {\bibfield
  {journal} {\bibinfo  {journal} {Phys. Scr.}\ }\textbf {\bibinfo {volume}
  {1999}},\ \bibinfo {pages} {48} (\bibinfo {year} {1999})}\BibitemShut
  {NoStop}%
\bibitem [{\citenamefont {Osborne}\ \emph {et~al.}(2000)\citenamefont
  {Osborne}, \citenamefont {Onorato},\ and\ \citenamefont
  {Serio}}]{osborne2000nonlinear}%
  \BibitemOpen
  \bibfield  {author} {\bibinfo {author} {\bibfnamefont {A.~R.}\ \bibnamefont
  {Osborne}}, \bibinfo {author} {\bibfnamefont {M.}~\bibnamefont {Onorato}},\
  and\ \bibinfo {author} {\bibfnamefont {M.}~\bibnamefont {Serio}},\ }\bibfield
   {title} {\bibinfo {title} {{The nonlinear dynamics of rogue waves and holes
  in deep-water gravity wave trains}},\ }\href@noop {} {\bibfield  {journal}
  {\bibinfo  {journal} {Phys. Lett. A}\ }\textbf {\bibinfo {volume} {275}},\
  \bibinfo {pages} {386} (\bibinfo {year} {2000})}\BibitemShut {NoStop}%
\bibitem [{\citenamefont {Osborne}(2010{\natexlab{a}})}]{osborne2010nonlinear}%
  \BibitemOpen
  \bibfield  {author} {\bibinfo {author} {\bibfnamefont {A.}~\bibnamefont
  {Osborne}},\ }\href@noop {} {\emph {\bibinfo {title} {{Nonlinear Ocean Waves
  and the Inverse Scattering Transform}}}}\ (\bibinfo  {publisher} {Academic
  Press},\ \bibinfo {year} {2010})\BibitemShut {NoStop}%
\bibitem [{\citenamefont {Shrira}\ and\ \citenamefont
  {Geogjaev}(2010)}]{shrira2010makes}%
  \BibitemOpen
  \bibfield  {author} {\bibinfo {author} {\bibfnamefont {V.~I.}\ \bibnamefont
  {Shrira}}\ and\ \bibinfo {author} {\bibfnamefont {V.~V.}\ \bibnamefont
  {Geogjaev}},\ }\bibfield  {title} {\bibinfo {title} {{What makes the
  peregrine soliton so special as a prototype of freak waves?}},\ }\href@noop
  {} {\bibfield  {journal} {\bibinfo  {journal} {J. Eng. Math.}\ }\textbf
  {\bibinfo {volume} {67}},\ \bibinfo {pages} {11} (\bibinfo {year}
  {2010})}\BibitemShut {NoStop}%
\bibitem [{\citenamefont {Kivshar}\ and\ \citenamefont
  {Agrawal}(2003)}]{kivshar2003optical}%
  \BibitemOpen
  \bibfield  {author} {\bibinfo {author} {\bibfnamefont {Y.~S.}\ \bibnamefont
  {Kivshar}}\ and\ \bibinfo {author} {\bibfnamefont {G.}~\bibnamefont
  {Agrawal}},\ }\href@noop {} {\emph {\bibinfo {title} {{Optical solitons: from
  fibers to photonic crystals}}}}\ (\bibinfo  {publisher} {Academic press,
  London},\ \bibinfo {year} {2003})\BibitemShut {NoStop}%
\bibitem [{\citenamefont {Kharif}\ \emph {et~al.}(2009)\citenamefont {Kharif},
  \citenamefont {Pelinovsky},\ and\ \citenamefont
  {Slunyaev}}]{pelinovsky2008book}%
  \BibitemOpen
  \bibfield  {author} {\bibinfo {author} {\bibfnamefont {C.}~\bibnamefont
  {Kharif}}, \bibinfo {author} {\bibfnamefont {E.}~\bibnamefont {Pelinovsky}},\
  and\ \bibinfo {author} {\bibfnamefont {A.}~\bibnamefont {Slunyaev}},\
  }\href@noop {} {\emph {\bibinfo {title} {{Rogue waves in the ocean,
  observation, theories and modeling}}}}\ (\bibinfo  {publisher} {Advances in
  Geophysical and Environmental Mechanics and Mathematics Series, Springer,
  Heidelberg},\ \bibinfo {year} {2009})\BibitemShut {NoStop}%
\bibitem [{\citenamefont {Osborne}(2010{\natexlab{b}})}]{OsborneBook2010}%
  \BibitemOpen
  \bibfield  {author} {\bibinfo {author} {\bibfnamefont {A.}~\bibnamefont
  {Osborne}},\ }\href@noop {} {\emph {\bibinfo {title} {{Nonlinear ocean
  waves}}}}\ (\bibinfo  {publisher} {Academic Press},\ \bibinfo {year}
  {2010})\BibitemShut {NoStop}%
\bibitem [{\citenamefont {Peregrine}(1983)}]{peregrine1983water}%
  \BibitemOpen
  \bibfield  {author} {\bibinfo {author} {\bibfnamefont {D.~H.}\ \bibnamefont
  {Peregrine}},\ }\bibfield  {title} {\bibinfo {title} {{Water waves, nonlinear
  Schr{\"o}dinger equations and their solutions}},\ }\href@noop {} {\bibfield
  {journal} {\bibinfo  {journal} {J. Aust. Math. Soc. Series B, Appl. Math.}\
  }\textbf {\bibinfo {volume} {25}},\ \bibinfo {pages} {16} (\bibinfo {year}
  {1983})}\BibitemShut {NoStop}%
\bibitem [{\citenamefont {Akhmediev}\ \emph
  {et~al.}(2009{\natexlab{a}})\citenamefont {Akhmediev}, \citenamefont
  {Ankiewicz},\ and\ \citenamefont {Soto-Crespo}}]{akhmediev2009rogue}%
  \BibitemOpen
  \bibfield  {author} {\bibinfo {author} {\bibfnamefont {N.}~\bibnamefont
  {Akhmediev}}, \bibinfo {author} {\bibfnamefont {A.}~\bibnamefont
  {Ankiewicz}},\ and\ \bibinfo {author} {\bibfnamefont {J.~M.}\ \bibnamefont
  {Soto-Crespo}},\ }\bibfield  {title} {\bibinfo {title} {{Rogue waves and
  rational solutions of the nonlinear Schr{\"o}dinger equation}},\ }\href@noop
  {} {\bibfield  {journal} {\bibinfo  {journal} {Phys. Rev. E}\ }\textbf
  {\bibinfo {volume} {80}},\ \bibinfo {pages} {026601} (\bibinfo {year}
  {2009}{\natexlab{a}})}\BibitemShut {NoStop}%
\bibitem [{\citenamefont {Akhmediev}\ and\ \citenamefont
  {Korneev}(1986)}]{akhmediev1986modulation}%
  \BibitemOpen
  \bibfield  {author} {\bibinfo {author} {\bibfnamefont {N.~N.}\ \bibnamefont
  {Akhmediev}}\ and\ \bibinfo {author} {\bibfnamefont {V.~I.}\ \bibnamefont
  {Korneev}},\ }\bibfield  {title} {\bibinfo {title} {{Modulation instability
  and periodic solutions of the nonlinear Schr{\"o}dinger equation}},\
  }\href@noop {} {\bibfield  {journal} {\bibinfo  {journal} {Teoret. Mat.
  Fiz.}\ }\textbf {\bibinfo {volume} {69}},\ \bibinfo {pages} {1089} (\bibinfo
  {year} {1986})}\BibitemShut {NoStop}%
\bibitem [{\citenamefont {Kuznetsov}(1977)}]{kuznetsov1977solitons}%
  \BibitemOpen
  \bibfield  {author} {\bibinfo {author} {\bibfnamefont {E.~A.}\ \bibnamefont
  {Kuznetsov}},\ }\bibfield  {title} {\bibinfo {title} {{Solitons in a
  parametrically unstable plasma}},\ }\href@noop {} {\bibfield  {journal}
  {\bibinfo  {journal} {DoSSR}\ }\textbf {\bibinfo {volume} {236}},\ \bibinfo
  {pages} {575} (\bibinfo {year} {1977})}\BibitemShut {NoStop}%
\bibitem [{\citenamefont {Kawata}\ and\ \citenamefont
  {Inoue}(1978)}]{kawata1978inverse}%
  \BibitemOpen
  \bibfield  {author} {\bibinfo {author} {\bibfnamefont {T.}~\bibnamefont
  {Kawata}}\ and\ \bibinfo {author} {\bibfnamefont {H.}~\bibnamefont {Inoue}},\
  }\bibfield  {title} {\bibinfo {title} {{Inverse scattering method for the
  nonlinear evolution equations under nonvanishing conditions}},\ }\href@noop
  {} {\bibfield  {journal} {\bibinfo  {journal} {J. Phys. Soc. Jpn.}\ }\textbf
  {\bibinfo {volume} {44}},\ \bibinfo {pages} {1722} (\bibinfo {year}
  {1978})}\BibitemShut {NoStop}%
\bibitem [{\citenamefont {Ma}(1979)}]{ma1979perturbed}%
  \BibitemOpen
  \bibfield  {author} {\bibinfo {author} {\bibfnamefont {Y.-C.}\ \bibnamefont
  {Ma}},\ }\bibfield  {title} {\bibinfo {title} {{The perturbed plane-wave
  solutions of the cubic Schr{\"o}dinger equation}},\ }\href@noop {} {\bibfield
   {journal} {\bibinfo  {journal} {Stud. Appl. Math.}\ }\textbf {\bibinfo
  {volume} {60}},\ \bibinfo {pages} {43} (\bibinfo {year} {1979})}\BibitemShut
  {NoStop}%
\bibitem [{\citenamefont {Zakharov}\ and\ \citenamefont
  {Shabat}(1972)}]{zakharov1972exact}%
  \BibitemOpen
  \bibfield  {author} {\bibinfo {author} {\bibfnamefont {V.~E.}\ \bibnamefont
  {Zakharov}}\ and\ \bibinfo {author} {\bibfnamefont {A.~B.}\ \bibnamefont
  {Shabat}},\ }\bibfield  {title} {\bibinfo {title} {{Exact theory of
  two-dimensional self-focusing and one-dimensional self-modulation of waves in
  nonlinear media}},\ }\href@noop {} {\bibfield  {journal} {\bibinfo  {journal}
  {Sov. Phys. JETP}\ }\textbf {\bibinfo {volume} {34}},\ \bibinfo {pages} {62}
  (\bibinfo {year} {1972})}\BibitemShut {NoStop}%
\bibitem [{\citenamefont {Novikov}\ \emph {et~al.}(1984)\citenamefont
  {Novikov}, \citenamefont {Manakov}, \citenamefont {Pitaevskii},\ and\
  \citenamefont {Zakharov}}]{novikov1984theory}%
  \BibitemOpen
  \bibfield  {author} {\bibinfo {author} {\bibfnamefont {S.}~\bibnamefont
  {Novikov}}, \bibinfo {author} {\bibfnamefont {S.~V.}\ \bibnamefont
  {Manakov}}, \bibinfo {author} {\bibfnamefont {L.~P.}\ \bibnamefont
  {Pitaevskii}},\ and\ \bibinfo {author} {\bibfnamefont {V.~E.}\ \bibnamefont
  {Zakharov}},\ }\href@noop {} {\emph {\bibinfo {title} {{Theory of solitons:
  the inverse scattering method}}}}\ (\bibinfo  {publisher} {Springer Science
  \& Business Media, New York},\ \bibinfo {year} {1984})\BibitemShut {NoStop}%
\bibitem [{\citenamefont {Suret}\ \emph {et~al.}(2024)\citenamefont {Suret},
  \citenamefont {Randoux}, \citenamefont {Gelash}, \citenamefont {Agafontsev},
  \citenamefont {Doyon},\ and\ \citenamefont {El}}]{suret2024soliton}%
  \BibitemOpen
  \bibfield  {author} {\bibinfo {author} {\bibfnamefont {P.}~\bibnamefont
  {Suret}}, \bibinfo {author} {\bibfnamefont {S.}~\bibnamefont {Randoux}},
  \bibinfo {author} {\bibfnamefont {A.}~\bibnamefont {Gelash}}, \bibinfo
  {author} {\bibfnamefont {D.}~\bibnamefont {Agafontsev}}, \bibinfo {author}
  {\bibfnamefont {B.}~\bibnamefont {Doyon}},\ and\ \bibinfo {author}
  {\bibfnamefont {G.}~\bibnamefont {El}},\ }\bibfield  {title} {\bibinfo
  {title} {{Soliton gas: Theory, numerics, and experiments}},\ }\href@noop {}
  {\bibfield  {journal} {\bibinfo  {journal} {Phys. Rev. E}\ }\textbf {\bibinfo
  {volume} {109}},\ \bibinfo {pages} {061001} (\bibinfo {year}
  {2024})}\BibitemShut {NoStop}%
\bibitem [{\citenamefont {Soto-Crespo}\ \emph {et~al.}(2016)\citenamefont
  {Soto-Crespo}, \citenamefont {Devine},\ and\ \citenamefont
  {Akhmediev}}]{soto2016integrable}%
  \BibitemOpen
  \bibfield  {author} {\bibinfo {author} {\bibfnamefont {J.~M.}\ \bibnamefont
  {Soto-Crespo}}, \bibinfo {author} {\bibfnamefont {N.}~\bibnamefont
  {Devine}},\ and\ \bibinfo {author} {\bibfnamefont {N.}~\bibnamefont
  {Akhmediev}},\ }\bibfield  {title} {\bibinfo {title} {{Integrable turbulence
  and rogue waves: breathers or solitons?}},\ }\href@noop {} {\bibfield
  {journal} {\bibinfo  {journal} {Phys. Rev. Lett.}\ }\textbf {\bibinfo
  {volume} {116}},\ \bibinfo {pages} {103901} (\bibinfo {year}
  {2016})}\BibitemShut {NoStop}%
\bibitem [{\citenamefont {Akhmediev}\ \emph {et~al.}(2016)\citenamefont
  {Akhmediev}, \citenamefont {Soto-Crespo},\ and\ \citenamefont
  {Devine}}]{akhmediev2016breather}%
  \BibitemOpen
  \bibfield  {author} {\bibinfo {author} {\bibfnamefont {N.}~\bibnamefont
  {Akhmediev}}, \bibinfo {author} {\bibfnamefont {J.~M.}\ \bibnamefont
  {Soto-Crespo}},\ and\ \bibinfo {author} {\bibfnamefont {N.}~\bibnamefont
  {Devine}},\ }\bibfield  {title} {\bibinfo {title} {{Breather turbulence
  versus soliton turbulence: Rogue waves, probability density functions, and
  spectral features}},\ }\href@noop {} {\bibfield  {journal} {\bibinfo
  {journal} {Phys. Rev. E}\ }\textbf {\bibinfo {volume} {94}},\ \bibinfo
  {pages} {022212} (\bibinfo {year} {2016})}\BibitemShut {NoStop}%
\bibitem [{\citenamefont {Grinevich}\ and\ \citenamefont
  {Santini}(2018{\natexlab{a}})}]{grinevich2018finite}%
  \BibitemOpen
  \bibfield  {author} {\bibinfo {author} {\bibfnamefont {P.~G.}\ \bibnamefont
  {Grinevich}}\ and\ \bibinfo {author} {\bibfnamefont {P.~M.}\ \bibnamefont
  {Santini}},\ }\bibfield  {title} {\bibinfo {title} {{The finite gap method
  and the analytic description of the exact rogue wave recurrence in the
  periodic NLS Cauchy problem. 1}},\ }\href@noop {} {\bibfield  {journal}
  {\bibinfo  {journal} {Nonlinearity}\ }\textbf {\bibinfo {volume} {31}},\
  \bibinfo {pages} {5258} (\bibinfo {year} {2018}{\natexlab{a}})}\BibitemShut
  {NoStop}%
\bibitem [{\citenamefont {Grinevich}\ and\ \citenamefont
  {Santini}(2018{\natexlab{b}})}]{grinevich2018exact}%
  \BibitemOpen
  \bibfield  {author} {\bibinfo {author} {\bibfnamefont {P.~G.}\ \bibnamefont
  {Grinevich}}\ and\ \bibinfo {author} {\bibfnamefont {P.~M.}\ \bibnamefont
  {Santini}},\ }\bibfield  {title} {\bibinfo {title} {{The exact rogue wave
  recurrence in the NLS periodic setting via matched asymptotic expansions, for
  1 and 2 unstable modes}},\ }\href@noop {} {\bibfield  {journal} {\bibinfo
  {journal} {Phys. Lett. A}\ }\textbf {\bibinfo {volume} {382}},\ \bibinfo
  {pages} {973} (\bibinfo {year} {2018}{\natexlab{b}})}\BibitemShut {NoStop}%
\bibitem [{\citenamefont {Grinevich}\ and\ \citenamefont
  {Santini}(2019)}]{grinevich2019finite}%
  \BibitemOpen
  \bibfield  {author} {\bibinfo {author} {\bibfnamefont {P.~G.}\ \bibnamefont
  {Grinevich}}\ and\ \bibinfo {author} {\bibfnamefont {P.~M.}\ \bibnamefont
  {Santini}},\ }\bibfield  {title} {\bibinfo {title} {{The finite-gap method
  and the periodic NLS Cauchy problem of anomalous waves for a finite number of
  unstable modes}},\ }\href@noop {} {\bibfield  {journal} {\bibinfo  {journal}
  {Russ. Math. Surv.}\ }\textbf {\bibinfo {volume} {74}},\ \bibinfo {pages}
  {211} (\bibinfo {year} {2019})}\BibitemShut {NoStop}%
\bibitem [{\citenamefont {Gelash}\ \emph {et~al.}(2019)\citenamefont {Gelash},
  \citenamefont {Agafontsev}, \citenamefont {Zakharov}, \citenamefont {El},
  \citenamefont {Randoux},\ and\ \citenamefont {Suret}}]{gelash2019bound}%
  \BibitemOpen
  \bibfield  {author} {\bibinfo {author} {\bibfnamefont {A.}~\bibnamefont
  {Gelash}}, \bibinfo {author} {\bibfnamefont {D.}~\bibnamefont {Agafontsev}},
  \bibinfo {author} {\bibfnamefont {V.}~\bibnamefont {Zakharov}}, \bibinfo
  {author} {\bibfnamefont {G.}~\bibnamefont {El}}, \bibinfo {author}
  {\bibfnamefont {S.}~\bibnamefont {Randoux}},\ and\ \bibinfo {author}
  {\bibfnamefont {P.}~\bibnamefont {Suret}},\ }\bibfield  {title} {\bibinfo
  {title} {{Bound state soliton gas dynamics underlying the noise-induced
  modulational instability}},\ }\href@noop {} {\bibfield  {journal} {\bibinfo
  {journal} {Phys. Rev. Lett.}\ }\textbf {\bibinfo {volume} {123}},\ \bibinfo
  {pages} {234102} (\bibinfo {year} {2019})}\BibitemShut {NoStop}%
\bibitem [{\citenamefont {Gelash}\ \emph {et~al.}(2021)\citenamefont {Gelash},
  \citenamefont {Agafontsev}, \citenamefont {Suret},\ and\ \citenamefont
  {Randoux}}]{gelash2021solitonic}%
  \BibitemOpen
  \bibfield  {author} {\bibinfo {author} {\bibfnamefont {A.}~\bibnamefont
  {Gelash}}, \bibinfo {author} {\bibfnamefont {D.}~\bibnamefont {Agafontsev}},
  \bibinfo {author} {\bibfnamefont {P.}~\bibnamefont {Suret}},\ and\ \bibinfo
  {author} {\bibfnamefont {S.}~\bibnamefont {Randoux}},\ }\bibfield  {title}
  {\bibinfo {title} {{Solitonic model of the condensate}},\ }\href@noop {}
  {\bibfield  {journal} {\bibinfo  {journal} {Phys. Rev. E}\ }\textbf {\bibinfo
  {volume} {104}},\ \bibinfo {pages} {044213} (\bibinfo {year}
  {2021})}\BibitemShut {NoStop}%
\bibitem [{\citenamefont {Agafontsev}\ \emph {et~al.}(2024)\citenamefont
  {Agafontsev}, \citenamefont {Gelash}, \citenamefont {Randoux},\ and\
  \citenamefont {Suret}}]{agafontsev2024multisoliton}%
  \BibitemOpen
  \bibfield  {author} {\bibinfo {author} {\bibfnamefont {D.}~\bibnamefont
  {Agafontsev}}, \bibinfo {author} {\bibfnamefont {A.}~\bibnamefont {Gelash}},
  \bibinfo {author} {\bibfnamefont {S.}~\bibnamefont {Randoux}},\ and\ \bibinfo
  {author} {\bibfnamefont {P.}~\bibnamefont {Suret}},\ }\bibfield  {title}
  {\bibinfo {title} {{Multisoliton interactions approximating the dynamics of
  breather solutions}},\ }\href@noop {} {\bibfield  {journal} {\bibinfo
  {journal} {Stud. Appl. Math.}\ }\textbf {\bibinfo {volume} {152}},\ \bibinfo
  {pages} {810} (\bibinfo {year} {2024})}\BibitemShut {NoStop}%
\bibitem [{\citenamefont {Agafontsev}\ \emph {et~al.}(2023)\citenamefont
  {Agafontsev}, \citenamefont {Gelash}, \citenamefont {Mullyadzhanov},\ and\
  \citenamefont {Zakharov}}]{agafontsev2023bound}%
  \BibitemOpen
  \bibfield  {author} {\bibinfo {author} {\bibfnamefont {D.~S.}\ \bibnamefont
  {Agafontsev}}, \bibinfo {author} {\bibfnamefont {A.~A.}\ \bibnamefont
  {Gelash}}, \bibinfo {author} {\bibfnamefont {R.~I.}\ \bibnamefont
  {Mullyadzhanov}},\ and\ \bibinfo {author} {\bibfnamefont {V.~E.}\
  \bibnamefont {Zakharov}},\ }\bibfield  {title} {\bibinfo {title}
  {{Bound-state soliton gas as a limit of adiabatically growing integrable
  turbulence}},\ }\href@noop {} {\bibfield  {journal} {\bibinfo  {journal}
  {Chaos Solitons Fractals}\ }\textbf {\bibinfo {volume} {166}},\ \bibinfo
  {pages} {112951} (\bibinfo {year} {2023})}\BibitemShut {NoStop}%
\bibitem [{\citenamefont {Gelash}\ and\ \citenamefont
  {Agafontsev}(2018)}]{gelash2018strongly}%
  \BibitemOpen
  \bibfield  {author} {\bibinfo {author} {\bibfnamefont {A.~A.}\ \bibnamefont
  {Gelash}}\ and\ \bibinfo {author} {\bibfnamefont {D.~S.}\ \bibnamefont
  {Agafontsev}},\ }\bibfield  {title} {\bibinfo {title} {{Strongly interacting
  soliton gas and formation of rogue waves}},\ }\href@noop {} {\bibfield
  {journal} {\bibinfo  {journal} {Phys. Rev. E}\ }\textbf {\bibinfo {volume}
  {98}},\ \bibinfo {pages} {042210} (\bibinfo {year} {2018})}\BibitemShut
  {NoStop}%
\bibitem [{\citenamefont {Randoux}\ \emph {et~al.}(2016)\citenamefont
  {Randoux}, \citenamefont {Suret},\ and\ \citenamefont
  {El}}]{randoux2016inverse}%
  \BibitemOpen
  \bibfield  {author} {\bibinfo {author} {\bibfnamefont {S.}~\bibnamefont
  {Randoux}}, \bibinfo {author} {\bibfnamefont {P.}~\bibnamefont {Suret}},\
  and\ \bibinfo {author} {\bibfnamefont {G.}~\bibnamefont {El}},\ }\bibfield
  {title} {\bibinfo {title} {{Inverse scattering transform analysis of rogue
  waves using local periodization procedure}},\ }\href@noop {} {\bibfield
  {journal} {\bibinfo  {journal} {Sci. Rep.}\ }\textbf {\bibinfo {volume} {6}}
  (\bibinfo {year} {2016})}\BibitemShut {NoStop}%
\bibitem [{\citenamefont {Agafontsev}\ and\ \citenamefont
  {Zakharov}(2015)}]{agafontsev2015integrable}%
  \BibitemOpen
  \bibfield  {author} {\bibinfo {author} {\bibfnamefont {D.~S.}\ \bibnamefont
  {Agafontsev}}\ and\ \bibinfo {author} {\bibfnamefont {V.~E.}\ \bibnamefont
  {Zakharov}},\ }\bibfield  {title} {\bibinfo {title} {{Integrable turbulence
  and formation of rogue waves}},\ }\href@noop {} {\bibfield  {journal}
  {\bibinfo  {journal} {Nonlinearity}\ }\textbf {\bibinfo {volume} {28}},\
  \bibinfo {pages} {2791} (\bibinfo {year} {2015})}\BibitemShut {NoStop}%
\bibitem [{\citenamefont {Kraych}\ \emph {et~al.}(2019)\citenamefont {Kraych},
  \citenamefont {Agafontsev}, \citenamefont {Randoux},\ and\ \citenamefont
  {Suret}}]{kraych2019statistical}%
  \BibitemOpen
  \bibfield  {author} {\bibinfo {author} {\bibfnamefont {A.~E.}\ \bibnamefont
  {Kraych}}, \bibinfo {author} {\bibfnamefont {D.}~\bibnamefont {Agafontsev}},
  \bibinfo {author} {\bibfnamefont {S.}~\bibnamefont {Randoux}},\ and\ \bibinfo
  {author} {\bibfnamefont {P.}~\bibnamefont {Suret}},\ }\bibfield  {title}
  {\bibinfo {title} {{Statistical properties of nonlinear stage of modulation
  instability in fiber optics}},\ }\href@noop {} {\bibfield  {journal}
  {\bibinfo  {journal} {Phys. Rev. Lett.}\ }\textbf {\bibinfo {volume} {123}},\
  \bibinfo {pages} {093902} (\bibinfo {year} {2019})}\BibitemShut {NoStop}%
\bibitem [{\citenamefont {Ablowitz}\ and\ \citenamefont
  {Segur}(1981)}]{ablowitz1981solitons}%
  \BibitemOpen
  \bibfield  {author} {\bibinfo {author} {\bibfnamefont {M.~J.}\ \bibnamefont
  {Ablowitz}}\ and\ \bibinfo {author} {\bibfnamefont {H.}~\bibnamefont
  {Segur}},\ }\href@noop {} {\emph {\bibinfo {title} {{Solitons and the inverse
  scattering transform}}}},\ Vol.~\bibinfo {volume} {4}\ (\bibinfo  {publisher}
  {SIAM, Philadelphia},\ \bibinfo {year} {1981})\BibitemShut {NoStop}%
\bibitem [{\citenamefont {Landau}\ and\ \citenamefont
  {Lifshitz}(1958)}]{landau1958quantum}%
  \BibitemOpen
  \bibfield  {author} {\bibinfo {author} {\bibfnamefont {L.~D.}\ \bibnamefont
  {Landau}}\ and\ \bibinfo {author} {\bibfnamefont {E.~M.}\ \bibnamefont
  {Lifshitz}},\ }\href@noop {} {\emph {\bibinfo {title} {{Quantum Mechanics:
  Non-relativistic Theory. V. 3 of Course of Theoretical Physics}}}}\ (\bibinfo
   {publisher} {Pergamon Press},\ \bibinfo {year} {1958})\BibitemShut {NoStop}%
\bibitem [{\citenamefont {Lewis}(1985)}]{lewis1985semiclassical}%
  \BibitemOpen
  \bibfield  {author} {\bibinfo {author} {\bibfnamefont {Z.}~\bibnamefont
  {Lewis}},\ }\bibfield  {title} {\bibinfo {title} {{Semiclassical solutions of
  the Zaharov-Shabat scattering problem for phase modulated potentials}},\
  }\href@noop {} {\bibfield  {journal} {\bibinfo  {journal} {Phys. Lett. A}\
  }\textbf {\bibinfo {volume} {112}},\ \bibinfo {pages} {99} (\bibinfo {year}
  {1985})}\BibitemShut {NoStop}%
\bibitem [{\citenamefont {Jenkins}\ and\ \citenamefont
  {McLaughlin}(2014)}]{jenkins2014semiclassical}%
  \BibitemOpen
  \bibfield  {author} {\bibinfo {author} {\bibfnamefont {R.}~\bibnamefont
  {Jenkins}}\ and\ \bibinfo {author} {\bibfnamefont {K.~D. T.-R.}\ \bibnamefont
  {McLaughlin}},\ }\bibfield  {title} {\bibinfo {title} {{Semiclassical limit
  of focusing NLS for a family of square barrier initial data}},\ }\href@noop
  {} {\bibfield  {journal} {\bibinfo  {journal} {Commun. Pure Appl. Math.}\
  }\textbf {\bibinfo {volume} {67}},\ \bibinfo {pages} {246} (\bibinfo {year}
  {2014})}\BibitemShut {NoStop}%
\bibitem [{\citenamefont {Matveev}\ and\ \citenamefont
  {Salle}(1991)}]{matveev1991darboux}%
  \BibitemOpen
  \bibfield  {author} {\bibinfo {author} {\bibfnamefont {V.~B.}\ \bibnamefont
  {Matveev}}\ and\ \bibinfo {author} {\bibfnamefont {M.~A.}\ \bibnamefont
  {Salle}},\ }\href@noop {} {\emph {\bibinfo {title} {{Darboux transformations
  and solitons}}}}\ (\bibinfo  {publisher} {Springer-Verlag, Berlin},\ \bibinfo
  {year} {1991})\BibitemShut {NoStop}%
\bibitem [{\citenamefont {Akhmediev}\ and\ \citenamefont
  {Mitzkevich}(1991)}]{akhmediev1991extremely}%
  \BibitemOpen
  \bibfield  {author} {\bibinfo {author} {\bibfnamefont {N.~N.}\ \bibnamefont
  {Akhmediev}}\ and\ \bibinfo {author} {\bibfnamefont {N.~V.}\ \bibnamefont
  {Mitzkevich}},\ }\bibfield  {title} {\bibinfo {title} {{Extremely high degree
  of N-soliton pulse compression in an optical fiber}},\ }\href@noop {}
  {\bibfield  {journal} {\bibinfo  {journal} {IEEE J. Quantum Electron.}\
  }\textbf {\bibinfo {volume} {27}},\ \bibinfo {pages} {849} (\bibinfo {year}
  {1991})}\BibitemShut {NoStop}%
\bibitem [{\citenamefont {Zakharov}\ and\ \citenamefont
  {Mikhailov}(1978)}]{zakharov1978relativistically}%
  \BibitemOpen
  \bibfield  {author} {\bibinfo {author} {\bibfnamefont {V.~E.}\ \bibnamefont
  {Zakharov}}\ and\ \bibinfo {author} {\bibfnamefont {A.~V.}\ \bibnamefont
  {Mikhailov}},\ }\bibfield  {title} {\bibinfo {title} {{Relativistically
  invariant two-dimensional models of field theory which are integrable by
  means of the inverse scattering problem method}},\ }\href@noop {} {\bibfield
  {journal} {\bibinfo  {journal} {Sov. Phys. JETP}\ }\textbf {\bibinfo {volume}
  {47}},\ \bibinfo {pages} {1017} (\bibinfo {year} {1978})}\BibitemShut
  {NoStop}%
\bibitem [{\citenamefont {Akhmediev}\ \emph
  {et~al.}(2009{\natexlab{b}})\citenamefont {Akhmediev}, \citenamefont
  {Soto-Crespo},\ and\ \citenamefont {Ankiewicz}}]{akhmediev2009extreme}%
  \BibitemOpen
  \bibfield  {author} {\bibinfo {author} {\bibfnamefont {N.}~\bibnamefont
  {Akhmediev}}, \bibinfo {author} {\bibfnamefont {J.~M.}\ \bibnamefont
  {Soto-Crespo}},\ and\ \bibinfo {author} {\bibfnamefont {A.}~\bibnamefont
  {Ankiewicz}},\ }\bibfield  {title} {\bibinfo {title} {{Extreme waves that
  appear from nowhere: on the nature of rogue waves}},\ }\href@noop {}
  {\bibfield  {journal} {\bibinfo  {journal} {Phys. Lett. A}\ }\textbf
  {\bibinfo {volume} {373}},\ \bibinfo {pages} {2137} (\bibinfo {year}
  {2009}{\natexlab{b}})}\BibitemShut {NoStop}%
\bibitem [{\citenamefont {Gelash}\ and\ \citenamefont
  {Zakharov}(2014)}]{gelash2014superregular}%
  \BibitemOpen
  \bibfield  {author} {\bibinfo {author} {\bibfnamefont {A.~A.}\ \bibnamefont
  {Gelash}}\ and\ \bibinfo {author} {\bibfnamefont {V.~E.}\ \bibnamefont
  {Zakharov}},\ }\bibfield  {title} {\bibinfo {title} {{Superregular solitonic
  solutions: a novel scenario for the nonlinear stage of modulation
  instability}},\ }\href@noop {} {\bibfield  {journal} {\bibinfo  {journal}
  {Nonlinearity}\ }\textbf {\bibinfo {volume} {27}},\ \bibinfo {pages} {R1}
  (\bibinfo {year} {2014})}\BibitemShut {NoStop}%
\bibitem [{\citenamefont {Tarasova}\ and\ \citenamefont
  {Slunyaev}(2023)}]{tarasova2023properties}%
  \BibitemOpen
  \bibfield  {author} {\bibinfo {author} {\bibfnamefont {T.~V.}\ \bibnamefont
  {Tarasova}}\ and\ \bibinfo {author} {\bibfnamefont {A.~V.}\ \bibnamefont
  {Slunyaev}},\ }\bibfield  {title} {\bibinfo {title} {{Properties of
  synchronous collisions of solitons in the Korteweg--de Vries equation}},\
  }\href@noop {} {\bibfield  {journal} {\bibinfo  {journal} {Commun. Nonlinear
  Sci. Numer. Simul.}\ }\textbf {\bibinfo {volume} {118}},\ \bibinfo {pages}
  {107048} (\bibinfo {year} {2023})}\BibitemShut {NoStop}%
\bibitem [{\citenamefont {Its}\ \emph {et~al.}(1988)\citenamefont {Its},
  \citenamefont {Rybin},\ and\ \citenamefont {Sall}}]{its1988exact}%
  \BibitemOpen
  \bibfield  {author} {\bibinfo {author} {\bibfnamefont {A.~R.}\ \bibnamefont
  {Its}}, \bibinfo {author} {\bibfnamefont {A.~V.}\ \bibnamefont {Rybin}},\
  and\ \bibinfo {author} {\bibfnamefont {M.~A.}\ \bibnamefont {Sall}},\
  }\bibfield  {title} {\bibinfo {title} {{Exact integration of nonlinear
  Schr{\"o}dinger equation}},\ }\href@noop {} {\bibfield  {journal} {\bibinfo
  {journal} {Theor. Math. Phys.}\ }\textbf {\bibinfo {volume} {74}},\ \bibinfo
  {pages} {20} (\bibinfo {year} {1988})}\BibitemShut {NoStop}%
\bibitem [{\citenamefont {Tajiri}\ and\ \citenamefont
  {Watanabe}(1998)}]{tajiri1998breather}%
  \BibitemOpen
  \bibfield  {author} {\bibinfo {author} {\bibfnamefont {M.}~\bibnamefont
  {Tajiri}}\ and\ \bibinfo {author} {\bibfnamefont {Y.}~\bibnamefont
  {Watanabe}},\ }\bibfield  {title} {\bibinfo {title} {{Breather solutions to
  the focusing nonlinear Schr{\"o}dinger equation}},\ }\href@noop {} {\bibfield
   {journal} {\bibinfo  {journal} {Phys. Rev. E}\ }\textbf {\bibinfo {volume}
  {57}},\ \bibinfo {pages} {3510} (\bibinfo {year} {1998})}\BibitemShut
  {NoStop}%
\bibitem [{\citenamefont {Xu}\ \emph {et~al.}(2019)\citenamefont {Xu},
  \citenamefont {Gelash}, \citenamefont {Chabchoub}, \citenamefont {Zakharov},\
  and\ \citenamefont {Kibler}}]{xu2019breather}%
  \BibitemOpen
  \bibfield  {author} {\bibinfo {author} {\bibfnamefont {G.}~\bibnamefont
  {Xu}}, \bibinfo {author} {\bibfnamefont {A.}~\bibnamefont {Gelash}}, \bibinfo
  {author} {\bibfnamefont {A.}~\bibnamefont {Chabchoub}}, \bibinfo {author}
  {\bibfnamefont {V.}~\bibnamefont {Zakharov}},\ and\ \bibinfo {author}
  {\bibfnamefont {B.}~\bibnamefont {Kibler}},\ }\bibfield  {title} {\bibinfo
  {title} {{Breather wave molecules}},\ }\href@noop {} {\bibfield  {journal}
  {\bibinfo  {journal} {Phys. Rev. Lett.}\ }\textbf {\bibinfo {volume} {122}},\
  \bibinfo {pages} {084101} (\bibinfo {year} {2019})}\BibitemShut {NoStop}%
\bibitem [{\citenamefont {Agafontsev}\ \emph {et~al.}(2015)\citenamefont
  {Agafontsev}, \citenamefont {Kuznetsov},\ and\ \citenamefont
  {Mailybaev}}]{agafontsev2015development}%
  \BibitemOpen
  \bibfield  {author} {\bibinfo {author} {\bibfnamefont {D.~S.}\ \bibnamefont
  {Agafontsev}}, \bibinfo {author} {\bibfnamefont {E.~A.}\ \bibnamefont
  {Kuznetsov}},\ and\ \bibinfo {author} {\bibfnamefont {A.~A.}\ \bibnamefont
  {Mailybaev}},\ }\bibfield  {title} {\bibinfo {title} {{Development of high
  vorticity structures in incompressible 3D Euler equations}},\ }\href@noop {}
  {\bibfield  {journal} {\bibinfo  {journal} {Phys. Fluids}\ }\textbf {\bibinfo
  {volume} {27}},\ \bibinfo {pages} {085102} (\bibinfo {year}
  {2015})}\BibitemShut {NoStop}%
\bibitem [{\citenamefont {Chiu}\ \emph {et~al.}(2013)\citenamefont {Chiu},
  \citenamefont {Stoyan}, \citenamefont {Kendall},\ and\ \citenamefont
  {Mecke}}]{chiu2013stochastic}%
  \BibitemOpen
  \bibfield  {author} {\bibinfo {author} {\bibfnamefont {S.~N.}\ \bibnamefont
  {Chiu}}, \bibinfo {author} {\bibfnamefont {D.}~\bibnamefont {Stoyan}},
  \bibinfo {author} {\bibfnamefont {W.~S.}\ \bibnamefont {Kendall}},\ and\
  \bibinfo {author} {\bibfnamefont {J.}~\bibnamefont {Mecke}},\ }\href@noop {}
  {\emph {\bibinfo {title} {{Stochastic geometry and its applications}}}}\
  (\bibinfo  {publisher} {John Wiley \& Sons},\ \bibinfo {year}
  {2013})\BibitemShut {NoStop}%
\bibitem [{\citenamefont {Agafontsev}\ and\ \citenamefont
  {Gelash}(2021)}]{agafontsev2021rogue}%
  \BibitemOpen
  \bibfield  {author} {\bibinfo {author} {\bibfnamefont {D.~S.}\ \bibnamefont
  {Agafontsev}}\ and\ \bibinfo {author} {\bibfnamefont {A.~A.}\ \bibnamefont
  {Gelash}},\ }\bibfield  {title} {\bibinfo {title} {{Rogue waves with rational
  profiles in unstable condensate and its solitonic model}},\ }\href@noop {}
  {\bibfield  {journal} {\bibinfo  {journal} {Front. Phys.}\ }\textbf {\bibinfo
  {volume} {9}},\ \bibinfo {pages} {610896} (\bibinfo {year}
  {2021})}\BibitemShut {NoStop}%
\bibitem [{\citenamefont {Bertola}\ and\ \citenamefont
  {Tovbis}(2017)}]{bertola2017maximal}%
  \BibitemOpen
  \bibfield  {author} {\bibinfo {author} {\bibfnamefont {M.}~\bibnamefont
  {Bertola}}\ and\ \bibinfo {author} {\bibfnamefont {A.}~\bibnamefont
  {Tovbis}},\ }\bibfield  {title} {\bibinfo {title} {{Maximal amplitudes of
  finite-gap solutions for the focusing nonlinear Schr{\"o}dinger equation}},\
  }\href@noop {} {\bibfield  {journal} {\bibinfo  {journal} {Commun. Math.
  Phys.}\ }\textbf {\bibinfo {volume} {354}},\ \bibinfo {pages} {525} (\bibinfo
  {year} {2017})}\BibitemShut {NoStop}%
\end{thebibliography}
\end{document}